\documentclass[twocolumn,english,aps,prx,floatfix,amssymb,superscriptaddress,longbibliography]{revtex4}
\usepackage[latin9]{inputenc}
\usepackage{verbatim}
\usepackage{float}
\usepackage{amsmath}
\usepackage{amssymb}
\usepackage{graphicx}
\usepackage{color}
\usepackage{xcolor}
\usepackage{tikz}
\usepackage[normalem]{ulem}
\newcommand{\ket}[1]{\ensuremath{\left| #1 \right>}}

\usepackage{bbold}

\usepackage[bookmarks=false,linkcolor=blue,urlcolor=blue,colorlinks,citecolor=blue]{hyperref}
\makeatletter

\newcommand{\be}{\begin{equation}}
\newcommand{\ee}{\end{equation}}
\newcommand{\bea}{\begin{eqnarray}}
\newcommand{\eea}{\end{eqnarray}}

\begin{document}

\title{Complete Boundary Phase Diagram of the Spin-$\frac{1}{2}$ XXZ Chain with Boundary Fields in the Anti-Ferromagnetic Gapped Regime}
\author{Parameshwar R. Pasnoori}
\thanks{These two authors contributed equally}
\email{pparmesh@umd.edu}
\affiliation{Department of Physics, University of Maryland, College Park, MD 20742, United
States of America}
\affiliation{Laboratory for Physical Sciences, 8050 Greenmead Dr, College Park, MD 20740,
United States of America}
\author{Yicheng Tang}
\thanks{These two autors contributed equally}
\email{yt291@physics.rutgers.edu}
\affiliation{Department of Physics and Astronomy, Center for Materials Theory, Rutgers University,
Piscataway, NJ 08854, United States of America}
\author{Junhyun Lee}
\affiliation{Department of Physics and Astronomy, Center for Materials Theory, Rutgers University,
Piscataway, NJ 08854, United States of America}
\author{J. H. Pixley}
\affiliation{Department of Physics and Astronomy, Center for Materials Theory, Rutgers University,
Piscataway, NJ 08854, United States of America}
\affiliation{Center for Computational Quantum Physics, Flatiron Institute, 162 5th Avenue, New York, NY 10010, USA}
\author{Patrick Azaria}
\affiliation{Laboratoire de Physique Th\'orique de la Mati\`ere Condens\'ee, Sorbonne Universit\'e and CNRS, 4 Place Jussieu, 75252 Paris, France}
\date{\today}
\author{Natan Andrei}
\affiliation{Department of Physics and Astronomy, Center for Materials Theory, Rutgers University,
Piscataway, NJ 08854, United States of America}

\begin{abstract}
We consider the spin $\frac{1}{2}$ XXZ chain with diagonal boundary fields and solve it exactly using Bethe ansatz in the gapped anti-ferromagnetic regime and obtain the complete phase boundary diagram. Depending on the values of the boundary fields, the system exhibits several phases which can be categorized based on the ground state exhibited by the system and also based on the number of bound states localized at the boundaries. We show that the Hilbert space is comprised of a certain number of towers whose number depends on the number of boundary bound states exhibited by the system. The system undergoes boundary phase transitions when boundary fields are varied across certain critical values. There exist two types of phase transitions. In the first type the ground state of the system undergoes a change. In the second type, named the `Eigenstate phase transition', the number of towers of the Hilbert space changes, which is again associated with the change in the number of boundary bound states exhibited by the system. We use the DMRG and exact diagonalization techniques to probe the signature of the Eigenstate phase transition and the ground state phase transition by analyzing the spin profiles in each eigenstate.   
\end{abstract}
\maketitle

\section{Introduction}

Symmetry is one of the most important aspects in physics. Traditionally various phases exhibited by a system were characterized based on whether a symmetry of the Hamiltonian or Lagrangian is respected by the ground state \cite{Landau1937}. If the ground state is not invariant under a certain symmetry exhibited by the Hamiltonian, then the system is said to exhibit spontaneous symmetry breaking. In the case of the classical Ising antiferromagnet, the system exhibits two degenerate Neel-ordered ground states corresponding to the two symmetry broken sectors. Adding `spin exchange' terms to the classical Ising antiferromagnet, one obtains the spin $1/2$ XXZ antiferromagnetic chain, which is one of the fundamental models describing quantum magnets. In the isotropic limit, it corresponds to the celebrated Heisenberg spin chain, which has been first solved by Bethe \cite{Bethe1931}. The solution was later extended to include anisotropy along the z-direction \cite{Orbach,Walker,YangYang1,YangYang2,YangYang3,BABELON}. It has been proposed to realize this system using ultra-cold atoms in optical lattices \cite{Demlerultracold} and using superconducting circuits in \cite{LarkinSC}, and recently it has been realized experimentally \cite{fu1,fu2}. In the gapped regime, it exhibits a discrete $\mathbb{Z}_2$ spin flip symmetry which is spontaneously broken \cite{Olav}, and in the thermodynamic limit, the system exhibits two degenerate symmetr -broken ground states \cite{Takahashi}. The Bethe ansatz method to include the boundaries was developed in \cite{Cherednik}\cite{Sklyanin}, using which the ground state and boundary excitations in various bulk phases exhibited by the XXZ spin chain were found in \cite{skorik,kapustinxxz,xxzbound2019,Nassar_1998,XXXpaper}. Recently new band structures in the spectrum at large anisotropies have been found \cite{Sharma}. There have been numerous studies in understanding the dynamics \cite{Wensho,Mitra,Pozsgay,Mesty,Yuzbashyan,Santos,Mallick} and recently, Bethe-Boltzmann hydrodynamic equations have been formulated \cite{Bertini,Collura,Moore} to understand the heat and spin transport. The system has been studied in the presence of impurities and also dephasing, where new phases are shown to emerge \cite{KondoXXX,Kattel_2023,kattel2024spin,Kattel2024}. In the presence of disorder, the XXZ spin chain exhibits many-body localization \cite{Prosen,Canovi,Elgart} and is argued to exhibit pairing in the spectrum at strong disorder \cite{Huse,Bauer,Vishwanath}.

Recently it was shown in \cite{Fendley} that the excitations built on top of the two symmetry broken ground states form two towers of eigenstates and there exists a strong zero energy operator which maps every eigenstate corresponding to one tower with a respective eigenstate corresponding to the other tower with corrections that vanish exponentially with the system size. It was recently shown in \cite{pasnooriXXZ} that in the low-energy sector the spin 1/2 XXZ chain exhibits a spin fractionalization, where the system hosts spin $S^z=\frac{1}{4}$ quantum numbers at the boundaries in the ground state and the mid-gap states, and generalized in \cite{kattelspinS} for the spin-S case.

In this work, we consider the system in the gapped regime and apply boundary magnetic fields which explicitly break the $\mathbb{Z}_2$ spin flip symmetry. We find that as the boundary magnetic fields are varied, the system exhibits several phases which are characterized either by the properties of the ground state exhibited by the system or by the number of boundary bound states localized at the edges, and if their energy is less than the mass gap (which is the minimum energy allowed for a spinon) or greater than the maximum energy of a single spinon $m$. Hilbert space is comprised of towers of excited states whose number depends on the number of bound states at the edges. There exist two types of phase transitions in the system- one in which the ground state of the system changes and the other one in which the ground state of the system does not change but the system undergoes a change in the number of towers of the Hilbert space, which is associated with the change in the number of boundary bound states. Using density matrix renormalization (DMRG) and exact diagonalization methods, we find a signature of this phase transition in the ground state by analyzing the spin profiles.

The paper is organized as follows: main results are summarized in section (\ref{sec:mainresults}). The numerical calculations are provided in the section (\ref{sec:numerical}). The Bethe ansatz equations are provided in the section (\ref{sec:betheequations}), and the description of the Bethe ansatz solution is provided in section (\ref{sec:summary}). The details of the Bethe ansatz calculation are provided in the appendix (\ref{appendix}). 

\section{Main results}
\label{sec:mainresults}

The Hamiltonian of the spin $\frac{1}{2}$ XXZ chain is given by
\bea
H=\sum_{j=1}^{N-1} \left[\sigma^x_j\sigma^x_{j+1}+\sigma^y_j\sigma^y_{j+1}+\Delta (\sigma^z_j\sigma^z_{j+1}-1)\right]\\+h_L\sigma^z_1+h_R\sigma^z_N \label{ch4hamiltonian}\eea

Here $\sigma^{\alpha}_j$, $\alpha=x,y,z$ are the Pauli matrices and $\Delta$ is the anisotropy along the $z$ direction. $h_L,h_R$ are the boundary magnetic fields. In the limit $\Delta\rightarrow\infty$, the system corresponds to the Ising antiferromagnet. We can introduce new parameters $\gamma$, $h_{c1}$, $h_{c2}$ such that

\bea \Delta=\cosh \gamma, \gamma>0,\;\;\;\;\; h_{c1}=\Delta-1, \;\;\;\;\; h_{c2}=\Delta+1. \eea

The Hamiltonian has a global $U(1)$ symmetry which corresponds to the conservation of the $z$-component of the total spin. In addition, in the absence of the boundary magnetic fields $h_{L}$ and $h_R$, the system has a global discrete $\mathbb{Z}_2$ symmetry, under the action of the global spin flip operator

\be
\tau=\prod_{j=1}^N \sigma^x_j,\;\; \tau^2=1,\label{gsf}
\ee
which flips all the spins in the system. Before discussing the ground state structure of the system with boundary magnetic fields, let us briefly discuss the system in the presence of periodic boundary conditions. 

\subsection{Periodic boundary conditions}
When $\Delta > 1$ the ground state
$|g\rangle$ displays antiferromagnetic order with non-zero staggered magnetization 

\be\label{smag} \sigma= \lim_{N\rightarrow \infty}N^{-1} \sum_{j=1}^N (-1)^j \langle g|\sigma_j^z |g\rangle.\ee 

The system exhibits two quasi-degenerate ground states for both odd and even number of sites chain. For an odd number of sites chain, the total $z$-component of the spin in the two ground states is $S^z=\pm \frac{1}{2}$ and can respectively be labeled by $\ket{\pm \frac{1}{2}}$. For a spin chain with even number of sites, both the ground states have total spin $S^z=0$, and can be labeled by $\ket{0}$ and $\ket{0'}$. Under the action of the global spin flip operator (\ref{gsf}) the ground states tranform into each other

\bea
\ket{\frac{1}{2}}=\tau \ket{-\frac{1}{2}}, \;\; \ket{0}=\tau \ket{0'}. \eea

On top of the ground states, the first excited state corresponds to adding two spinons, where each spinon carries spin $\pm\frac{1}{2}$. The energy of a spinon takes the following form 

\bea E_{\theta}=\sinh\gamma\sum_{\omega=-\infty}^{\infty}\frac{\cos(\theta\omega)}{\cosh(\gamma\omega)},\eea

where $\theta$ is the rapidity of the spinon. The energy of the spinon lies in the range 
$m<E_{\theta}<M$, where

\begin{align} m= \sinh\gamma\sum_{\omega=-\infty}^{\infty}\frac{(-1)^{\omega}}{\cosh(\gamma\omega)}, \\ M= \sinh\gamma\sum_{\omega=-\infty}^{\infty}\frac{1}{\cosh(\gamma\omega)}.\end{align}

Here $m$ which is the minimum energy of a single spinon is called the mass gap and corresponds to the rapidity $\theta=\pi$ of the spinon, and $M$, which is the highest energy of a single spinon that we refer to as \textit{band height} corresponds to the rapidity $\theta=0$ of the spinon. In the following, we describe the ground state structure in the presence of boundary magnetic fields. As we shall see, the ground state exhibited by the system depends on the values of the magnetic fields applied at the boundaries, and also depends on the number of sites in the spin chain being even or odd.

\subsection{Ground states: Odd number of sites}

 The ground state structure for an odd number of sites chain is depicted in (Fig \ref{ch4gsXXZodd}), where the axes represent the values of the boundary magnetic fields. In the region shown in yellow, which corresponds to the range of the values of the boundary magnetic fields: $h_R>|h_L|$ for  $h_L<0$, and $|h_R|<h_L$ for $h_R<0$ and the entire region of $h_L,h_R>0$, the ground state contains spin accumulation oriented in the negative $z$ direction localized at both edges that sums to $-\frac{1}{2}$, which is also equal to the total spin $S^z$ of the ground state.

In the region shown in green, which corresponds to the range of the values of the boundary magnetic fields: $h_R<|h_L|$  for $h_L<0$, and  $|h_R|>h_L$ for $h_R<0$ and the entire region of $h_L,h_R<0$, the ground state contains spin accumulation oriented in the positive $z$ direction localized at both edges that sums to $+\frac{1}{2}$, resulting in the ground state with total spin $S^z=\frac{1}{2}$.

In the region shown in red, where the boundary magnetic fields at the left and right edges point in the negative and positive $z$ directions respectively, and take absolute values greater than $h_{c2}$ (to be defined below), the spin orientation at both edges in the ground state is opposite to the magnetic fields, and sums to zero. There exists a spinon whose spin orientation can point either be in the positive or negative $z$ direction, which leads to a two fold degenerate ground state with total spin $S^z=\pm\frac{1}{2}$. 

Similarly, in the region shown in blue, where the boundary magnetic fields at the left and right edges point in the positive and negative $z$ directions respectively, and take absolute values greater than $h_{c2}$, the spin orientation at both edges is opposite to the magnetic fields and sums to zero. There exists a spinon whose spin orientation can point either be in the positive or negative $z$ direction, which leads to a two fold degenerate ground state with total spin $S^z=\pm\frac{1}{2}$. 

 \subsection{Ground states: Even number of sites}
\begin{center}
\begin{figure}[!h]
\includegraphics[width=1\columnwidth]{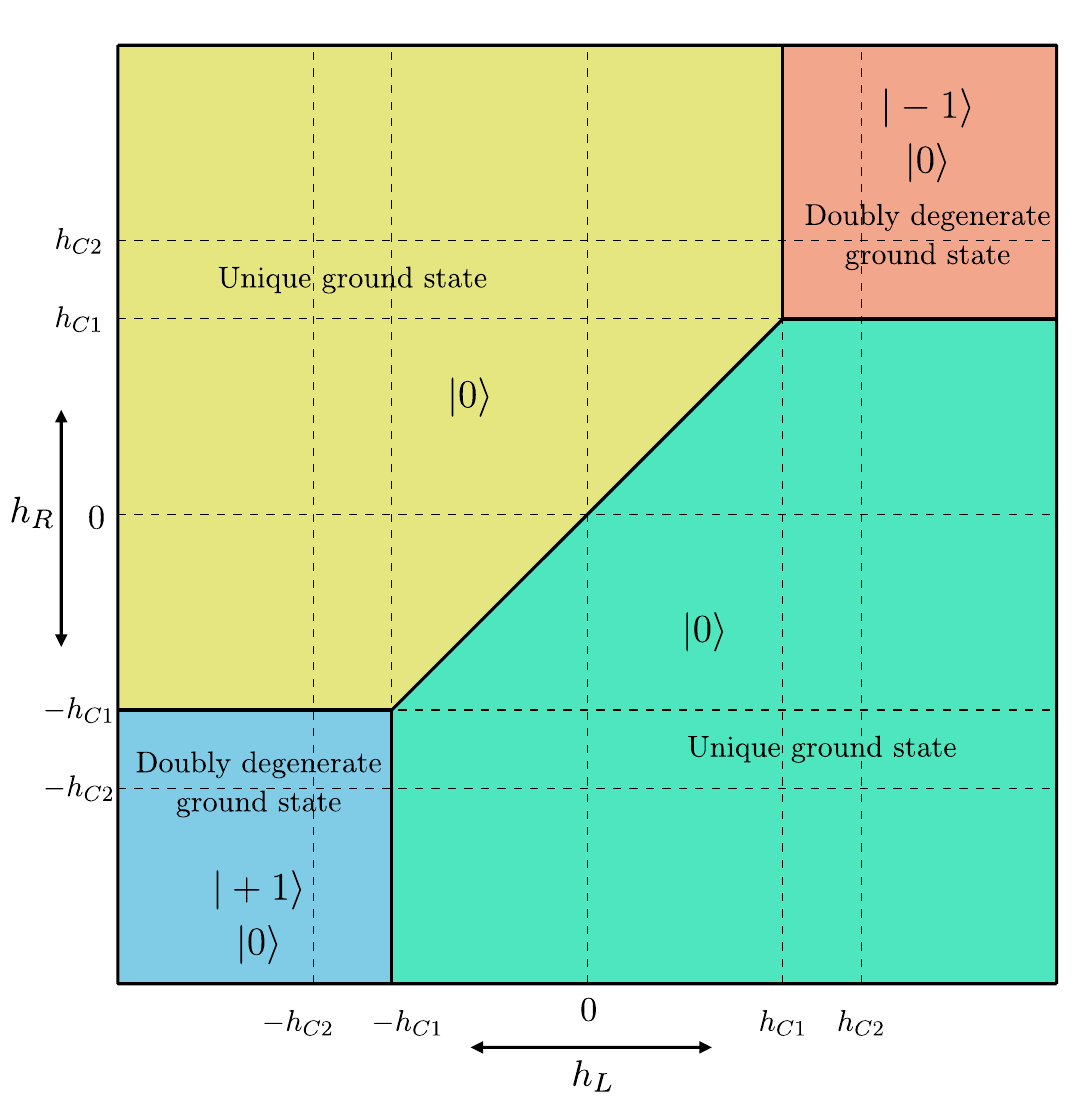}
 \caption{The figure shows the ground states exhibited by the even number of sites spin chain for different values of the boundary magnetic fields. In the red region the spin accumulation at both the left and right edges is along the negative $z$ direction, and it contains a spinon with spin orientation either in the positive or negative $z$ direction giving rise to a two fold degenerate ground state. In the blue region the spin accumulation at both the left and right edges is along the positive $z$ direction, and it contains a spinon with spin orientation either in the positive or negative $z$ direction giving rise to a two fold degenerate ground state. In contrast to the above regions, the green and yellow regions have no spinons in the ground state. In the yellow region the spin accumulation at the left and right edges is along the positive and negative $z$ directions respectively but equal in magnitude, and hence the ground state has $S^z=0$. In the green region the spin accumulation at the left and right edges is along the negative and positive $z$ directions respectively but equal in magnitude, and hence the ground state has $S^z=0$.}
\label{ch4gsXXZeven}
\end{figure}
\end{center}

The ground state structure for even number of sites chain is shown in (Fig \ref{ch4gsXXZeven}).  In the region shown in yellow, which corresponds to the range of the values of the boundary magnetic fields: $h_R>h_L$, for $h_L>0$, and $|h_R|<h_L$ for $h_L<0$ and the entire region where $h_L<0, h_R>0$, the ground state exhibits spin accumulation localized at both edges. The spin accumulations are oriented in the positive and negative $z$ directions at the left and right edges respectively, and sum to zero, resulting in the ground state with total spin $S^z=0$.

Similarly, in the region shown in green, which corresponds to the range of the values of the boundary magnetic fields: $h_R<h_L$ for $h_L>0$, and $|h_R|>h_L$ for $h_L<0$ and the entire region where $h_L>0, h_R<0$, the ground state exhibits spin accumulation localized at both edges. These spins accumulations are oriented in the positive and negative $z$ directions at the right and left edges respectively, and sum to zero, resulting in the ground state with total spin $S^z=0$. Note that the orientation of the spin accumulations at each boundary in this region is exactly equal and opposite to that in the region depicted by yellow.

In the region shown in red, where both the boundary magnetic fields point in positive $z$ direction and take absolute values greater than $h_{c2}$, the ground state exhibits spin accumulation localized at both edges which is in the negative $z$ direction, and sums to $-\frac{1}{2}$. There exists a spinon whose spin orientation can point in the positive or negative $z$ direction, leading to a two-fold degenerate ground state with total spin $S^z=-1,0$.

Similarly, in the region shown in blue, where both boundary magnetic fields point in the negative $z$ direction and take absolute values greater than $h_{c2}$, the ground state exhibits spin accumulation localized at both edges, which is in the positive $z$ direction, and sums to $+\frac{1}{2}$. There exists a spinon whose spin orientation can point in the positive or negative $z$ direction, leading to a two-fold degenerate ground state with total spin $S^z=+1,0$.
 
\subsection{Bound state structure}

\begin{center}
\begin{figure}[!h]
\includegraphics[width=1\columnwidth]{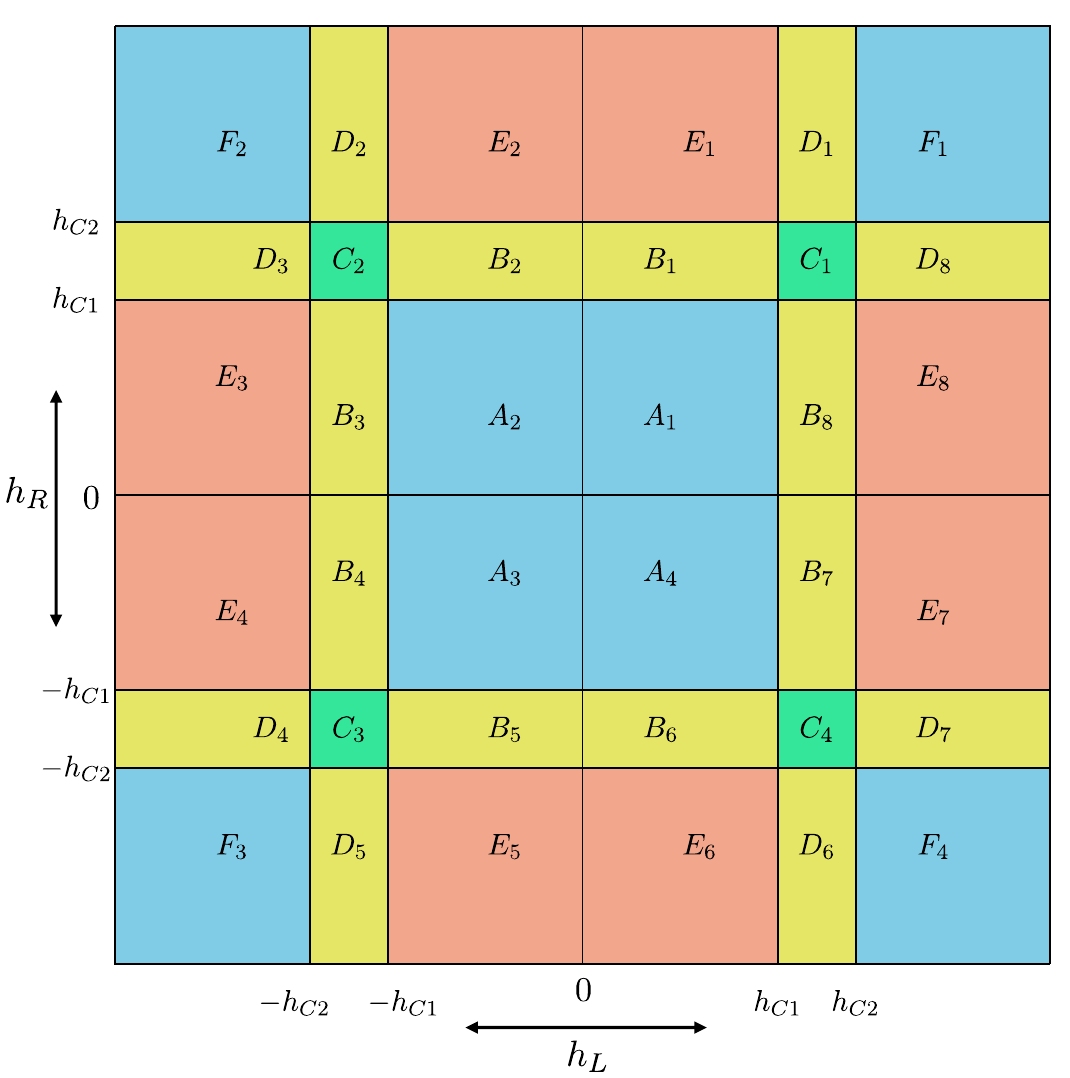}
 \caption{Qualitative phase diagram of the gapped XXZ spin $\frac{1}{2}$ chain. The phase diagram is divided into 36 regions depending on the values of the boundary magnetic fields as shown in the figure. In regions $A_i$, $E_i$ and $F_i$ both the boundaries contain bound states, in the regions $B_i$, $D_i$ only one of the boundaries contains a bound state and in the regions $C_i$ none of the boundaries contain bound states. The energy of the bound states is less than the mass gap in regions $A_i, B_i$, whereas it is above the gap in regions $F_i, D_i$.  The energy of one of the bound states is less than the mass gap and the other is above the band in the regions $E_i$.}
\label{ch4pd1}
\end{figure}
\end{center}

  The phase diagram can be divided into several regimes based on the values of the boundary magnetic fields as shown in (Fig \ref{ch4pd1}). There exist two bound states in the phases labeled by $A,E$ and $F$. In the phases labeled by $A$, both bound states have energy less than $m$, which is the mass gap, whereas in the phases labeled by $F$, both bound states have energy greater than $M$, which is the band height. In the phases labeled by $E$, one of the bound states has energy less than $m$ whereas the other bound state has energy greater than $M$. There exist on an average, spin $\frac{1}{4}$ exponentially localized at both edges. Each edge may contain spin $\pm \frac{1}{4}$ and hence there exist four states corresponding to four combinations of the $\pm\frac{1}{4}$ spins, and correspond to the four states- one without bound states, two with a bound state at either the left or the right edge, and one state with bound states at both the edges. In addition to these bound states at the edges, one can construct excited states in the bulk, and the Hilbert space is comprised of four towers, each corresponding to a certain combination of the spin $\pm \frac{1}{4}$ at the edges in the respective lowest energy state. As discussed above, each of these phases may be further classified into different sub-phases based on the ground state of the system. As the direction and magnitude of the boundary magnetic fields is changed, the spin accumulation at the edges varies, which results in the system exhibiting a different ground state. The phases labeled by $E$, which exhibit the same ground state, can be further classified based on which edge contains the bound state whose energy is less than the mass gap of the bulk.
  
  \begin{center}
\begin{figure}[!h]
\includegraphics[width=1\columnwidth]{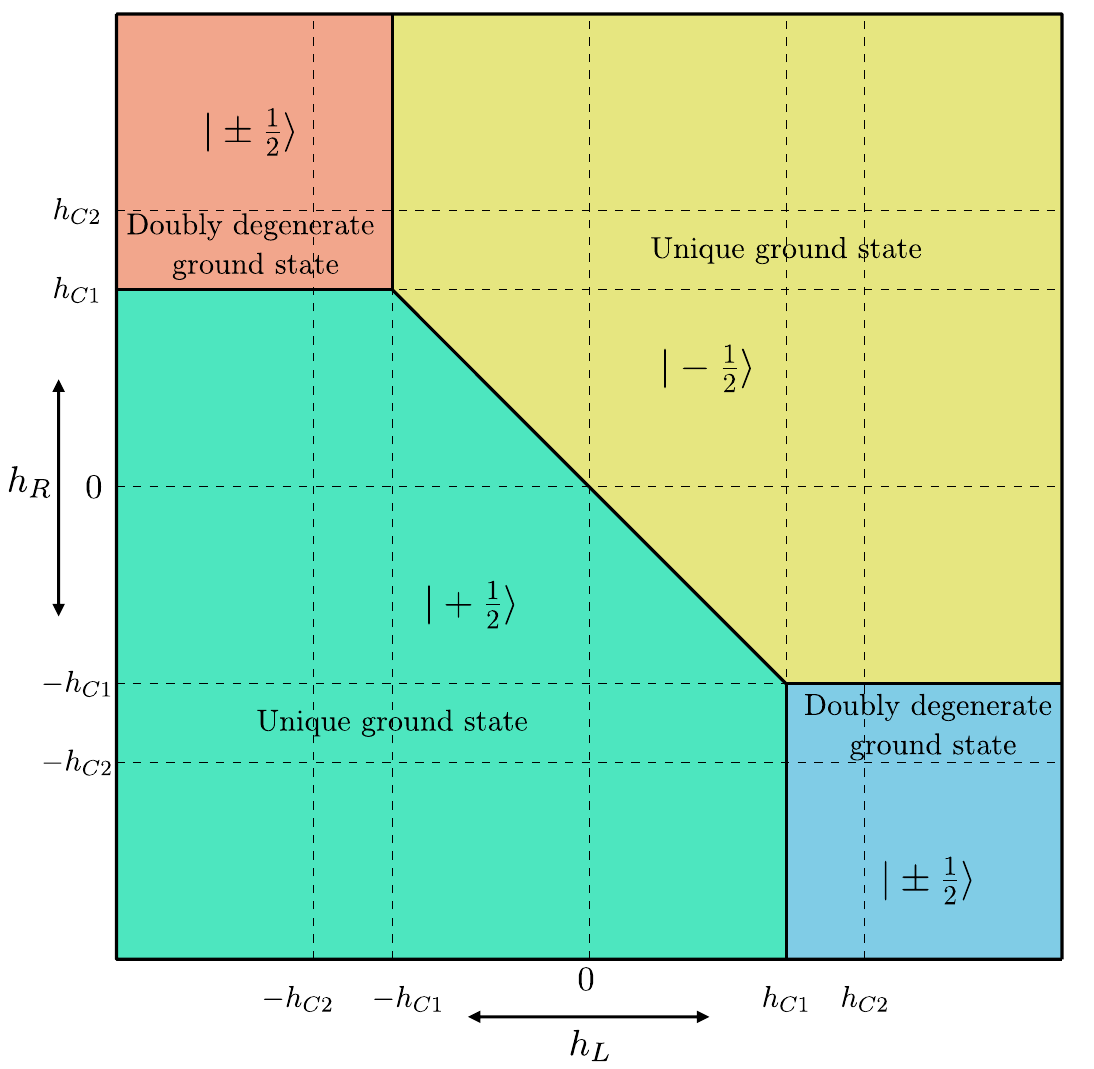}
 \caption{The figure shows the ground state exhibited by the odd number of sites spin chain for different values of the boundary magnetic fields. The ground states in the red and blue regions have the equal values of $S^z$ corresponding to the spin orientation of the spinon, but they differ in the orientation of spin accumulation at the edges, which is along the negative and positive $z$ direction at the left and right edges in the blue region and along the positive and negative $z$ direction at the left and right edges in the red region. The green and yellow regions have no spinons in the ground state and the difference in the spin accumulation at the edges gives rise to different values of $S^z$. The spin accumulation at both the edges is along the negative and positive $z$ directions in the yellow and green regions respectively.}
\label{ch4gsXXZodd}
\end{figure}
\end{center}

  In $B,D$ phases there exists only one bound state either at the left or the right edge and it has energy lesser than $m$ and greater than $M$ in $B$ and $D$ phases respectively. Since there exits only one bound state, the complete set of spin $\pm \frac{1}{4}$ does not exist at the edges and the Hilbert space in these phases is comprised of only two towers. Just as in the previously discussed phases, $B$ and $D$ phases can be further classified into different sub-phases based on the ground state. There exist more than one sub-phase in $B$ and $D$ phases which exhibit the same ground state but they differ in whether the bound state exists at the left edge or the right edge. In the $C$ phases there exists no bound states and the Hilbert space is comprised of a single tower.  The $C$ phase can be further divided into four sub-phases which exhibit different ground states. 
  
 There exists eigenstate phase transitions separating the phases named with a different alphabet, where the number of towers of the Hilbert space changes. Across these phase transitions, even though the total spin of the ground state remains the same as mentioned above, we find using DMRG that the spin profile undergoes a significant change. This will be discussed in detail in the next section. Within each of these phases labeled by an alphabet, as discussed above, there exists sub-phases which may exhibit different ground states, and the phase transitions separating these sub-phases are first order phase transitions involving level crossings. For even number of sites chain, when both the magnetic fields have equal values, the ground state is two fold degenerate in the $A_1, A_3$ phases with spin $+\frac{1}{4},-\frac{1}{4}$ and $\frac{1}{4},+\frac{1}{4}$ at the edges and form two towers of degenerate eigenstates. For odd number of sites chain, when both the magnetic fields have equal and opposite values, the ground state is two fold degenerate in the $A_2, A_4$ phases with spin $+\frac{1}{4},+\frac{1}{4}$ and $-\frac{1}{4},-\frac{1}{4}$ at the edges and form two towers of degenerate eigenstates. 

 The fractional spin $\pm \frac{1}{4}$ mentioned above are average expectation values of the spin $S^z$ at the boundaries. In order for these to be quantum observables, the associated variance has to vanish \cite{pasnooriXXZ,JackiwRebbi}. In the following section, we use DMRG to obtain the spin profiles and calculate the variance associated with the boundary spins.


\section{Numerical Calculation}
\label{sec:numerical}

We use the density-matrix renormalization group (DMRG) to obtain the ground state of the XXZ spin chain in the presence of boundary fields with open boundary conditions [Eq.~\eqref{ch4hamiltonian}]. 
%
One central quantity we are interested in this work is the local magnetization $S^z_i = \frac{1}{2}\langle \sigma^z_i \rangle$. 
By analyzing the magnetization, we observe the spin fractionalization at the boundary and confirm that the spin configuration is as expected from the Bethe Ansatz calculation. 
The DMRG calculations in this paper are performed using the TeNPy library \cite{tenpy2024},
with a truncation error up to $\sim10^{-10}$ from the maximum bond dimension $\chi_{max} = 200$. Before computing  the local magnetization  let us benchmark our numerical calculations against exact results from the Bethe Ansatz.

\subsection{Boundary States}
\begin{figure*}
    \centering
    \includegraphics[width= \linewidth]{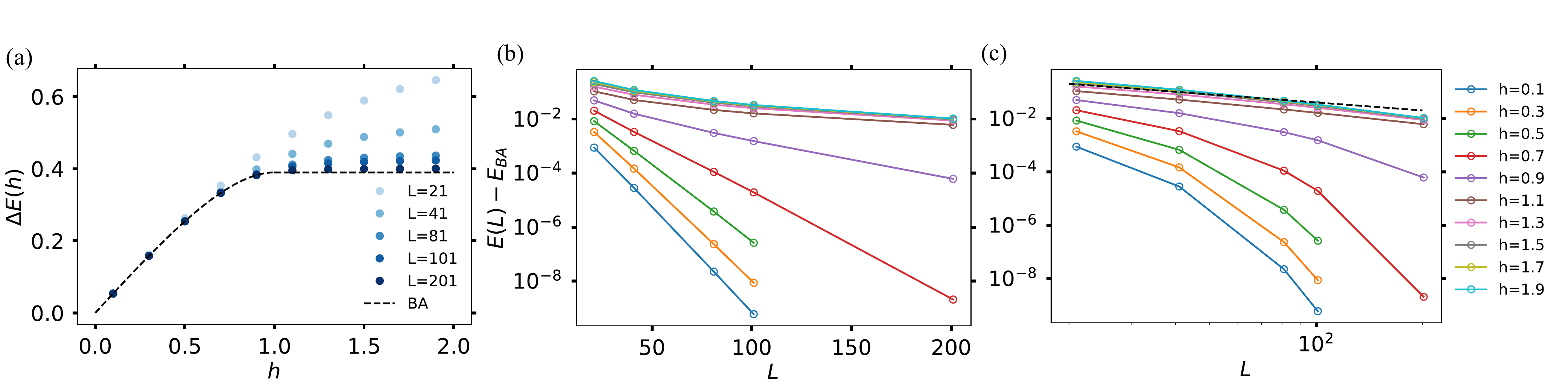}
    \caption{(a) DMRG calculation comparing with Bethe Ansatz calculation for the energy gap $\Delta E$ at $\Delta =2$ with different system size $L$ with field configuration $h_0=h_L=h$. , we show the difference between the finite DMRG result and the BA result for various boundary fields (h), indicating (b) when $h<h_{c_1}$, the first excited state contains bound states exponentially localized at the boundaries and hence has the finite energy scaling $E(L)-E_{BA}\sim e^{-mL}$ and (c) when $h>h_{c_1}$, the first excited state contains a spinon excitation, and hence has the finite energy scaling $E(L)-E_{BA}\sim L^{-1}$, with dashed line being $4L^{-1}$.}
    \label{energyyy}
\end{figure*}

To this end, we consider the boundary bound state energy whose
exact  expression  as a function of boundary field $h$ is  obtained from Bethe Ansatz (see Eq.(\ref{ch4boundenA1}) below), which can be written as
\begin{equation}
    E_B(h) = \sum_{n=0}^\infty 2(-1)^n \sinh(\gamma an)e^{-\gamma n}/\cosh(\gamma n),
    \label{eq:bound}
\end{equation}
here $\gamma=\cosh^{-1}(\Delta)$ and $h$ is implicit in $a$ by the relation $h=\sinh(g)\tanh(\frac{ga}{2})$. As mentioned above, when $h<h_{C1}$, $E_B$ is less than the mass gap.
However, when $h>h_{C1}$, the boundary bound state energy becomes larger than that of the mass gap and merges with the continuum of excited states.
With the field configuration,  $h_L=h_R = h$ and an odd number of sites, we computed the ground state energy and the energy of the first excited state which hosts two boundary bound states, one at each edge of the chain. As mentioned above, these two states have total spin $S^z=\pm 1/2$ depending on the sign of $h$. When performing the DMRG calculation, we can restrict the system to be in a certain total spin sector, and thus by calculating the energy difference of the ground states in the $\pm 1/2$ sectors we numerically obtain the energy of the boundary bound state.
The results are shown in Fig.~\ref{energyyy} (dots), and match with the analytic calculation (Eq.~\eqref{eq:bound}, solid line) very well. Note that $h_{C1}=1$ for $\Delta = 2$, where the analytic line ends.

\subsection{Edge Spin Fractionalization}

We calculate the boundary spin accumulation that arises on each edge due to the presence of the boundary fields. We define the spin accumulation at the left boundary as~\cite{Jackiw,XXXpaper}:
\begin{equation}
    S^z \equiv \lim_{\alpha\to0}\lim_{L\to\infty}\langle S^z(L,\alpha)\rangle =\sum_{x_i<L} \left\langle S^z(x_i) \right\rangle e^{-\alpha x_i},
    \label{eq:sz}
\end{equation}
where $\alpha$ is the cutoff scale.
Notice the order of limit is important, and one should take the thermodynamic limit first.
Otherwise, $\alpha\to 0$ limit will remove the cutoff and the result becomes merely the sum of local $S^z(x_i)$'s of the system. 
In other words, to make the cutoff to be meaningful, it must satisfy $\alpha \gg L^{-1}$. 

In Fig.~\ref{fig:Sz} we show $S^z(\alpha) = \lim_{L\to\infty}\langle S^z(L,\alpha)\rangle$ for $\Delta=$5, which is in the gapped phase of the XXZ model. 
The three panels are for different boundary fields $h \equiv h_L = h_R$ (which corresponds to $A$, $C$, and $F$ phases, respectively) and we plot four different system sizes. 
To obtain an estimate of the 
spin accumulation in the thermodynamic limit, we extrapolate our results to $L\rightarrow \infty$ by computing the ground state across a large range of system sizes. We find that the data converges in system size for $L \gg 1/\alpha$
to a straight line and deviates strongly from this in the opposite regime $L \ll 1/\alpha$ due to the finite size of the system. Based on the converged results we extrapolate this to the thermodynamic limit by extending it all the way to $\alpha =0$.
From this analysis, we confirm the Bethe ansatz prediction of fractionalized $1/4$-spin for all phases: $h \ge h_{c2}$ ($F$ phase), $h \le h_{c1}$ ($A$ phase) and $h_{c1} \le h \le h_{c2}$ ($C$ phase). 

\begin{figure*}
    \centering
    \includegraphics[width=\linewidth]{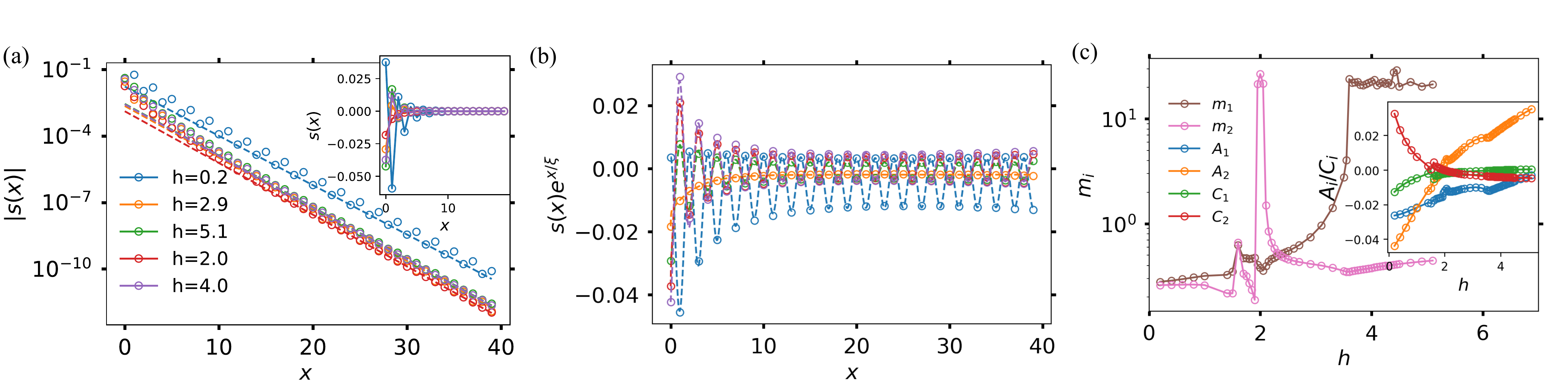}
    \caption{Fitting the spin profile for $\Delta=3$ with various boundary fields $h_L=h_R=h$ and system size $L=401$. (a) We show the boundary deviation from the anti-ferromagnetic bulk decays exponentially $s(x)$ in the inset and its absolute value in the linear-log scale to illustrate $|s(x)|\sim e^{-x/\xi}$ for $x\gg 1$. (b) Sharing the same label as (a), we fit the boundary deviation of the spin profile with the ansatz proposed in Eq.\ref{fitansatz}. (c) Showing different fitting parameters as a function of boundary field $h$, $m_i=m_1,m_2$ and $A_i=A_1,A_2$, $C_i=C_1,C_2$ in the inset.}.
    \label{VD}%
\end{figure*}

We now analyze the boundary spin profile.   Fig.\ref{VD} (a) illustrates the edge spin accumulation after subtracting off the bulk antiferromagnetic order \begin{equation}
    s(x_i) \equiv \langle S^z(x_i)\rangle -\sigma(-1)^{x_i}.
\end{equation}
This quantity shows the boundary effect decays exponentially into the bulk. On top of the exponential decay, we found an ansatz that fits the boundary oscillation well, which takes the form
\begin{equation}
\begin{split}
\label{fitansatz}
    &s(x_i)=g(x_i)e^{-x_i/\xi}
    \\& g(x_i)=(A_1e^{-m_1x_i}+C_1)+(-1)^x(A_2e^{-m_2x_i}+C_2).
\end{split}
\end{equation}
In Fig.\ref{VD} (b,c) we show the fitting of $g(x_i)$ as well as the corresponding fitting parameters for different boundary fields.

\begin{figure*}
    \centering
    \includegraphics[width=\linewidth]{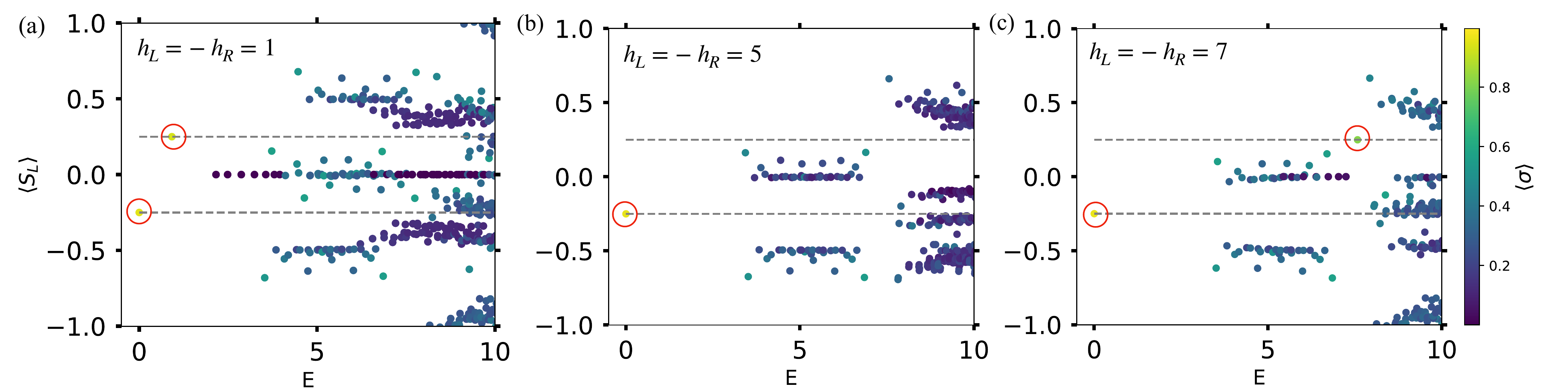}
    \caption{Exact diagonalization calculation for the manybody spectrum with the expectation value for edge spin operator $\langle S_L\rangle$ and the Neel order parameter $\langle S^z \rangle$ for $\Delta = 5$ and (a) $h_0=-h_L=1$, (b) $h_0=-h_L=5$, (c) $h_0=-h_L=7$. The two states with spin $1/4$ at the boundary are circled in red.}
    \label{ED}
\end{figure*}

To demonstrate that the system has two critical boundary fields, we use exact diagonalization (ED) to compute the full spectrum and detect the bound state and confirm the BA results. In Fig.\ref{ED}, we show that when $h_{C_1}<h<h_{C_2}$, only the ground state has a $1/4$  edge spin accumulation. When $h<h_{C_1}$, the first excited state and the ground state have $1/4$  edge spin accumulation. When $h>h_{C_2}$, both the ground state and the state which hosts bound states with energy higher than the band height have $1/4$ edge spin accumulation.

\subsection{Spin Variance}

To verify that the edge spin operator defined in Eq.\ref{eq:sz} is a sharp quantum operator in the ground state of our system, we calculate its variance and show that it vanishes in the thermodynamic limit.
We define the spin variance at the edge for a finite system and cutoff as 
\begin{equation}
    \delta S^2(L,\alpha) = \langle S^z(L,\alpha)^2\rangle -\langle S^z(L,\alpha)\rangle^2.
\end{equation}
The thermodynamic limit of the spin variance is defined through the same limit as in Eq.~\eqref{eq:sz}, and we are going to show that this quantity vanishes:

\begin{equation}
    \delta S^2 \equiv \lim_{\alpha\to\infty}\lim_{L\to\infty}\delta S^2(L,\alpha)=0.
\end{equation}

Taking the $L\rightarrow\infty$ is challenging, and we circumvent this issue by assuming an ansatz relating $\delta S^2(L,\alpha)$ and $\delta S^2(\infty,\alpha)$ \cite{XXXpaper}.

\begin{figure*}
    \includegraphics[width=\linewidth]{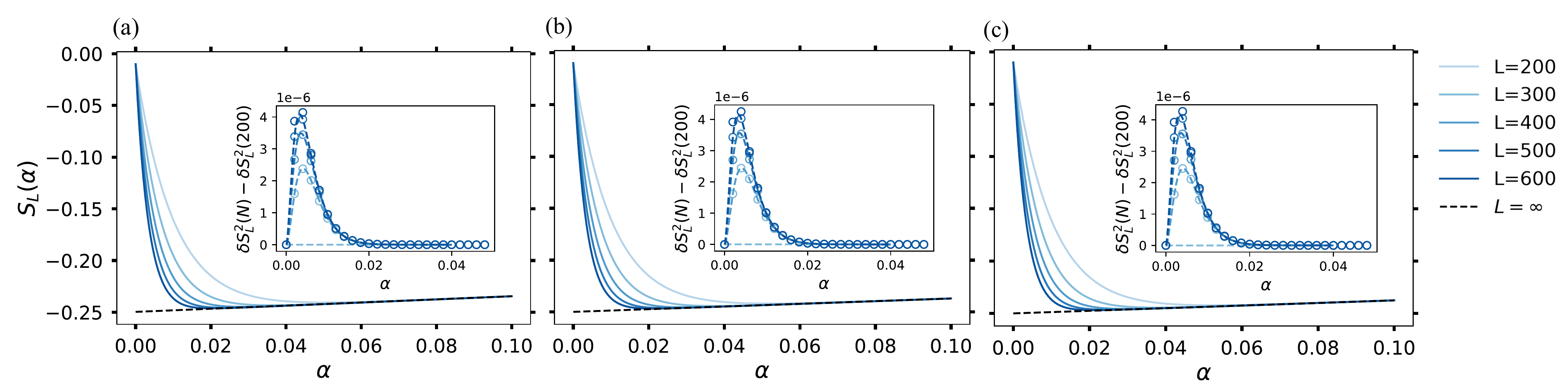}
\caption{Spin accumulation $S^z(\alpha)$ and Spin variance $\delta S^2(L,\alpha)$ in insets for anisotropy $\Delta=5$ and boundary field (a) $h_L=-h_R=0.2$, (b) $h_L=-h_R=5$, (c)  $h_L=-h_R=10$ with different system size $L$. 
All data shows that the edge spin accumulates to $\frac{1}{4}$ and variance vanishes in the scaling limit $\alpha\to 0$. }%
\label{fig:Sz}%
\end{figure*}

As we are focusing on the static properties of the ground state we utilize the Ornstein-Zernicke form of a two-dimensional  correlation function to construct the ansatz for the corrections of the variance of the spin correlations
\begin{equation}
     \delta S^2(L,\alpha)= \delta S^2(\infty,\alpha)-\frac{A}{\Delta}\alpha e^{-B\alpha L}.
\end{equation}
Then, $\delta S^2 = \lim_{\alpha\rightarrow0}(\infty,\alpha) = S^2(L,0)$ which allows us to calculate $\delta S^2$ in the thermodynamic limit. 
We verify this ansatz by taking the difference of $\delta S^2(L,\alpha)$ for different $L$ as shown in Fig.~\ref{fig:Sz}.
Therefore, in the thermodynamic limit, the variance does vanish $ \delta S^2=0$, and the ground state is an eigenstate of the boundary operator $\mathcal{S}^z$. 
The fitted parameter $B\approx 2$ is nearly independent of the boundary field, while $A$ takes a non-universal value.
    

\section{Bethe equations}
\label{sec:betheequations}

 The Bethe equations can be obtained by following the method of coordinate or algebraic Bethe ansatz \cite{Sklyanin,Alcaraz,kapustinxxz,xxzbound2019}. One obtains the following Bethe equations for reference state with all spin up \footnote{We use the notation of \cite{xxzbound2019})}

\begin{widetext}

\bea\label{ch4be1}
\left(\frac{\sin\frac{1}{2}(\lambda_j-i\gamma)}{\sin\frac{1}{2}(\lambda_j+i\gamma)}\right)^{2N}\prod_{\alpha}^{L,R}\left(\frac{\sin\frac{1}{2}(\lambda_j+i\gamma(1+\epsilon_{\alpha}))}{\sin\frac{1}{2}(\lambda_j+i\gamma(1+\epsilon_{\alpha}))}\right)=\prod_{\sigma=\pm}\prod_{k=1}^M\left(\frac{\sin\frac{1}{2}(\lambda_j+\sigma\lambda_k-2i\gamma)}{\sin\frac{1}{2}(\lambda_j+\sigma\lambda_k+2i\gamma)} \right)
\eea

where \bea h_\alpha= -\sinh\gamma\coth(\frac{\epsilon_{\alpha}\gamma}{2}), \;\;\;\;\;\;\;\;\;   \epsilon_\alpha=\tilde{\epsilon}_{\alpha}+i\delta_\alpha \frac{\pi}{\gamma},\;\;\;\;\;\;\;\; \delta_{\alpha} =\begin{cases}  \gamma & |h_{\alpha}|<\sinh\gamma \\ 0 & |h_{\alpha}|>\sinh\gamma \end{cases} \eea

\end{widetext}

 Note that $h_{c1}<\sinh\gamma<h_{c2}$. The Bethe equations for reference state with all spin down can be obtained by the transformation $\epsilon_{\alpha}\rightarrow -\epsilon_{\alpha}$ \cite{ODBA}. The energy of a state described by the set of Bethe roots $\lambda_j$ is given by

\bea\nonumber E=\frac{1}{2}\left[(N-1)\cosh\gamma+h_L+h_R\right]\\-2\sinh\gamma\sum_{j=1}^M\frac{\sinh\gamma}{\cosh\gamma-\cos\lambda_j} .\label{ch4energy}\eea

The boundary magnetic fields break the $\mathbb{Z}_2$ spin flip symmetry. Under the spin flip of all the sites, the bulk remains invariant but the boundary terms remain invariant only after the direction of both the magnetic fields is reversed, hence we have the following isometry 
\bea \label{ch4z2} \prod_{i=1}^{N}\sigma^{x}_i H \sigma^{x}_i, \;\;h_L\rightarrow-h_L, \;\; h_R\rightarrow-h_R.\eea


The detailed solution to the Bethe equations is provided in the Appendix. In the following section, we describe these results.

\section{Summary of the Bethe ansatz solution}
\label{sec:summary}
\subsection{$\it{A}$ phases}
\label{ch4sec:Aphase}
We start with the $A$ phases where two boundary bound-states are stabilized.
The four $A_{j=(1,2,3,4)}$ sub-phases  corresponds to the domains of boundary fields 
 $(h_{L}\le h_{c1},h_{R}\le h_{c1}), (h_{L} \ge -h_{c1},h_{R}\le h_{c1}) ,(h_{L} \ge -h_{c1},h_{R}\ge -h_{c1})$ and $(h_{L} \le h_{c1},h_{R}\ge -h_{c1})$ respectively. In the following we shall distinguish  between odd end even chains and discuss separately 
 the sub-phases $A_{j=(1,3)}$ and $A_{j=(2,4)}$.
 
\subsubsection{\textbf{Odd number of sites}}
\label{ch4sec:Aoddnumber}
\paragraph{The $A_1$ and $A_3$ sub-phases.}
In  these cases  both  boundary magnetic fields point towards the same direction: along  the positive  $z$ axis for  the  $A_1$ sub-phase and negative $z$ axis for the  $A_3$ sub-phase. Both cases are related by the isometry (\ref{ch4z2}). Qualitatively speaking, in the sub-phases $A_{1,3}$ and for $N$ odd, the boundary magnetic fields are not frustrating in the sense that in the Ising  limit of (\ref{ch4hamiltonian})  the ground-state would exhibit perfect antiferromagnetic order.

In the $A_1$ phase we find that the ground-state is unique and has a total spin $S^z=-\frac{1}{2}$.
We accordingly label the ground-state in this phase by
\bea\label{ch4gsA1odd} |-\frac{1}{2}\rangle
\eea

and denote by $E_0$ its energy. We notice that due to the presence of the boundary fields the spin $-\frac{1}{2}$ of the ground-state is the consequence of a static spin density
distribution. One can build up excitations in the bulk on top of this ground state by adding an arbitrary {\it even} number of spinons, bulk strings and quartets. These bulk excitations built on top of the state $|-\frac{1}{2}\rangle$  form a tower of excited states that we shall denote the ground-state tower.

As said above in the $A$ phases there exists two boundary bound-state solutions exponentially localized at either the  left or the right edge. In the language of the Bethe ansatz they correspond to purely imaginary roots solutions of (\ref{ch4boundstringA}). These bound-states carry a spin $\frac{1}{2}$, whose spin orientation is along the boundary fields
 at each edge, and have an energy

 \bea \label{ch4boundenA1}
 m_{\beta}=\sinh\gamma \sum_{\omega=-\infty}^{\infty} (-1)^{\omega}\frac{e^{-\gamma(1-\tilde{\epsilon}_{\beta})|\omega|}}{\cosh\gamma|\omega|}, \hspace{3mm} \beta=L,R
 \eea
Since the bound-states carry  a spin half, in order  to add a bound-state  to the ground-state 
one also needs to add a spinon.  This spinon may have  spin $+\frac{1}{2}$ or $-\frac{1}{2}$ and an arbitrary 
 rapidity $\theta$. The energy cost in the process is
 \be
 \label{ch4bsspinonA1odd}
 E_0+m_{L,R}+E_{\theta},
 \ee
 and is minimal when $\theta \rightarrow \pi$.
 The corresponding states  
 \be
 \label{ch4LRtowersA1odd}
|\pm \frac{1}{2}\rangle_{L}\; {\rm and}\; |\pm \frac{1}{2}\rangle_{R}\; ,
 \ee
 have total spins $S^z=\pm \frac{1}{2}$ and energies $E_0+m_{L}$ and  $E_0+m_{R}$. 
 The lowest excited states above (\ref{ch4LRtowersA1odd}) consist of spinon branches with energies given by (\ref{ch4bsspinonA1odd}) and $\theta \neq \pi$. On top of these, the states (\ref{ch4LRtowersA1odd})
generate, each, a tower of excited states obtained by  adding an arbitrary {\it even} number of spinons, bulk strings and quartets. In both the left and right towers, built upon (\ref{ch4LRtowersA1odd}), a localized bound-state at the left and the right edge is present and the number of spinon excitations is always odd. 

On top of the above three towers there exists a fourth one which correspond to states which host two bound-states. The state with the lowest energy consists into adding to the ground-state (\ref{ch4gsA1odd}) a localized bound-state at each, left and right, edge. Since in the process the total spin of the state is shifted by $1$, no spinon is required. The resulting state 
\be\label{ch4LRtowerA1odd}
| +\frac{1}{2}\rangle_{LR}\; ,
\ee
which has a total spin $S^z=\frac{1}{2}$ and an energy 
$E_0+m_{L}+m_{R}$, generates a tower of excited states
that comprises an arbitrary even number of spinons, bulk strings and quartets. The number of spinon states in the whole tower is always even. We thus see that, in the $A_1$ sub-phase, the whole Hilbert space can be split into four towers generated by the states given in Eqs (\ref{ch4gsA1odd}), (\ref{ch4LRtowersA1odd}) and (\ref{ch4LRtowerA1odd}). On top of the ground-state tower which governs the low-energy physics,
the remaining three towers contain at least one bound-state at the edges and are higher energy states.
In particular, we notice that in the $A_1$ sub-phase, a single spinon excitation costs at least a boundary gap $m_L$ or $m_R$.

The situation in the $A_3$ sub-phase can be described in the very same way as above. Using the isometry
(\ref{ch4z2}), we can obtain all the states in the sub-phase $A_3$ starting from the states in the sub-phase $A_1$ by reversing the sign of the total spin $S^z$ of the states.
We obtain so four towers of states in the sub-phase $A_3$ generated by the states $| +\frac{1}{2}\rangle$, $| \pm \frac{1}{2}\rangle_{L,R}$ and $| -\frac{1}{2}\rangle_{LR}$
at energies $E_0, E_0+m_{L,R}$ and $E_0+m_{L} +m_{R}$.

\paragraph{The  $A_2$ and $A_4$  phases}
In these cases the boundary fields are frustrating for $N$ odd  in the sense discussed above. 
As we shall see in these sub-phases the Hilbert space is also split into four towers of states corresponding to the presence of  boundary bound-states.  
However, since  the boundary magnetic fields at the two edges point toward opposite directions, the nature of these towers differ from the ones described above. 
Consider for instance the $A_2$ sub-phase in which 
the left boundary field points towards the negative $z$ axis  while the one at the right boundary points in the opposite direction. Just as in the phase $A_1$, there exists two boundary bound-state solutions one at each edge.  The bound state's spin is always oriented along  the boundary magnetic field. Hence, in the sub-phase $A_2$ the bound-state localized at the left edge has spin $-\frac{1}{2}$ whereas the bound-state localized at the right edge has spin $+\frac{1}{2}$. We find that for $|h_R| \le |h_L|$, the ground-state contains a bound state at the right edge and has total spin $S^z=+\frac{1}{2}$. For $|h_R|\ge |h_L|$, the ground-state contains a bound state at the left edge and has total spin $S^z=-\frac{1}{2}$. These two states are represented by 

\bea \label{ch4gstowerA2odd}|\pm\frac{1}{2}\rangle_{L/R} \eea

The excitations on top of these two states are generated by adding an arbitrary even number of spinons, bulk strings, higher order boundary strings and quartets.

 We find that in order to remove the bound-state at either the left or the right edge with spin $\mp\frac{1}{2}$  one has to add a
 spinon with rapidity $\theta$, whose minimum energy occurs at $\theta=\pi$. The resulting state has total spin $S^z=\pm\frac{1}{2}$, which depends on the spin orientation of the spinon, and has energy $E_0+m$. It is represented by
 
 \bea \label{ch4noboundtowerA2odd} |\pm\frac{1}{2}\rangle. \eea
 
  The lowest excited states above (\ref{ch4noboundtowerA2odd}) consist of spinon branches with $\theta \neq \pi$. On top of these, the states (\ref{ch4noboundtowerA2odd}) generate, each, a tower of excited states obtained by  adding an arbitrary {\it even} number of spinons, bulk strings and quartets. 
 
  Finally, the fourth tower is obtained by adding a bound-state at each edge to the two states (\ref{ch4noboundtowerA2odd}). The total spin of the resulting state 
 does not change since the two, left and right, bound-states have opposite spins. We obtain the states
 \be\label{ch4LandRA2towersodd}
 |\pm \frac{1}{2}\rangle_{LR} ,
 \ee
which have an energy $E_0+m_L+m_R+m$.  The lowest excited states above (\ref{ch4LandRA2towersodd}) consist of spinon branches with $\theta \neq \pi$. On top of this the states (\ref{ch4LandRA2towersodd}) generate, each, a tower of excited state by adding an arbitrary even number of spinons, bulk strings, higher order boundary strings and quartets. 

Using the isometry (\ref{ch4z2}), we can obtain all the states in the sub-phase $A_4$ from the states in the sub-phase $A_2$  by reversing their spins.
The Hilbert space in the 
sub-phase $A_4$ can be similarly sorted out in terms of four towers of states built upon the states 
$|\pm\frac{1}{2}\rangle$, $|+\frac{1}{2}\rangle_L,  |-\frac{1}{2}\rangle_R$ and $|\pm \frac{1}{2}\rangle_{LR}$
with energies $E_0, E_0+m_{L,R}$ and $E_0+m_{L} +m_{R}$.

\subsubsection{\textbf{Even number of sites}}
\label{ch4sec:Aevennumber}
\paragraph{The $A_1$ and $A_3$ sub-phases.}

In the phase $A_1$ we find that for $|h_R| \le |h_L|$, the ground-state contains a bound state at the right edge and has total spin $S^z=0$. For $|h_R|\ge |h_L|$, the ground-state contains a bound state at the left edge and has total spin $S^z=0$. These two states are represented by 

\bea \label{ch4gstowerA1even}|0\rangle_{L/R}, \eea

and have energy $E_0+m_{L/R}$ respectively.

The ground state generates a tower of excited states, which are obtained by adding an arbitrary even number of spinons, bulk strings and quartets. In this tower the number of spinon states is always even.

Starting from one of the two ground-states (\ref{ch4gstowerA1even}), one may remove the bound-state 
at either the left or the right edge by adding a $\pm \frac{1}{2}$ spinon with rapidity $\theta$. The lowest energy of this spinon corresponds to $\theta\rightarrow\pi$. As a result, we end up with two states of  total spin $S^z=0,-1$ which are denoted by
\be\label{ch4nobstowerA1even}
|0\rangle \; {\rm and}\; |-1\rangle,
\ee
and both have energies $E_0+m$ in thermodynamic limit. The lowest excited states above (\ref{ch4nobstowerA1even}) consist of spinon branches with $\theta \neq \pi$. On top of these, the states (\ref{ch4nobstowerA1even}) generate, each, a tower of excited states obtained by adding an arbitrary {\it even} number of spinons, bulk strings and quartets.

The fourth tower is obtained from the states (\ref{ch4nobstowerA1even}) by adding a bound-state at each edge. 
Since the change of total spin is $1$ there is no need to add or remove a spinon. In the process we obtain two degenerate states, with total spins $S^z=1$ and $S^z=0$
and energy  $E_0+m+m_L+m_R$, 
\be\label{ch4LRtowerA1LRteven}
|1\rangle_{LR} \; {\rm and}\; |0\rangle_{LR},
\ee
that host spin $\pm \frac{1}{2}$ spinons with rapidity $\theta\rightarrow \pi$ as in the ground-states. The fourth tower of excited states comprises spin $\pm \frac{1}{2}$ spinon states. These states have energy $E_0+m_{L} +m_{R}+E_{\theta}$ and are gapped high energy states. The remaining states of this tower 
are then built up  by adding an even number of spinons, bulk strings, higher order boundary strings and quartets, and hence the number of spinons is always odd in this tower.

Similar to the odd number of sites case, using the isometry (\ref{ch4z2}), we can obtain all the states in the phase $A_3$ starting from the states in the phase $A_1$ described above. We obtain
$|0\rangle_{L/R}$, ($0\rangle, |1\rangle$), ($0\rangle_{LR}, |-1\rangle_{LR}$) with energies $E_0+m_{L/R}$, $E_0+m$ and $E_0+m+m_L+m_R+m$ respectively.

\paragraph{The $A_2$ and $A_4$ sub-phases.}

In the $A_2$ phase we find that the ground-state is unique and has a total spin $S^z=0$.
We accordingly label the ground-state in this phase by
\bea\label{ch4gsA2even}|0\rangle
\eea

and denote by $E_0$ its energy. One can build up excitations in the bulk on top of this ground state by adding an arbitrary {\it even} number of spinons, bulk strings and quartets. These bulk excitations built on top of the state $|0\rangle$  form a tower of excited states that we shall denote the ground-state tower.

One can add a bound state with spin $S^z=-\frac{1}{2}$ at the left edge to the ground state by adding a spinon with rapidity $\theta$ and spin $S^z=\pm\frac{1}{2}$, resulting in a state with total spin $S^z=0,-1$. The lowest energy of the spinon corresponds to $\theta\rightarrow\pi$. We denote this state by

\bea \label{ch4LtowerA2even} |0\rangle_L, \hspace{2mm} |-1\rangle_L \eea

which has energy $E_0+m_L+m$. Similarly, one can also add a bound state with spin $S^z=+\frac{1}{2}$ at the right edge by adding a spinon, resulting in a state with total spin $S^z=0,1$. This state with the lowest energy is represented by 

\bea \label{ch4RtowerA2even} |0\rangle_L, \hspace{2mm} |1\rangle_L \eea

and has energy $E_0+m_R+m$.

 The lowest excited states above (\ref{ch4LtowerA2even}) and (\ref{ch4RtowerA2even}) consist of spinon branches with $\theta \neq \pi$. On top of these, the states (\ref{ch4LtowerA2even}) and (\ref{ch4RtowerA2even}) generate, each, a tower of excited states obtained by  adding an arbitrary {\it even} number of spinons, bulk strings and quartets. In both the left and right towers, built upon (\ref{ch4LtowerA2even}) and (\ref{ch4RtowerA2even}), a localized bound-state at either the left or the right edge is present and the number of spinons in the excitated state is always odd. 

On top of the above three towers there exists a fourth one which correspond to states which host two bound-states. The state with the lowest energy consists into adding to the ground-state (\ref{ch4gsA2even}) a localized bound-state at each, left and right, edge. Since in the process the total spin of the state is shifted by $1$, no spinon is required. The resulting state 
\be\label{ch4LRtowerA2even}
|0\rangle_{LR}\; ,
\ee
which has a total spin $S^z=0$ and an energy 
$E_0+m_{L}+m_{R}$, generates a tower of excited states
that comprises an arbitrary even number of spinons, bulk strings and quartets. The number of spinon states in the whole tower is always even. We thus see that, in the $A_2$ sub-phase, the whole Hilbert space can be split  into four towers generated by the states (\ref{ch4gsA2even}), (\ref{ch4LtowerA2even}), (\ref{ch4RtowerA2even}) and (\ref{ch4LRtowerA2even}). On top of the ground-state tower which governs the low-energy physics, the remaining three towers contain at least one bound-state at the edges and are high-energy states. In particular, we notice that in the $A_2$ sub-phase,
although the system is massless, a single spinon excitation costs at least a boundary gap $m_L$ or $m_R$.

The situation in the $A_4$ sub-phase can be described in the very same way as above. Using the isometry
(\ref{ch4z2}), we can obtain all the states in the sub-phase $A_4$ starting from the states in the sub-phase $A_2$ by reversing the sign of the total spin $S^z$ of the states.
We obtain so four towers of states in the sub-phase $A_3$ generated by the states $| 0\rangle$, ($|0\rangle_{L}$, $|1\rangle_{L}$), ($|0\rangle_{R}$, $|-1\rangle_{R}$) and $|0\rangle_{LR}$
at energies $E_0$, $E_0+m_{L}+m$, $E_0+m_{R}+m$ and $E_0+m_{L} +m_{R}$.

\subsection{$\it{F}$ phases}
\label{ch4sec:Fphase}
In the $F$ phases the two boundary bound-states are stabilized, but unlike in the $A$ phases, these bound states are high energy states with energy above the maximum energy $M$ of a spinon, that is the band height. 

The four $F_{j=(1,2,3,4)}$ sub-phases  corresponds to the domains of boundary fields 
 $(h_{L}\ge h_{c2},h_{R}\ge h_{c2}), (h_{L} \le -h_{c2},h_{R}\ge h_{c2}) ,(h_{L} \le -h_{c2},h_{R}\le -h_{c2})$ and $(h_{L} \ge h_{c2},h_{R}\le -h_{c2})$ respectively. In the following we shall distinguish  between odd and even chains and discuss separately the sub-phases $F_{j=(1,3)}$ and $F_{j=(2,4)}$.
 
\subsubsection{\textbf{Odd number of sites}}

\paragraph{The $F_1$ and $F_3$ sub-phases.}
\label{ch4sec:F1oddnumber}
In  these cases  both  boundary magnetic fields point
towards the same direction: along  the positive  $z$ axis
for  the  $F_1$ sub-phase and negative $z$ axis for 
the  $F_3$ sub-phase. Both cases are related by the isometry (\ref{ch4z2}). Qualitatively speaking, in the sub-phases $F_{1,3}$, same as in $A_{1,3}$ phases, for $N$ odd, the boundary magnetic fields are not frustrating in the sense that in the Ising  limit of (\ref{ch4hamiltonian})  the ground-state would exhibit perfect antiferromagnetic order.

In the $F_1$ phase we find that the ground-state is unique and has a total spin $S^z=-\frac{1}{2}$.
We accordingly label the ground-state in this phase   by
\be\label{ch4gsF1odd}
|-\frac{1}{2}\rangle,
\ee
and denote by $E_0$ its energy. We notice that due to the presence of the boundary fields, just as in the case of $A_1$ phase, the spin $-\frac{1}{2}$ of the ground-state is not carried by a spinon in contrast with the periodic chain with $N$ odd, it is rather due to the static spin distribution. Similarly  to the case of periodic boundary conditions, one can build up excitations in the bulk on top of this ground state by adding an arbitrary {\it even} number of spinons, bulk strings and quartets. These bulk excitations  built on top of the state $|-\frac{1}{2}\rangle$  form a tower of excited states  that we shall denote the ground-state tower.

As said above in the $F$ phases there exists two boundary bound-state solutions exponentially localized at either the  left or the right edge. These bound-states carry  a spin $\frac{1}{2}$, whose spin orientation is along the boundary fields at each edge, and have an energy
 \bea \label{ch4bounden2}
  m'_{\beta}=\sinh\gamma \sum_{\omega=-\infty}^{\infty} \frac{e^{-\gamma(1-\tilde{\epsilon}_{\beta})|\omega|}}{\cosh\gamma|\omega|}, \hspace{3mm} \beta=L,R 
 \eea
Since the bound-states carry  a spin half, in order  to add a bound-state  to the ground-state 
one also needs to add a spinon.  This spinon may have  spin $+\frac{1}{2}$ or $-\frac{1}{2}$ and an arbitrary 
 rapidity $\theta$. The energy cost in the process is
 \be
 \label{ch4bsspinonF1}
 E_0+m'_{L,R}+E_{\theta},
 \ee
 and is minimal when $\theta \rightarrow \pi$.
 The corresponding states  
 \be
 \label{ch4LRtowersF1odd}
|\pm \frac{1}{2}\rangle_{L}\; {\rm and}\; |\pm \frac{1}{2}\rangle_{R}\; ,
 \ee
 have total spins $S^z=\pm \frac{1}{2}$ and energies $E_0+m'_{L}+m$ and  $E_0+m'_{R}+m$. 
 The lowest excited states above (\ref{ch4LRtowersF1odd}) consist of spinon branches with energies given by (\ref{ch4bsspinonF1}) and $\theta \neq \pi$. On top of these, the states (\ref{ch4LRtowersF1odd}) generate, each, a tower of excited states obtained by  adding an arbitrary {\it even} number of spinons, bulk strings, higher order boundary strings and quartets. In both the left and right towers, built upon (\ref{ch4LRtowersF1odd}), a localized bound-state at the left and the right edge is present and the number of spinon excitations is always odd. 

On top of the above three towers there exists a fourth one which correspond to states which host two bound-states. The state with the lowest energy consists into adding to the ground-state (\ref{ch4gsF1odd}) a localized bound-state at each, left and right, edge. Since in the process the total spin of the state is shifted by $1$, no spinon is required. The resulting state 
\be\label{ch4LRtowerF1odd}
| +\frac{1}{2}\rangle_{LR}\; ,
\ee
which has a total spin $S^z=\frac{1}{2}$ and an energy 
$E_0+m'_{L}+m'_{R}$, generates a tower of excited states
that comprises an arbitrary even number of spinons, bulk strings, higher order boundary strings and quartets. The number of spinon states in the whole tower is always even. We thus see that, in the $F_1$ sub-phase, the whole Hilbert space can be split  into four towers generated by the states (\ref{ch4gsF1odd}), (\ref{ch4LRtowersF1odd}) and (\ref{ch4LRtowerF1odd}). On top of the ground-state tower which governs the low-energy physics, the remaining three towers contain at least one bound-state at the edges and are high-energy states.

The situation in the $F_3$ sub-phase can be described in the very same way as above. Using the isometry (\ref{ch4z2}), we can obtain all the states in the sub-phase $F_3$ starting from the states in the sub-phase $F_1$ by reversing the sign of the total spin $S^z$ of the states. We obtain four towers of states in the sub-phase $F_3$ generated by the states $| +\frac{1}{2}\rangle$, $| \pm \frac{1}{2}\rangle_{L,R}$ and $| -\frac{1}{2}\rangle_{LR}$ at energies $E_0, E_0+m'_{L,R}+m$ and $E_0+m'_{L} +m'_{R}$.

\paragraph{The  $F_2$ and $F_4$  phases.}
\label{ch4sec:F2oddnumber}
In these cases the boundary fields are frustrating for $N$ odd  in the sense discussed above.  As we shall see in these sub-phases the Hilbert space is also split into four towers of states corresponding to the presence of  boundary bound-states. However, since  the boundary magnetic fields at the two edges point toward opposite directions, the nature of these towers differ from the ones described above. Just as in the case of $A_2$ sub-phase, in the $F_2$ sub-phase in which the left boundary field points towards the negative $z$ axis  while  the one at the right boundary points in the opposite direction. In this case we find that the ground-state is two-fold degenerated, each one  containing a spinon (but no bound-state)  with spin 
$\pm\frac{1}{2}$ and rapidity $\theta\rightarrow\pi$.
These two states, i.e:
\be\label{ch4gsF2odd}
|\pm\frac{1}{2}\rangle,
\ee
 have energy $E_0+m$ and total spin $S^z=\pm\frac{1}{2}$ corresponding to the spin  of the spinon, and generate a tower of excited states. It is obtained by adding an arbitrary even number of spinons, bulk strings and quartets on top of the two  spin $\pm\frac{1}{2}$ spinon  branches with spectrum (\ref{ch4energyhole}) and rapidity $\theta \neq \pi$. In contrast with the $F_{1,3}$ sub-phases the ground-state tower contains an odd number of spinons. 

Just as in the phase $F_1$, there exists two boundary bound-state solutions one at each edge.  The bound state's spin is always oriented along  the boundary magnetic field. Hence, in the sub-phase $F_2$ the bound-state localized at the left edge has spin $-\frac{1}{2}$ whereas the bound-state localized at the right edge has spin $+\frac{1}{2}$. We find that in order to add the bound-state at the left edge with spin $-\frac{1}{2}$  one has to remove the spinon with spin $-\frac{1}{2}$ and rapidity $\theta=\pi$ in the $|-\frac{1}{2}\rangle$ ground-state (\ref{ch4gsF2odd}).
 The resulting state  has  total spin $S^z=-\frac{1}{2}$ and energy $E_0+m'_L$. Similarly adding a spin $+\frac{1}{2}$ bound-state at the right edge requires to remove the  spin $\frac{1}{2}$ spinon from the ground-state $|+\frac{1}{2}\rangle$  (\ref{ch4gsF2odd}). 
 The resulting state  has  total spin $S^z=+\frac{1}{2}$ and energy $E_0+m'_R$. The two states with a bound-state at either the left or right edge
 \be\label{ch4LRF2towersodd}
 |-\frac{1}{2}\rangle_L \; {\rm and}\; |+\frac{1}{2}\rangle_R,
 \ee
 generate, each, a tower of excited states upon adding  an arbitrary even number of spinons, bulk strings, higher order boundary strings and quartets. In these two towers the number of spinons in every state is always even. Finally, the fourth tower is obtained by adding a bound-state at each edge to  the two ground-states (\ref{ch4gsF2odd}). The total spin of the resulting state does not change since the two, left and right, bound-states have opposite spins. We obtain the states
 \be\label{ch4LandRF2towersodd}
 |\pm \frac{1}{2}\rangle_{LR} ,
 \ee
which have an energy $E_0+m'_L+m'_R+m$ and generate a tower of excited states. This tower contains two spin  $\pm \frac{1}{2}$ spinon states with dispersion $E_0+m'_L+m'_R +E_{\theta} $ and arbitrary even number of spinons, bulk strings, higher order boundary strings and quartets. 

Using the isometry (\ref{ch4z2}), we can obtain all the states in the sub-phase $F_4$ from the states in the sub-phase $F_2$  by reversing their spins. The Hilbert space in the sub-phase $F_4$ can be similarly sorted out in terms of four towers of states built upon the states $|\pm\frac{1}{2}\rangle$, $|+\frac{1}{2}\rangle_L,  |-\frac{1}{2}\rangle_R$ and $|\pm \frac{1}{2}\rangle_{LR}$
with energies $E_0+m, E_0+m'_{L,R}$ and $E_0+m'_{L} +m'_{R}+m$.

\subsubsection{\textbf{Even number of sites}}
\label{ch4sec:F1evennumber}
When the number of sites is even the frustrating effect of the magnetic fields is reversed as compared to the $N$ odd case, just as in the case of $A$ phases. The boundary fields are frustrating in sub-phases $F_{1,3}$ while non-frustrating in the sub-phases $F_{2,4}$.

In the phase $F_1$ we find that the ground-state is two-fold degenerated. It does not contain bound-states but contains spinons with  rapidity 
$\theta \rightarrow \pi$ and spins $\pm \frac{1}{2}$. Despite this, since $N$ is even, the total spin of the two degenerate ground-states has to be an integer. Indeed, as it comes out from our exact solution the two ground-states have total spins $S^z=0$ and $S^z=-1$. Our interpretation of this fact is that the two ground-states contain a spin $+\frac{1}{2}$ and  a spin $- \frac{1}{2}$ spinon on top of a static background spin $- \frac{1}{2}$ distribution in the ground-state as it is the case for the $F_1$ sub-phase  when $N$ is odd. In the following we denote these two ground-states by
\be\label{ch4evenF1gseven}
|0\rangle\;  {\rm and} \; |-1\rangle. 
\ee
The ground-state tower of excitated states comprises of spin $\pm \frac{1}{2}$ spinon states with energy $E_0+E_{\theta}$ and finite rapidity $\theta \neq \pi$. The rest of the tower is then obtained by adding an arbitrary even number of spinons, bulk strings and quartets. In this tower the number of spinon states is always odd.

Starting from one of the two ground-states (\ref{ch4evenF1gseven}), one may add a  bound-state at either the left or the right edge. To this end one needs to remove the spin $\pm \frac{1}{2}$  spinon. The resulting total spin is then the sum of the bound-state spin  $+\frac{1}{2}$ with that of the static background spin $- \frac{1}{2}$ distribution mentioned above. As a result, we end up with two states of  total spin $S^z=0$. The corresponding states with the bound-state at the left or the right  edge are denoted
\be\label{ch4evenF1bstowers}
|0\rangle_L \; {\rm and}\; |0\rangle_R,
\ee
and have energies $E_0+m'_{L}$ and $E_0+ m'_{R}$. Each of these two states generates a tower of excited states. In these towers the number of spinon states is always even. 

The fourth tower is obtained from the ground-states (\ref{ch4evenF1gseven}) by adding a bound-state at each edge. Since the change of total spin is $1$ there is no need to add or remove a spinon. In the process we obtain two degenerate states, with total spins $S^z=1$ and $S^z=0$ and energy  $E_0+m'_L+m'_R+m$, 
\be\label{ch4evenF1LRtowers}
|1\rangle_{LR} \; {\rm and}\; |0\rangle_{LR},
\ee
that host spin $\pm \frac{1}{2}$ spinons with rapidity $\theta=\pi$ as in the ground-states. The fourth tower of excited states comprises, as in the ground-state tower, spin $\pm \frac{1}{2}$  spinon states. These states  have energy $E_0+m'_{L} +m'_{R}+E_{\theta}$ and are gapped high energy states. The remaining states of this tower are then built  by adding an even number of spinons, bulk strings, higher order boundary strings and quartets, and hence the number of spinons is always odd in this tower.

Similar to the odd number of sites case, using the symmetry (\ref{ch4z2}), we can obtain all the states in the phase $F_3$ starting from the states in the phase $F_1$ described above. We obtain
($|0\rangle, |1\rangle$), $|0\rangle_{L/R}$ and ($|-1\rangle_{LR}, |0\rangle_{LR}$) with energies $E_0+m$, $E_0+m'_{L/R}$ and $E_0+m+m'_{L}+m'_R$ respectively

\paragraph{The   $F_2$ and $F_4$ sub-phases.}
\label{ch4sec:F2evnnumber}
 
In the sub-phase $F_2$  we find that the ground-state is non-degenerated
\be\label{ch4F2evengs}
|0\rangle,
\ee
 and has total spin $S^z=0$ with  energy $E_0$. Starting from this ground state we can add a bound-state at the left edge whose spin is $-\frac{1}{2}$. As already emphasized one also need to add a spinon, with rapidity $\theta=\pi$, for the total spin shift to be an integer.  Depending on the spinon spin, which can be either $\pm \frac{1}{2}$, one ends up with two states
 \be\label{ch4F2evenLtower}
 |-1\rangle_{L},|0\rangle_{L},
 \ee
 which have total spins $S^z=-1$ and $S^z=0$ and energy 
 $E_0+m'_L+m$. One may repeat the same line of arguments with the right edge, paying attention to the orientation of the bound-sate spin, which in this case is $+\frac{1}{2}$. The resulting two states
 \be\label{ch4F2evenRtower}
 |1\rangle_{R},|0\rangle_{R},
 \ee
 hosting a bound-state at the right edge have total spins $S^z=1$ and $S^z=0$ and energy $E_0+m'_R+m$. Each state with bound state at either the left (\ref{ch4F2evenLtower}) or the right (\ref{ch4F2evenRtower}) edges generates a tower of excited states that comprise odd number of spinons along with bulk strings, higher order boundary strings and quartets.
 
The forth tower is obtained from the ground-state (\ref{ch4F2evengs}) by adding a bound-state with spin $- \frac{1}{2}$ at the left edge and spin $+ \frac{1}{2}$ at the right edge. No spinons are needed
in the process and one ends up with a single state
\be\label{ch4F2evenLRtower}
 |0\rangle_{LR},
 \ee
with total spin $S^z=0$ and energy $E_0+m'_R+m'_L$. The latter state generates also a tower of states including any pairs of spinons, bulk strings, higher order boundary strings and quartets.

Using the symmetry (\ref{ch4z2}), similar to the odd number of sites case, we can obtain all the states in the sub-phase $F_4$ starting from the states in the sub-phase $F_2$ described above. We obtain 
$| 0\rangle$, ($|0\rangle_{L}$, $|1\rangle_{L}$), ($|0\rangle_{R}$, $|-1\rangle_{R}$) and $|0\rangle_{LR}$ at energies $E_0$, $E_0+m'_{L}+m$, $E_0+m'_{R}+m$ and $E_0+m'_{L} +m'_{R}$.

\subsection{$\it{E}$ phases}
\label{ch4sec:Ephase}
In the $E$ phases the two boundary bound-states are stabilized, but unlike in the $A$ and $F$ phases, one of the bound states is a high energy state with energy above the maximum energy $M$ of a spinon, that is the band height, while the other is a low energy state with energy below the mass gap $m$. The eight $E_{j=(1...8)}$ sub-phases correspond to the domains of boundary fields shown in table 
 
  \begin{table}[h!]
\centering
\caption{Values of the boundary fields corresponding to eight $E$ phases}
\begin{tabular}{|c|c|c|}
 \hline
 \hline
 Phase&$h_L$&$h_R$\\
 \hline
 $E_1$&$(0,h_{c1})$&$(h_{c2},\infty)$\\
 $E_8$&$(h_{c2},\infty)$&$(0,h_{c1})$\\ 
$E_2$&$(-h_{c1},0)$&$(h_{c2},\infty)$\\
$E_7$&$(h_{c2},\infty)$&$(-h_{c1},0)$\\ 
  $E_3$&$(-\infty,-h_{c2})$&$(0,h_{c1})$\\   
   $E_4$&$(-\infty,-h_{c2})$&$(-h_{c1},0)$\\   
   $E_5$&$(-h_{c1},0)$&$(-\infty,-h_{c2})$\\   
   $E_6$&$(0,h_{c1})$&$(-\infty,-h_{c2})$\\    
     \hline 
      \end{tabular}
 \end{table}
 
 In the following we shall distinguish  between odd end even chains and discuss separately 
 the sub-phases $E_{j=(1,5,8,4)}$ and $F_{j=(2,6,3,7)}$.
 
\subsubsection{\textbf{Odd number of sites}}

\paragraph{The ($E_1$,$E_5$) and ($E_8$,$E_4$) sub-phases.}
\label{ch4sec:E1oddnumber}
In  these cases  both  boundary magnetic fields point towards the same direction: along  the positive  $z$ axis for  the  $E_1,E_8$ sub-phases and negative $z$ axis for the $E_5,E_4$ sub-phases. The pairs of sub-phases $E_1,E_5$ and $E_8,E_4$ are related by the isometry (\ref{ch4z2}). Qualitatively speaking, in the sub-phases $E_{1,8}$, same as in $A_{1,3}$ and $F_{1,3}$ phases, for $N$ odd, the boundary magnetic fields are not frustrating.

In the $E_1$ phase we find that the ground-state is unique and has a total spin $S^z=-\frac{1}{2}$. We accordingly label the ground-state in this phase by
\be\label{ch4gsE1odd}
|-\frac{1}{2}\rangle,
\ee
and denote by $E_0$ its energy. We notice that due to the presence of the boundary fields, just as in the case of $A_1$ phase, the spin $-\frac{1}{2}$ of the ground-state is not carried by a spinon in contrast with the periodic chain with $N$ odd, it is rather due to the static spin distribution. Similarly  to the case of periodic boundary conditions, one can build up excitations in the bulk on top of this ground state by adding an arbitrary {\it even} number of spinons, bulk strings and quartets. These bulk excitations  built on top of the state $|-\frac{1}{2}\rangle$  form a tower of excited states  that we shall denote the ground-state tower.

As said above in the $E$ phases there exists two boundary bound-state solutions exponentially localized at either the  left or the right edge. These bound-states carry  a spin $\frac{1}{2}$, whose spin orientation is along the boundary fields at each edge. Since the bound-states carry a spin half, in order  to add a bound-state  to the ground-state one also needs to add a spinon.  This spinon may have  spin $+\frac{1}{2}$ or $-\frac{1}{2}$ and an arbitrary rapidity $\theta$. 

The bound state at the left edge is a low energy state and the energy cost of having this is
 \be
 \label{ch4bsspinonE1L}
 E_0+m_{L}+E_{\theta},
 \ee
 and is minimal when $\theta \rightarrow \pi$.  The corresponding states  
 \be
 \label{ch4LtowersE1odd}
|\pm \frac{1}{2}\rangle_{L},
 \ee
 have total spins $S^z=\pm \frac{1}{2}$ and energies $E_0+m_{L}+m$. 
 
 The bound state at the right edge is a high energy state and the energy cost of having this is
 \be
 \label{ch4bsspinonE1R}
 E_0+m'_{R}+E_{\theta},
 \ee
 and is minimal when $\theta \rightarrow \pi$.  The corresponding states  
 \be
 \label{ch4RtowersE1odd}
|\pm \frac{1}{2}\rangle_{R},
 \ee
 have total spins $S^z=\pm \frac{1}{2}$ and energies $E_0+m'_{R}+m$.

 The lowest excited states above (\ref{ch4LtowersE1odd}) and  (\ref{ch4RtowersE1odd}) consist of spinon branches with energies given by (\ref{ch4bsspinonE1L}) and (\ref{ch4bsspinonE1R}) and $\theta \neq \pi$. On top of these, the states (\ref{ch4LtowersE1odd}) and (\ref{ch4RtowersE1odd}) generate, each, a tower of excited states obtained by  adding an arbitrary {\it even} number of spinons, bulk strings, higher order boundary strings and quartets. In both the left and right towers, built upon (\ref{ch4LtowersE1odd}) and (\ref{ch4RtowersE1odd}) respectively, a localized bound-state at the left and the right edge is present and the number of spinon excitations is always odd. 

On top of the above three towers there exists a fourth one which correspond to states which host two bound-states. The state with the lowest energy consists into adding to the ground-state (\ref{ch4gsE1odd}) a localized bound-state at each, left and right, edge. Since in the process the total spin of the state is shifted by $1$, no spinon is required. The resulting state 
\be\label{ch4LRtowerE1odd}
| +\frac{1}{2}\rangle_{LR}\; ,
\ee
which has a total spin $S^z=\frac{1}{2}$ and an energy 
$E_0+m_{L}+m'_{R}$, generates a tower of excited states
that comprises an arbitrary even number of spinons, bulk strings, higher order boundary strings and quartets. The number of spinon states in the whole tower is always even. We thus see that, in the $E_1$ sub-phase, the whole Hilbert space can be split into four towers generated by the states (\ref{ch4gsE1odd}), (\ref{ch4LtowersE1odd}), (\ref{ch4RtowersE1odd}) and (\ref{ch4LRtowerE1odd}). 

The situation in the $E_5$ sub-phase can be described in the very same way as above. Using the isometry (\ref{ch4z2}), we can obtain all the states in the sub-phase $E_5$ starting from the states in the sub-phase $E_1$ by reversing the sign of the total spin $S^z$ of the states. We obtain four towers of states in the sub-phase $E_5$ generated by the states $| +\frac{1}{2}\rangle$, $| \pm \frac{1}{2}\rangle_{L,R}$ and $| -\frac{1}{2}\rangle_{LR}$ at energies $E_0, E_0+m_{L}+m, E_0+m'_{R}+m$ and $E_0+m_{L} +m'_{R}$.

The states in the sub-phases $E_8$ and $E_4$ are obtained from the states in the sub-phases $E_1$ and $E_5$ respectively by the transformation $L\leftrightarrow R$.

\paragraph{The ($E_2$,$E_6$) and ($E_3$,$E_7$) sub-phases.}
\label{ch4sec:E2oddnumber}
In these cases the boundary fields are frustrating for $N$ odd  in the sense discussed above.  As we shall see in these sub-phases the Hilbert space is also split into four towers of states corresponding to the presence of  boundary bound-states. However, since  the boundary magnetic fields at the two edges point toward opposite directions, the nature of these towers differ from the ones described above. Just as in the case of $A_2$ sub-phase, in the $E_2$ sub-phase in which the left boundary field points towards the negative $z$ axis  while  the one at the right boundary points in the opposite direction. The ground-state contains a bound state at the left edge and has total spin $S^z=-\frac{1}{2}$ and is represented by 

\bea \label{ch4gstowerE2odd}|-\frac{1}{2}\rangle_{L}. \eea

The energy of this state is $E_0+m_L$. The state which contains a bound state at the right edge has total spin $S^z=+\frac{1}{2}$ and is is represented by 

\bea \label{ch4RtowerE2odd}|\frac{1}{2}\rangle_{R}. \eea

The energy of this state is $E_0+m'_R$. The excitations on top of the states (\ref{ch4gstowerE2odd}) and (\ref{ch4RtowerE2odd}) are generated by adding an arbitrary even number of spinons, bulk strings, higher order boundary strings and quartets.

In order to remove the bound-state at either the left edge or the right edge with spin $\mp\frac{1}{2}$ respectively, one has to add a spinon with rapidity $\theta$, whose minimum energy occurs at $\theta=\pi$. The resulting state has total spin $S^z=\pm 1/2$, which depends on the spin orientation of the spinon, and has energy $E_0+m$. It is represented by
 
 \bea \label{ch4noboundtowerE2odd} |\pm\frac{1}{2}\rangle. \eea
 
  The lowest excited states above (\ref{ch4noboundtowerE2odd}) consist of spinon branches with $\theta \neq \pi$. On top of these, the states (\ref{ch4noboundtowerE2odd}) generate, each, a tower of excited states obtained by  adding an arbitrary {\it even} number of spinons, bulk strings and quartets.

  Finally, the fourth tower is obtained by adding a bound-state at each edge to the two states (\ref{ch4noboundtowerE2odd}). The total spin of the resulting state 
 does not change since the two, left and right, bound-states have opposite spins. We obtain the states
 \be\label{ch4LandRE2towersodd}
 |\pm\frac{1}{2}\rangle_{LR}
 \ee
which have an energy $E_0+m_L+m'_R+m$.  The lowest excited states above (\ref{ch4LandRE2towersodd}) consist of spinon branches with $\theta \neq \pi$. On top of this the states (\ref{ch4LandRE2towersodd}) generate, each, a tower of excited state by adding an arbitrary even number of spinons, bulk strings, higher order boundary strings and quartets.

Using the isometry (\ref{ch4z2}), we can obtain all the states in the sub-phase $E_6$ starting from the states in the sub-phase $E_2$ by reversing the sign of the total spin $S^z$ of the states. We obtain four towers of states in the sub-phase $E_5$ generated by the states $| +\frac{1}{2}\rangle_L$, $|- \frac{1}{2}\rangle_{R}$ and $|\pm\frac{1}{2}\rangle$ and $|\pm\frac{1}{2}\rangle_{LR}$ at energies $E_0+m_L, E_0+m'_{R}$, $E_0+m$ and $E_0+m_{L} +m'_{R}+m$ respectively.

The states in the sub-phases $E_3$ and $E_7$ are obtained from the states in the sub-phases $E_2$ and $E_6$ respectively by the transformation $L\leftrightarrow R$.

\subsubsection{\textbf{Even number of sites}}

\paragraph{The ($E_1$,$E_5$) and ($E_8$,$E_4$) sub-phases.}
\label{ch4sec:E1evennumber}

In the sub-phase $E_1$, the ground-state contains a bound state at the left edge and has total spin $S^z=0$ and is represented by 

\bea \label{ch4gstowerE1even}|0\rangle_{L}. \eea

The energy of this state is $E_0+m_L$. The state which contains a bound state at the right edge has total spin $S^z=0$ and is is represented by 

\bea \label{ch4RtowerE1even}|0\rangle_{R}. \eea

The energy of this state is $E_0+m'_R$. The excitations on top of the states (\ref{ch4gstowerE1even}) and (\ref{ch4RtowerE1even}) are generated by adding an arbitrary even number of spinons, bulk strings, higher order boundary strings and quartets.

In order to remove the bound-state at either the left edge or the right edge with spin $\mp\frac{1}{2}$ respectively, one has to add a
 spinon with rapidity $\theta$, whose minimum energy occurs at $\theta=\pi$. The resulting state has total spin $S^z=0,-1$, which depends on the spin orientation of the spinon, and has energy $E_0+m$. It is represented by
 
 \bea \label{ch4noboundtowerE1even} |0\rangle, |-1\rangle. \eea
 
  The lowest excited states above (\ref{ch4noboundtowerE1even}) consist of spinon branches with $\theta \neq \pi$. On top of these, the states (\ref{ch4noboundtowerE1even}) generate, each, a tower of excited states obtained by  adding an arbitrary even number of spinons, bulk strings and quartets.

  Finally, the fourth tower is obtained by adding a bound-state at each edge to the two states (\ref{ch4noboundtowerE1even}). The total spin of the resulting state 
 does not change since the two, left and right, bound-states have opposite spins. We obtain the states
 \be\label{ch4LandRE2towerseven}
 |0\rangle_{LR} , |-1\rangle_{LR}
 \ee
which have an energy $E_0+m_L+m'_R+m$.  The lowest excited states above (\ref{ch4LandRE2towerseven}) consist of spinon branches with $\theta \neq \pi$. On top of this the states (\ref{ch4LandRE2towerseven}) generate, each, a tower of excited state by adding an arbitrary even number of spinons, bulk strings, higher order boundary strings and quartets. 

Similar to the odd number of sites case, using the isometry (\ref{ch4z2}), we can obtain all the states in the phase $E_5$ starting from the states in the phase $E_1$ described above. We obtain
$|0\rangle_L$, $|0\rangle_{R}$, ($|0\rangle, |1\rangle$) ($|1\rangle_{LR}, |0\rangle_{LR}$) with energies $E_0+m_L$, $E_0+m'_{R}$, $E_0+m$ and $E_0+m+m_L+m'_R$ respectively.

Just as in the odd number of sites case, the states in the sub-phases $E_8$ and $E_4$ are obtained from the states in the sub-phases $E_1$ and $E_5$ respectively by the transformation $L\leftrightarrow R$.

\paragraph{The ($E_2$,$E_6$) and ($E_3$,$E_7$) sub-phases.}
\label{ch4sec:E2evennumber}

In the $E_2$ phase we find that the ground-state is unique and has a total spin $S^z=0$. We accordingly label the ground-state in this phase by
\be\label{ch4gsE2even}
|0\rangle,
\ee
and denote by $E_0$ its energy. One can build up excitations in the bulk on top of this ground state by adding an arbitrary even number of spinons, bulk strings and quartets. These bulk excitations  built on top of the state $|0\rangle$  form a tower of excited states  that we shall denote the ground-state tower.

The bound state at the left edge is a low energy state and the energy cost of having this is
 \be
 \label{ch4bsspinonE2evenL}
 E_0+m_{L}+E_{\theta},
 \ee
 and is minimal when $\theta \rightarrow \pi$.  The corresponding states  
 \be
 \label{ch4LtowersE2even}
|0\rangle_{L}, |-1\rangle_{L}
 \ee
 have total spins $S^z=0,-1$ respectively which depends on the spin orientation of the spinon, and energies $E_0+m_{L}+m$.  The bound state at the right edge is a high energy state and the energy cost of having this is
 \be
 \label{ch4bsspinonE2evenR}
 E_0+m'_{R}+E_{\theta},
 \ee
 and is minimal when $\theta \rightarrow \pi$.  The corresponding states  
 \be
 \label{ch4RtowersE2even}
|0\rangle_{R}, |1\rangle_{R},
 \ee
 have total spins $S^z=0,1$ respectively which depends on the spin orientation of the spinon, and energies $E_0+m'_{R}+m$.

 The lowest excited states above (\ref{ch4LtowersE2even}) and (\ref{ch4RtowersE2even}) consist of spinon branches with energies given by (\ref{ch4bsspinonE2evenL}) and (\ref{ch4bsspinonE2evenR}) and $\theta \neq \pi$. On top of these, the states (\ref{ch4LtowersE2even}) and (\ref{ch4RtowersE2even}) generate, each, a tower of excited states obtained by  adding an arbitrary even number of spinons, bulk strings, higher order boundary strings and quartets. In both the left and right towers, built upon (\ref{ch4LtowersE2even}) and (\ref{ch4RtowersE2even}) respectively, a localized bound-state at the left and the right edge is present and the number of spinon excitations is always odd. 

On top of the above three towers there exists a fourth one which correspond to states which host two bound-states. The state with the lowest energy consists into adding to the ground-state (\ref{ch4gsE2even}) a localized bound-state at each, left and right, edge. Since in the process the total spin of the state is shifted by $1$, no spinon is required. The resulting state 
\be\label{ch4LRtowerE2even}
|0\rangle_{LR}\; ,
\ee
which has a total spin $S^z=0$ and an energy 
$E_0+m_{L}+m'_{R}$, generates a tower of excited states
that comprises an arbitrary even number of spinons, bulk strings, higher order boundary strings and quartets. The number of spinon states in the whole tower is always even. We thus see that, in the $E_2$ sub-phase, the whole Hilbert space can be split into four towers generated by the states (\ref{ch4gsE2even}), (\ref{ch4LtowersE2even}), (\ref{ch4RtowersE2even}) and (\ref{ch4LRtowerE2even}). 

The situation in the $E_6$ sub-phase can be described in the very same way as above. Using the isometry (\ref{ch4z2}), we can obtain all the states in the sub-phase $E_6$ starting from the states in the sub-phase $E_2$ by reversing the sign of the total spin $S^z$ of the states. We obtain four towers of states in the sub-phase $E_6$ generated by the states $|0$, $|0\rangle_{L}, |-1\rangle_{L}$, $|0\rangle_{R}, |1\rangle_{R}$ and $|0\rangle_{LR}$ at energies $E_0, E_0+m_{L}+m, E_0+m'_{R}+m$ and $E_0+m_{L} +m'_{R}$.

The states in the sub-phases $E_3$ and $E_7$ are obtained from the states in the sub-phases $E_2$ and $E_6$ respectively by the transformation $L\leftrightarrow R$.

\subsection{$\it{B}$ phases}
\label{ch4sec:Bphase}

In the $B$ phases only one boundary bound-state is stabilized with energy below the mass gap $m$. The eight $B_{j=(1...8)}$ sub-phases correspond to the domains of boundary fields shown in table 
 
  \begin{table}[h!]
\centering
\caption{Values of the boundary fields corresponding to eight $B$ phases}
\begin{tabular}{|c|c|c|}
 \hline
 \hline
 Phase&$h_L$&$h_R$\\
 \hline
 $B_1$&$(0,h_{c1})$&$(h_{c1},h_{c2})$\\
 $B_2$&$(-h_{c1},0)$&$(h_{c1},h_{c2})$\\
    $B_3$&$(-h_{c2},-h_{c1})$&$(0,h_{c1})$\\   
   $B_4$&$(-h_{c2},-h_{c1})$&$(-h_{c1},0)$\\   
   $B_5$&$(-h_{c1},0)$&$(-h_{c2},-h_{c1})$\\   
   $B_6$&$(0,h_{c1})$&$(-h_{c2},-h_{c1})$\\    
    $B_7$&$(h_{c1},h_{c2})$&$(-h_{c1},0)$\\      
    $B_8$&$(h_{c1},h_{c2})$&$(0,h_{c1})$\\ 
   
     \hline 
      \end{tabular}
 \end{table}
 
 In the following we shall distinguish between odd end even chains and discuss the sub-phases $B_{j=(1...8)}$.

\subsubsection{\textbf{Odd number of sites}}
\label{ch4sec:Boddnumber}

In the $B_1$ phase, the ground state has total spin $S^z=-\frac{1}{2}$  which corresponds to a static spin distribution and is represented by 
\bea \label{ch4gstowerB1odd}|-\frac{1}{2}\rangle.\eea 

The ground state (\ref{ch4gstowerB1odd}) generates a tower of excited states obtained by adding an arbitrary even number of spinons, bulk strings and quartets. Unlike in the $A$ phases, there exists only a single boundary bound state solution corresponding to the bound state at the left edge. Starting from the ground state, this bound state can be added which has spin $S^z=+\frac{1}{2}$, by adding a spinon with arbitrary rapidity $\theta$ whose spin orientation can be either in the positive or negative $z$ direction resulting in the state with total spin $S^z=\pm \frac{1}{2}$ respectively. This state has energy $E_0+E_L+E_{\theta}$, and hence the lowest energy corresponds to the limit $\theta\rightarrow\pi$. This state is represented by 

\bea \label{ch4LtowerB1odd}|\pm\frac{1}{2}\rangle_{L}.\eea

The lowest excited states above (\ref{ch4LtowerB1odd}) consist of a spinon branch with $\theta \neq \pi$. On top of this, the state (\ref{ch4LtowerB1odd}) generates a tower of excited states obtained by adding an arbitrary even number of spinons, bulk strings, higher order boundary strings and quartets.

\vspace{1mm}

In the phase $B_2$, the lowest energy state contains a bound state at the left edge which has total spin $S^z=-\frac{1}{2}$. This state has energy $E_0+E_L$ and is represented by 

\bea \label{ch4LtowerB2odd}|-\frac{1}{2}\rangle_L.\eea 

On top of this, the state (\ref{ch4LtowerB2odd}) generates a tower of excited states obtained by adding an arbitrary {\it even} number of spinons, bulk strings, higher order boundary strings and quartets. The bound state at the left edge can be removed by adding a spinon with rapidity $\theta$, whose spin orientation is either in the positive or negative $z$ direction. The lowest energy state corresponds to $\theta\rightarrow\pi$, and has energy $E_0$ which is represented by 

\bea \label{ch4gstowerB2odd}|\pm\frac{1}{2}\rangle.\eea

The lowest energy of this state above the state (\ref{ch4gstowerB2odd}) consists into a spinon branch with $\theta\neq\pi$, which has energy $E_0+E_{\theta}$. The ground state (\ref{ch4gstowerB2odd})
generates a tower of excited states obtained by adding an arbitrary even number of spinons, bulk strings and quartets.

\vspace{2mm}

 By using the transformation $L\rightarrow R$, the states in the phases $B_8$ and $B_7$ can be obtained by starting with the states in the phases $B_1$ and $B_2$ respectively.  By using the isometry (\ref{ch4z2}), the states in the phases $B_5,B_6,B_3$ and $B_4$ can be obtained from the states in the phases $B_1,B_2,B_7$ and $B_8$ respectively. The results are summarized in the table (\ref{ch4tableBodd}).

\begin{table}[h!]
\centering
\caption{Energies and local fermionic parities of the ground state and the lowest energy states corresponding to each tower in all the $B$ phases for odd number of sites is summarized below.}
\begin{tabular}{|c|c|ccc|}
\hline
\hline
\;\;Phase\;\;&\;\;State\;\;&\;\;Energy\;\;&\;\;$\mathcal{P}_L$\;\;&\;\;$\mathcal{P}_R$\;\;\\
\hline
$B_1$&$|-\frac{1}{2}\rangle$& $E_0$ \;(g.s)& 1 & 1\\
\;&$|\pm\frac{1}{2}\rangle_{L}$& $E_0+m_L+m$ & -1 & 1\\
\hline
$B_8$&$|-\frac{1}{2}\rangle$& $E_0$,\;(g.s) & 1 & 1\\
\;&$|\pm\frac{1}{2}\rangle_{R}$&  $E_0+m_R+m$& 1 & -1\\
\hline
$B_2$&$|-\frac{1}{2}\rangle_{L}$& $E_0+E_L$, (g.s)& -1 & 1\\
\;&$|\pm\frac{1}{2}\rangle$& $E_0+m$  & 1 & -1\\
\hline
$B_7$&$|\pm\frac{1}{2}\rangle$& $E_0+m$ & 1 & 1\\
\;&$|-\frac{1}{2}\rangle_{R}$&  $E_0+m_R$, \; (g.s)& 1 & -1\\
\hline
$B_4$&$|\frac{1}{2}\rangle$& $E_0$,\;(g.s)& 1 & 1\\
\;&$|\pm\frac{1}{2}\rangle_{R}$& $E_0+m_R+m$ & 1 & -1\\
\hline
$B_5$&$|\frac{1}{2}\rangle$& $E_0$,\;(g.s)& 1 & 1\\
\;&$|\pm\frac{1}{2}\rangle_{L}$& $E_0+m_L+m$  & -1 & 1\\
\hline
$B_3$&$|\pm\frac{1}{2}\rangle$& $E_0+m$ & 1 & 1\\
\;&$|\frac{1}{2}\rangle_{R}$&  $E_0+m_R, \; (g.s)$& 1 & -1\\
\hline
$B_6$&$|\pm\frac{1}{2}\rangle_{\theta}$& $E_0+m$ & 1 & 1\\
\;&$|\frac{1}{2}\rangle_{L}$&  $E_0+m_L, \; (g.s)$& -1 & 1\\
\hline
\hline
\end{tabular}
\label{ch4tableBodd}
\end{table}

\subsubsection{\textbf{Even number of sites}}
\label{ch4sec:Bevennumberb1}

In the phase $B_1$, the lowest energy state contains the bound state at the left edge and has total spin $S^z=0$ with energy $E_0+E_{L}$. This state is represented by

\bea \label{ch4LtowerB1even} |0\rangle_L. \eea

On top of this, the state (\ref{ch4LtowerB1even}) generates a tower of excited states obtained by adding an arbitrary {\it even} number of spinons, bulk strings, higher order boundary strings and quartets.

The bound state at the left edge can be removed by adding a spinon with rapidity $\theta$. The spin orientation of the spinon can be either in the positive or negative $z$ direction which results in the state to have total spin $S^z=0,-1$. The lowest energy of this state is obtained in the limit $\theta\rightarrow\pi$, and has energy $E_0+m$

\bea \label{ch4gstowerB1even} |0\rangle, \hspace{3mm} |-1\rangle.\eea

The lowest excited states above (\ref{ch4gstowerB1even}) consist of a spinon branch with $\theta \neq \pi$. On top of this, the state (\ref{ch4LtowerB1even}) generates a tower of excited states obtained by adding an arbitrary even number of spinons, bulk strings and quartets.

In the phase $B_2$, the state which does not contain bound state at either edge has total spin $S^z=0$ and has energy $E_0$. It is represented by (\ref{ch4gstowerB2even})

\bea \label{ch4gstowerB2even} |0\rangle.\eea  

The ground state (\ref{ch4gstowerB2even}) generates a tower of excited states obtained by adding an arbitrary even number of spinons, bulk strings and quartets. We can add the bound state at the left edge with spin $S^z=-\frac{1}{2}$ by adding a spinon with rapidity $\theta$, whose lowest energy corresponds to the limit $\theta\rightarrow\pi$.  This state has total spin $S^z=-1,0$ depending on the spin orientation of the spinon and has energy $E_0+E_L$.  It is represented by

\bea \label{ch4LtowerB2even}|-1\rangle_{L}, \hspace{3mm} |0\rangle_{L}.\eea

The lowest excited states above (\ref{ch4LtowerB2even}) consist of a spinon branch with energies given by (\ref{ch4energyhole}) and $\theta \neq \pi$. On top of this, the state (\ref{ch4LtowerB2even}) generates a tower of excited states obtained by adding an arbitrary even number of spinons, bulk strings, higher order boundary strings and quartets.

\vspace{2mm}

Similar to the odd number of sites case, the states in the phases $B_8$ and $B_7$ can be obtained by starting with the states in $B_1$ and $B_2$ respectively, by making the transformation $L\rightarrow R$. By using the isometry (\ref{ch4z2}), the states in the phases $B_5,B_6,B_3$ and $B_4$ can be obtained from the states in the phases $B_1,B_2,B_7$ and $B_8$ respectively.

\begin{table}[h!]
\centering
\caption{Energies and local fermionic parities of the ground state and the lowest energy states corresponding to each tower in all the $B$ phases for even number of sites is shown below.}
\begin{tabular}{|c|c|ccc|}
\hline
\hline
\;\;Phase\;\;&\;\;State\;\;&\;\;Energy\;\;&\;\;$\mathcal{P}_L$\;\;&\;\;$\mathcal{P}_R$\;\;\\
\hline
$B_1$&$|-1\rangle$, \;$|0\rangle$& $E_0+m$ & 1 & 1\\
\;&$|0\rangle_{L}$& $E_0+m_L$, \; (g.s) & -1 & 1\\
\hline
$B_8$&$|-1\rangle$, \;$|0\rangle$& $E_0+m$& 1 & 1\\
\;&$|0\rangle_{R}$& $E_0+m_R$, \; (g.s) & 1 & -1\\
\hline
$B_2$&$|-1\rangle_{L}$,\; $|0\rangle_{L}$& $E_0+m_L+m$& -1 & 1\\
\;&$|0\rangle$& $E_0$ , \;(g.s) & 1 & 1\\
\hline
$B_7$&$|-1\rangle_{R}$,\; $|0\rangle_{R}$& $E_0+m_R+m$& 1 & -1\\
\;&$|0\rangle$& $E_0$ , \;(g.s) & 1 & 1\\
\hline
$B_4$&$|1\rangle$, \;$|0\rangle$& $E_0+m$& 1 & 1\\
\;&$|0\rangle_{R}$& $E_0+m_R$, \; (g.s) & 1 & -1\\
\hline
$B_5$&$|1\rangle$, \;$|0\rangle$& $E_0+m$& 1 & 1\\
\;&$|0\rangle_{L}$& $E_0+m_L$, \; (g.s) & -1 & 1\\
\hline
$B_3$&$|1\rangle_{R}$,\; $|0\rangle_{R}$& $E_0+m_R+m$& 1 & -1\\
\;&$|0\rangle$& $E_0$ , \;(g.s) & 1 & 1\\
\hline
$B_6$&$|1\rangle_{L}$,\; $|0\rangle_{L}$& $E_0+E_L+m$& -1 & 1\\
\;&$|0\rangle$& $E_0$ , \;(g.s) & 1 & 1\\
\hline
\hline
\end{tabular}
\label{ch4tableBeven}
\end{table}  

Unlike in the $A$ phases where there exists bound states at both the edges, we have seen that in $B$ phases there exists only one bound state at either the left or the right edge. This leads to the Hilbert space in each $B$ phase breaking up into only two towers. 
The ground states and the lowest energy states corresponding to the two towers in all the $B$ phases are summarized in the tables (\ref{ch4tableBodd}), (\ref{ch4tableBeven}).

\subsection{$\it{D}$ phases}
\label{ch4sec:Dphase}

In the $D$ phases, just as in the $B$ phases, only one boundary bound-state is stabilized, but in contrast to $B$ phases, the bound state energy is above the maximum energy $M$ of a spinon. The eight $D_{j=(1...8)}$ sub-phases correspond to the domains of boundary fields shown in table 
 
  \begin{table}[h!]
\centering
\caption{Values of the boundary fields corresponding to eight $B$ phases}
\begin{tabular}{|c|c|c|}
 \hline
 \hline
 Phase&$h_L$&$h_R$\\
 \hline
 
 $D_1$&$(h_{c1},h_{c2})$&$(h_{c2},\infty)$\\ 
$D_2$&$(-h_{c2},-h_{c1})$&$(h_{c2},\infty)$\\  
$D_3$&$(-\infty,-h_{c2})$&$(h_{c1},h_{c2})$\\
 $D_4$&$(-\infty,-h_{c2})$&$(-h_{c2},-h_{c1})$\\   
  $D_5$&$(-h_{c2},-h_{c1})$&$(-\infty,-h_{c2})$\\   
 $D_6$&$(h_{c1},h_{c2})$&$(-\infty,-h_{c2})$\\ 
  $D_7$&$(h_{c2},\infty)$&$(-h_{c2},-h_{c1})$\\    
 $D_8$&$(h_{c2},\infty)$&$(h_{c1},h_{c2})$\\
   \hline 
      \end{tabular}
 \end{table}
 
 In the following we shall distinguish between odd end even chains and discuss the sub-phases $D_{j=(1...8)}$.

\subsubsection{\textbf{Odd number of sites}}
\label{ch4sec:Doddnumber}

In the $D_1$ phase, the ground state has total spin $S^z=-\frac{1}{2}$  which corresponds to a static spin distribution and is represented by 
\bea \label{ch4gstowerD1odd}|-\frac{1}{2}\rangle.\eea 

The ground state (\ref{ch4gstowerD1odd}) generates a tower of excited states obtained by adding an arbitrary even number of spinons, bulk strings and quartets. Unlike in the $A$ phases, there exists only a single boundary bound state solution corresponding to the bound state at the right edge. Starting from the ground state, this bound state can be added which has spin $S^z=+\frac{1}{2}$, by adding a spinon with arbitrary rapidity $\theta$ whose spin orientation can be either in the positive or negative $z$ direction resulting in the state with total spin $S^z=\pm \frac{1}{2}$ respectively. This state has energy $E_0+m'_R+E_{\theta}$, and hence the lowest energy corresponds to the limit $\theta\rightarrow\pi$. This state is represented by 

\bea \label{ch4RtowerD1odd}|\pm\frac{1}{2}\rangle_{R}\eea  

and has energy $E_0+m'_R+m$. The lowest excited states above (\ref{ch4RtowerD1odd}) consist of a spinon branch with $\theta \neq \pi$. On top of this, the state (\ref{ch4RtowerD1odd}) generates a tower of excited states obtained by adding an arbitrary {\it even} number of spinons, bulk strings, higher order boundary strings and quartets.

\vspace{1mm}

In the phase $D_3$, the lowest energy state contains a bound state at the left edge which has total spin $S^z=-\frac{1}{2}$. This state has energy $E_0+m'_L$ and is represented by 

\bea \label{ch4LtowerD3odd}|-\frac{1}{2}\rangle_L.\eea 

On top of this, the state (\ref{ch4LtowerD3odd}) generates a tower of excited states obtained by adding an arbitrary even number of spinons, bulk strings, higher order boundary strings and quartets.

The bound state at the left edge can be removed by adding a spinon with rapidity $\theta$, whose spin orientation is either in the positive or negative $z$ direction. The lowest energy state corresponds to $\theta\rightarrow\pi$, and has energy $E_0$ which is represented by 

\bea \label{ch4gstowerD3odd}|\pm\frac{1}{2}\rangle.\eea

The lowest energy of this state above the state (\ref{ch4gstowerD3odd}) consists into a spinon branch with $\theta\neq\pi$, which has energy $E_0+E_{\theta}$. The ground state (\ref{ch4gstowerD3odd})
generates a tower of excited states obtained by adding an arbitrary even number of spinons, bulk strings and quartets.

\vspace{2mm}

 By using the transformation $L\rightarrow R$, the states in the phases $D_8$ and $D_6$ can be obtained by starting with the states in the phases $D_1$ and $D_3$ respectively.  By using the isometry (\ref{ch4z2}), the states in the phases $D_5,D_7,D_2$ and $D_4$ can be obtained from the states in the phases $D_1,D_3,D_6$ and $D_8$ respectively. The results are summarized in the table (\ref{ch4tableDodd}).

\begin{table}[h!]
\centering
\caption{Energies and local fermionic parities of the ground state and the lowest energy states corresponding to each tower in all the $B$ phases for odd number of sites is summarized below.}
\begin{tabular}{|c|c|ccc|}
\hline
\hline
\;\;Phase\;\;&\;\;State\;\;&\;\;Energy\;\;&\;\;$\mathcal{P}_L$\;\;&\;\;$\mathcal{P}_R$\;\;\\
\hline
$D_8$&$|-\frac{1}{2}\rangle$& $E_0$ \;(g.s)& 1 & 1\\
\;&$|\pm\frac{1}{2}\rangle_{L}$& $E_0+m'_L+m$ & -1 & 1\\
\hline
$D_1$&$|-\frac{1}{2}\rangle$& $E_0$,\;(g.s) & 1 & 1\\
\;&$|\pm\frac{1}{2}\rangle_{R}$&  $E_0+m'_R+m$& 1 & -1\\
\hline
$D_3$&$|-\frac{1}{2}\rangle_{L}$& $E_0+m'_L$& -1 & 1\\
\;&$|\pm\frac{1}{2}\rangle$& $E_0+m$, \;(g.s)  & 1 & -1\\
\hline
$D_6$&$|\pm\frac{1}{2}\rangle$& $E_0+m$, \; (g.s) & 1 & 1\\
\;&$|-\frac{1}{2}\rangle_{R}$&  $E_0+m'_R$, & 1 & -1\\
\hline
$D_5$&$|\frac{1}{2}\rangle$& $E_0$,\;(g.s)& 1 & 1\\
\;&$|\pm\frac{1}{2}\rangle_{R}$& $E_0+m_R+m$ & 1 & -1\\
\hline
$D_4$&$|\frac{1}{2}\rangle$& $E_0$,\;(g.s)& 1 & 1\\
\;&$|\pm\frac{1}{2}\rangle_{L}$& $E_0+m'_L+m$  & -1 & 1\\
\hline
$D_2$&$|\pm\frac{1}{2}\rangle$& $E_0+m$,\; (g.s) & 1 & 1\\
\;&$|\frac{1}{2}\rangle_{R}$&  $E_0+m'_R, $& 1 & -1\\
\hline
$D_7$&$|\pm\frac{1}{2}\rangle_{\theta}$& $E_0+m$,\; (g.s) & 1 & 1\\
\;&$|\frac{1}{2}\rangle_{L}$&  $E_0+m_L$& -1 & 1\\
\hline
\hline
\end{tabular}
\label{ch4tableDodd}
\end{table}

\subsubsection{\textbf{Even number of sites}}
\label{ch4sec:Bevennumberd1}

In the phase $D_1$, the lowest energy state contains a spinon with rapidity $\theta\rightarrow\pi$. This state and has total spin $S^z=0,-1$ depending on the spin orientation of the spinon. This state has energy $E_0+m$ and is represented by

\bea \label{ch4gstowerD1even} |0\rangle, |-1\rangle. \eea

The lowest excited states above (\ref{ch4gstowerB1even}) consist of a spinon branch with $\theta \neq \pi$. On top of this, the state (\ref{ch4gstowerD1even}) generates a tower of excited states obtained by adding an arbitrary {\it even} number of spinons, bulk strings, higher order boundary strings and quartets.

We can add the bound state at the right edge with spin $S^z=+\frac{1}{2}$ to the ground state by removing the existing spinon. This state has total spin $S^z=0$ and is represented by

\bea \label{ch4RtowerD1even} |0\rangle_{R}.\eea

This state has energy $E_0+m'_R$. On top of this, the state (\ref{ch4RtowerD1even}) generates a tower of excited states obtained by adding an arbitrary {\it even} number of spinons, bulk strings and quartets.

In the phase $D_3$, the ground state does not contain a bound state at either edge, and has total spin $S^z=0$ and energy $E_0$. It is represented by \ref{ch4gstowerD3even}

\bea \label{ch4gstowerD3even} |0\rangle.\eea  

The ground state (\ref{ch4gstowerD3even}) generates a tower of excited states obtained by adding an arbitrary even number of spinons, bulk strings and quartets. We can add the bound state at the left edge with spin $S^z=-\frac{1}{2}$ by adding a spinon with rapidity $\theta$, whose lowest energy corresponds to the limit $\theta\rightarrow\pi$.  This state has total spin $S^z=-1,0$ depending on the spin orientation of the spinon and has energy $E_0+E_L+m$.  It is represented by

\bea \label{ch4LtowerD3even}|-1\rangle_{L}, \hspace{3mm} |0\rangle_{L}.\eea

The lowest excited states above (\ref{ch4LtowerD3even}) consist of a spinon branch with $\theta \neq \pi$. On top of this, the state (\ref{ch4LtowerD3even}) generates a tower of excited states obtained by adding an arbitrary even number of spinons, bulk strings, higher order boundary strings and quartets.

\vspace{2mm}

Similar to the odd number of sites case, phases $D_8$ and $D_6$ can be obtained by starting with the states in the phases $D_1$ and $D_3$ respectively.  By using the isometry (\ref{ch4z2}), the states in the phases $D_5,D_7,D_2$ and $D_4$ can be obtained from the states in the phases $D_1,D_3,D_6$ and $D_8$ respectively. The results are summarized in the table (\ref{ch4tableDeven}).

\begin{table}[h!]
\centering
\caption{Energies and local fermionic parities of the ground state and the lowest energy states corresponding to each tower in all the $B$ phases for even number of sites is shown below.}
\begin{tabular}{|c|c|ccc|}
\hline
\hline
\;\;Phase\;\;&\;\;State\;\;&\;\;Energy\;\;&\;\;$\mathcal{P}_L$\;\;&\;\;$\mathcal{P}_R$\;\;\\
\hline
$D_1$&$|-1\rangle$, \;$|0\rangle$& $E_0+m$ \; (g.s)& 1 & 1\\
\;&$|0\rangle_{R}$& $E_0+m'_R$ & 1 & -1\\
\hline
$D_8$&$|-1\rangle$, \;$|0\rangle$& $E_0+m$\; (g.s)& 1 & 1\\
\;&$|0\rangle_{L}$& $E_0+m'_L$,  & -1 & 1\\
\hline
$D_3$&$|-1\rangle_{L}$,\; $|0\rangle_{L}$& $E_0+m'_L+m$& -1 & 1\\
\;&$|0\rangle$& $E_0$ , \;(g.s) & 1 & 1\\
\hline
$D_2$&$|1\rangle_{R}$,\; $|0\rangle_{R}$& $E_0+m'_R+m$& 1 & -1\\
\;&$|0\rangle$& $E_0$ , \;(g.s) & 1 & 1\\
\hline
$D_5$&$|1\rangle$, \;$|0\rangle$& $E_0+m$\; (g.s) & 1 & 1\\
\;&$|0\rangle_{R}$& $E_0+m'_R$ & 1 & -1\\
\hline
$D_4$&$|1\rangle$, \;$|0\rangle$& $E_0+m$\; (g.s) & 1 & 1\\
\;&$|0\rangle_{L}$& $E_0+m'_L$ & -1 & 1\\
\hline
$D_6$&$|-1\rangle_{R}$,\; $|0\rangle_{R}$& $E_0+m'_R+m$& 1 & -1\\
\;&$|0\rangle$& $E_0$ , \;(g.s) & 1 & 1\\
\hline
$D_7$&$|1\rangle_{L}$,\; $|0\rangle_{L}$& $E_0+m'_L+m$& -1 & 1\\
\;&$|0\rangle$& $E_0$ , \;(g.s) & 1 & 1\\
\hline
\hline
\end{tabular}
\label{ch4tableDeven}
\end{table}

\subsection{$\it{C}$ phases}
\label{ch4sec:Cphase}
\subsubsection{\textbf{Odd number of sites}}
 In the phases $\it{C_1}, \it{C_3}$, the ground state has total spin $S^z=\mp\frac{1}{2}$ respectively, which corresponds to a static spin distribution. The ground states in $\it{C_1}, \it{C_3}$ are represented by 
 
 \bea \label{ch4gstowerC13odd}|\mp\frac{1}{2}\rangle \eea
 
 respectively. The energy of these states is $E_0$. On top of this, the state (\ref{ch4gstowerC13odd}) generates a tower of excited states obtained by adding an arbitrary {\it even} number of spinons, bulk strings, higher order boundary strings and quartets.

 In the phases $\it{C_2}, \it{C_4}$ the ground state is two fold degenerate and contains a spinon with rapidity $\theta\rightarrow\pi$. The spin orientation of the spinon dictates the total spin $S^z=\pm\frac{1}{2}$ of the state. They are represented by 
 
 \bea \label{ch4gstowerC24odd}|\pm\frac{1}{2}\rangle.\eea  
 
 and have energy $E_0+m$. The lowest excited states above the state (\ref{ch4gstowerC24odd}) consist of a spinon branch with $\theta \neq \pi$. On top of this, the state (\ref{ch4gstowerC24odd}) generates a tower of excited states obtained by adding an arbitrary even number of spinons, bulk strings, higher order boundary strings and quartets.

 \subsubsection{\textbf{Even number of sites}}
 In the phase $\it{C_1}$, the ground state contains a spinon with rapidity $\theta\rightarrow\pi$ on top of the static spin distribution of the ground state in the phase $C_1$ corresponding to odd number of sites case. It is two fold degenerate with energy $E_0+m$ and and have total spin $S^z=0$, $S^z=-1$ corresponding to the spin orientation of the spinon which is along the positive and negative $z$ direction respectively. The are represented by 
 
 \bea \label{ch4gstowerC1even}|0\rangle, \;|-1\rangle. \eea

 Similarly, in the phase $\it{C_3}$, the ground state contains a spinon with rapidity $\theta\rightarrow\pi$ on top of the static spin distribution of the ground state in the phase $C_3$ corresponding to odd number of sites case.  It is two fold degenerate with energy $E_0+m$ and has total spin $S^z=0, 1$ corresponding to the spin orientation of the spinon which is along the negative and positive $z$ direction respectively. The are represented by 
 
 \bea \label{ch4gstowerC3even}|0\rangle, \;|1\rangle. \eea
 
The lowest excited states above the state (\ref{ch4gstowerC1even}) and (\ref{ch4gstowerC3even}) consist of a spinon branch with $\theta \neq \pi$. On top of this, the states (\ref{ch4gstowerC1even}) and (\ref{ch4gstowerC3even}) generate towers of excited states obtained by adding an arbitrary even number of spinons, bulk strings, higher order boundary strings and quartets.
 
 In the phases $C_2,C_4$, the ground state has total spin $S^z=0$ and energy $E_0$. They are represented by 
 
 \bea \label{ch4gstowerC24even} |0\rangle. \eea

On top of this, the states (\ref{ch4gstowerC24even}) generate towers of excited states obtained by adding an arbitrary even number of spinons, bulk strings, higher order boundary strings and quartets.

\subsection{Boundary Eigenstate Phase Transition}
\label{sec:eigenstatept}

As we saw there exists two critical value of the edge fields $h_{c1}=\Delta-1$ and $h_{c2}=\Delta+1$, at each edge associated with the existence of an edge bound state. For $|h_{i=(L,R)}| < h_{c1}$ and $|h_{i=(L,R)}| > h_{c2}$, there exists an exponentially localized bound state at the corresponding edge  $i=(L,R)$, whose energy is less than the mass gap $m$ and greater than the band height $M$ respectively. For $h_{c1}<|h_{i=(L,R)}| < h_{c2}$, the bound state at the corresponding edge is absent.  

The three types of phases $A,E,F$ and $B,D$ and $C$ distinguish themselves by the number of bound states  they support, i.e: two, one and zero respectively. Independently of the parity of $N$ we showed that in the $A,E,F$-type phases the Hilbert space splits into four towers of excited states while there exists two towers in the $B,D$-type phases and only one tower in the $C$-type phases.
When compared to the ground state phase diagrams (see Figs.(\ref{ch4gsXXZeven},\ref{ch4gsXXZodd})) each quadrant splits into one $C$ sub-phase, two $B,D,E$ sub-phases and one $A,F$ sub-phase as displayed in  the Fig. \ref{ch4pd1}. At this point a natural question arises: what is the nature of the transition that occurs as one moves from an $A$ or $E$ sub-phase to a $B$ or from $F$ or $E$ to $D$ sub-phase or from a $B$ or $D$ sub-phase to a  $C$ sub-phase, and also from $A$ or $F$ sub-phase to a $C$ sub-phase by varying the edge fields. 

Without loss of generality let us fix on quadrant with $h_L >0$ and $h_R > 0$. Consider first the situation where 
both $h_{L,(R)} <h_{c1}$, that is one sits in the $A_1$  sub-phase. Let  the left boundary magnetic field $h_L$  be fixed while the right boundary fields $h_R$ is increased. As $h_R$  is increased above the critical value $h_{c1}$, we move into the sub-phase ${B}_1$. The two states which contain the bound state at the right edge no longer exist. On the boundary between the $A_1$ and $B_1$ sub-phases, the energy of the bound state and energy of the spinon with rapidity $\theta=\pi$ coincide $m_R=m=E_{\theta\rightarrow \pi}$. Hence it is natural to interpret that the bound state at the right edge leaks into the bulk by taking the form of a spinon with rapidity $\theta \sim \pi$. Similarly, moving from ${A}_1$ to ${B}_8$ (see Fig. \ref{ch4pd1}), the bound state corresponding to left boundary leaks into the bulk.  Similarly, moving from ${B}_1$ to ${C}_1$, the value of the left boundary field takes values greater than critical value $h_{c1}$, and hence the bound state present at the left edge leaks into the bulk in a similar way, resulting in $C_1$ having no bound states at either edge. The same phenomena of bound states leaking into the bulk occurs as one moves from any $A$ sub-phase into the respective $B$ and $C$ sub-phases. 

Now consider the situation where 
both $h_{L,(R)} >h_{c2}$, that is one sits in the $F_1$  sub-phase. Let the left boundary magnetic field $h_L$  be fixed while the right boundary fields $h_R$ is decreased. As $h_R$  is decreased below the critical value $h_{c2}$, we move into the sub-phase ${D}_8$. The two states which contain the bound state at the right edge no longer exist. On the boundary between the $F_1$ and $D_8$ sub-phases, the energy of the bound state and energy of the spinon with rapidity $\theta=0$ coincide $m_R=M=E_{\theta\rightarrow 0}$. Hence the bound state at the right edge leaks into the bulk by taking the form of a spinon with rapidity $\theta \sim 0$. Similarly, moving from ${F}_1$ to ${D}_1$ (see Fig. \ref{ch4pd1}), the bound state corresponding to left boundary leaks into the bulk.  Similarly, moving from ${D}_1$ to ${C}_1$, the value of the right boundary field takes values less than critical value $h_{c2}$, and hence the bound state present at the right edge leaks into the bulk in a similar way, resulting in $C_1$ having no bound states at either edge. The same phenomena of bound states leaking into the bulk occurs as one moves from any $F$ sub-phase into the respective $D$ and $C$ sub-phases.

Now consider the situation where 
both $h_{R} >h_{c2}$, $h_L<h_{c1}$, that is one sits in the $E_1$  sub-phase. Let the left boundary magnetic field $h_L$  be fixed while the right boundary fields $h_R$ is decreased. As $h_R$  is decreased below the critical value $h_{c2}$, we move into the sub-phase ${B}_1$. The two states which contain the bound state at the right edge no longer exist. On the boundary between the $E_1$ and $B_1$ sub-phases, the energy of the bound state and energy of the spinon with rapidity $\theta=0$ coincide $m_R=M=E_{\theta\rightarrow 0}$. Hence the bound state at the right edge leaks into the bulk by taking the form of a spinon with rapidity $\theta \sim 0$. Similarly, moving from ${E}_1$ to ${D}_1$ (see Fig. \ref{ch4pd1}), the bound state corresponding to left boundary leaks into the bulk where its rapidity is $\theta\sim\pi$. The same phenomena of bound states leaking into the bulk occurs as one moves from any $E$ sub-phase into the respective $B$ and $D$ sub-phases. Associated with the appearance or disappearance of localized bound states is the fact that when one goes from any sub-phase to another, the whole structure of the Hilbert space changes. The excited states organize themselves into towers whose number is different in the $A,E,F$ and $B,D$ and $C$ type phases.

The towers are labeled by the bound state parities 

\bea
\mathcal{P}_{L,R}=(-1)^{\mathcal{N}_{L,R}}
\eea
where $\mathcal{N}_{L,R}$ are number of bound states at the left and right edges. The four towers in $A,E,F$-type phases are labeled by $(\mathcal{P}_{L}, \mathcal{P}_{R})=(\pm 1, \pm 1)$, the two towers in the $B,D$-type phases
by $(\mathcal{P}_{L}, \mathcal{P}_{R})=(\pm 1, +1)$ and $(\mathcal{P}_{L}, \mathcal{P}_{R})=(+1, \pm 1)$ and the unique tower of the $C$-type phases by 
$(\mathcal{P}_{L}, \mathcal{P}_{R})=(+1, +1)$.

\section{Discussions}

 In this work we considered the spin $1/2$ XXZ chain in the gapped anti-ferromagnetic regime in the presence of boundary magnetic fields. We analyzed it using Bethe ansatz and extensive DMRG techniques. It is known that in the absence of boundary fields, the Hamiltonian has discrete $\mathbb{Z}_2$ spin flip symmetry which is  spontaneously broken and the system exhibits degenerate ground states. One can build up excitations on top of these two symmetry broken ground states and the system exhibits two degenerate towers of eigenstates. It is also known that there exists strong zero modes which map these pairs of states \cite{Fendley}. In this work we have applied boundary magnetic fields which explicitly break the $\mathbb{Z}_2$ spin flip symmetry and solved the system exactly using the method of Bethe Ansatz. We found that the system exhibits a very rich phase diagram with several phases characterized by the ground state the system exhibits and also by the number of possible bound states at both the edges and their energy. There exists two critical values of the boundary magnetic fields which dictate whether a bound state may or may not be present at the corresponding edge. The energy of the bound state depends on the value of the magnetic field at the corresponding boundary which plays a very important role in selecting the ground state exhibited by the system. The ground state exhibited by the system depends on whether the number of sites is even or odd and also depends on the value and the orientation of the boundary magnetic fields. 



 The boundary magnetic fields may drive the system through a phase transition where the ground state of the system changes. Such a phase transition may or may not be associated with a loss of the bound state at one of the edges. When a bound state is not lost, the phase transition is a first order phase transition where a level crossing occurs. In the case where the bound state is lost, it leaks into the bulk and turns into a spinon, and this phase transition is associated with the change in the number of towers of the Hilbert space and is termed `eigenstate phase transition' or `Hilbert space phase transition'. This new type of phase transition can also occur when the ground state of the system remains unchanged but the other towers containing the bound states are lost. This phase transition can be probed through dynamics that involve operators associated with boundaries at either zero or infinite temperature. We hope to address these questions in the future work.

\bibliography{refXXZ}

\appendix
\label{appendix}
\begin{widetext}

\section{Bethe ansatz Solution}
In this section we construct the ground state and boundary excitations in each region of the phase diagram for both odd and even number of sites.

\subsection{Region $A_1$: odd number of sites}
 The region $A_1$ corresponds to the following values of the boundary magnetic fields: $0<h_L,h_R< h_{c1}$. This corresponds to $\epsilon_{\alpha}=-\tilde{\epsilon}_{\alpha}+i\pi$, with $\tilde{\epsilon}_{\alpha}<1$, $\alpha=L,R$.

\smallskip

First consider the state with all real $\lambda$, which take values between $(-\pi,\pi]$. Applying logarithm to (\ref{ch4be1}) we obtain 

\bea\nonumber
2N \varphi(\lambda_j,1)-\sum_{\alpha=L,R}\varphi(\lambda_j,1-\tilde{\epsilon}_{\alpha})+\varphi(\lambda_j,1)+\varphi^{\prime}(\lambda_j,1)\\=2\pi I_j+ \sum_{\sigma=\pm}\sum_{k\neq j}\varphi(\lambda_j+\sigma \lambda_k,2)\label{ch4logbea11},
\eea
where \bea \varphi(x,y)=\ln \left(\frac{\sin \frac{1}{2}(x-i\gamma y)}{\sin \frac{1}{2}(x+i\gamma y)}\right), \;\;\;\;  \varphi^{\prime}(x,y)=\ln \left(\frac{\cos \frac{1}{2}(x-i\gamma y)}{\cos \frac{1}{2}(x+i\gamma y)}\right).\eea
We define the counting function $\nu(\lambda)$ such that $\nu(\lambda_j)=I_j$. Differentiating (\ref{ch4logbea11}) and using $\frac{d}{d\lambda}\nu(\lambda)=\rho(\lambda)$, we obtain

\bea\nonumber
(2N+1) a(\lambda,1)-\sum_{\alpha=L,R}a(\lambda-\pi,1-\tilde{\epsilon}_{\alpha})+a(\lambda-\pi,1)-2\pi\delta(\lambda)-2\pi\delta(\lambda-\pi)\\=2\pi\rho(\lambda)+\sum_{\sigma=\pm}\int a(\lambda+\sigma \mu,2)\rho(\mu)d\mu, \label{ch4logbea12}\eea
where we have removed the solutions $\lambda=0,\pi$ as they lead to a vanishing wavefunction \cite{ODBA}. Here

\bea a(x,y)=\frac{\sinh(\gamma y)}{\cosh(\gamma y)-\cos(\lambda)}.\eea
The above equation can be solved by applying Fourier transform

\bea f(x)= \sum_{k=-\infty}^{\infty}\hat{f}(\omega)e^{i\omega x}, \;\;\;\;\;\;\;\; \hat{f}(\omega)=\frac{1}{2\pi}\int_{-\pi}^{\pi}f(x)e^{-i\omega x}dx. \eea
Using $\hat{a}(\omega,y)=e^{-\gamma y |\omega|}$, we obtain the following density distribution for the state with all real roots

\bea\nonumber \hat{\rho}_{\ket{\frac{1}{2}}_{A_1}}(\omega)=\frac{(2N+1)e^{-\gamma|\omega|} + (-1)^{\omega}e^{-\gamma|\omega|}-(1+(-1)^{\omega})}{4\pi(1+e^{-2\gamma|\omega|})}\\-\frac{\sum_{\alpha=L,R}(-1)^{\omega}e^{-\gamma (1-\tilde{\epsilon}_{\alpha})|\omega|}}{4\pi(1+e^{-2\gamma|\omega|})}. \label{ch4denodda11}\eea
The reason for the subscripts will become evident when we find the spin $S^z$ of the state. The number of Bethe roots can be obtained by using the relation 
\bea \label{ch4nroots1}M= \int_{-\pi}^{\pi} \rho(\lambda)d\lambda. \eea 
The total spin $S^z$ of the state can be found using the relation $S^z=\frac{N}{2}-M$. Using (\ref{ch4denodda11}) in the above relations we find that the total spin $S^z$ of the state described by the distribution $\hat{\rho}_{\ket{\frac{1}{2}}_{A_1}}(\omega)$ is $S^z =\frac{1}{2}$.  We denote this state by $\ket{\frac{1}{2}}_{A_1}$

\smallskip

By starting with the Bethe equations corresponding to all spin down reference state we have

\bea\nonumber
(2N+1) a(\lambda,1)-\sum_{\alpha=L,R}a(\lambda-\pi,1+\tilde{\epsilon}_{\alpha})+a(\lambda-\pi,1)-2\pi\delta(\lambda)-2\pi\delta(\lambda-\pi)\\=2\pi\rho(\lambda)+\sum_{\sigma=\pm}\int a(\lambda+\sigma \mu,2)\rho(\mu)d\mu. \label{ch4logbea13}\eea
Following the same procedure as above, we obtain the following distribution for a state with all real $\lambda$ 

\bea\nonumber \hat{\rho}_{\ket{-\frac{1}{2}}_{A_1}}(\omega)=\frac{(2N+1)e^{-\gamma|\omega|} + (-1)^{\omega}e^{-\gamma|\omega|}-(1+(-1)^{\omega})}{4\pi(1+e^{-2\gamma|\omega|})}\\
-\frac{\sum_{\alpha=L,R}(-1)^{\omega}e^{-\gamma (1+\tilde{\epsilon}_{\alpha})|\omega|}}{4\pi(1+e^{-2\gamma|\omega|})}. \label{ch4denodda12}\eea
The total spin $S^z$ of this state is $S^z=-\frac{1}{2}$. We denote this state by $\ket{-\frac{1}{2}}_{A_1}$. Using (\ref{ch4energy}) we can calculate the energy difference between the two states $\ket{\frac{1}{2}}_{A_1}$ and $\ket{-\frac{1}{2}}_{A_1}$. We have \bea \label{ch4diffena11}E_{\ket{\frac{1}{2}}_{A_1}}-E_{\ket{-\frac{1}{2}}_{A_1}}= h_L+h_R-2\sinh\gamma\sum_{\alpha=L,R}\int_{-\pi}^{\pi}a(\lambda,1)\;\delta\rho_{\tiny{\ket{\frac{1}{2}},\ket{-\frac{1}{2}}}}(\lambda)d\lambda.
\eea
Here $\delta\rho_{\tiny{\ket{\frac{1}{2}},\ket{-\frac{1}{2}}}}(\lambda)$ is the difference in the density distributions of the states $\ket{\frac{1}{2}}_{A_1}$ and $\ket{-\frac{1}{2}}_{A_1}$. The expression (\ref{ch4diffena11}) can be written as
 
 \bea  E_{\ket{\frac{1}{2}}_{A_1}}-E_{\ket{-\frac{1}{2}}_{A_1}}= h_L+h_R+4\pi\sinh\gamma\sum_{\omega=-\infty}^{\infty} \hat{a}(\omega,1)\Delta\hat{\rho}_{\tiny{\ket{\frac{1}{2}},\ket{-\frac{1}{2}}}}(\omega). \eea Using (\ref{ch4denodda11}) and (\ref{ch4denodda12}) in the above expression we obtain
 \bea  E_{\ket{\frac{1}{2}}_{A_1}}-E_{\ket{-\frac{1}{2}}_{A_1}}= h_L+h_R+ \sinh\gamma\sum_{\alpha=L,R} \sum_{\omega=-\infty}^{\infty}(-1)^{\omega}\; \frac{\sinh(\gamma \tilde{\epsilon}_{\alpha}|\omega|)}{\cosh(\gamma\omega)}e^{-\gamma|\omega|}. \eea
This can be written as \bea E_{\ket{\frac{1}{2}}_{A_1}}-E_{\ket{-\frac{1}{2}}_{A_1}}=m_L+m_R,\eea where

\bea m_{\alpha}=h_{\alpha}+ \sinh\gamma \sum_{\omega=-\infty}^{\infty}(-1)^{\omega}\; \frac{\sinh(\gamma \tilde{\epsilon}_{\alpha}|\omega|)}{\cosh(\gamma\omega)}e^{-\gamma|\omega|}. \eea
Since $m_L,M_R>0$ in the region $A_1$, the ground state is $\ket{-\frac{1}{2}}_{A_1}.$

 \subsection{Region $A_1$: Even number of sites}
 
The Bethe equations corresponding to all spin up reference state have two boundary string solutions $\lambda_{bs \alpha}$, where

\bea \label{ch4boundstringA}\lambda_{bs \alpha}=\pi+\pm i\gamma(1-\tilde{\epsilon}_{\alpha}),\;\;\;\; \alpha=L,R.\eea 
Adding either of these two boundary strings to the Bethe equations (\ref{ch4be1}) and taking logarithm we obtain

\bea \nonumber 2N \varphi(\lambda_j,1)-\sum_{\alpha=L,R}\varphi(\lambda_j-\pi,1-\tilde{\epsilon}_{\alpha})+\varphi(\lambda_j,1)+\varphi^{\prime}(\lambda_j,1)-\varphi(\lambda,(3-\tilde{\epsilon}_{\beta}))\\-\varphi(\lambda,(1+\tilde{\epsilon}_{\beta}))=2\pi I_j+ \sum_{\sigma=\pm}\sum_{k\neq j}\varphi(\lambda_j+\sigma \lambda_k,2),
\eea
where $\beta$ is either $L$ or $R$. Differentiating the above equation with respect to $\lambda$ and taking the Fourier transform we obtain

\bea\nonumber\tilde{\rho}_{\ket{0}_{\beta A_1}} (\omega)=\tilde{\rho}_{\ket{\frac{1}{2}}_{A_1}}(\omega)+\Delta\tilde{\rho}_{\beta} (\omega), \\\Delta\tilde{\rho}_{\beta} (\omega)=-\frac{1}{4\pi}(-1)^{\omega}\frac{e^{-\gamma(3-\tilde{\epsilon}_{\beta})|\omega|}+e^{-\gamma(1+\tilde{\epsilon}_{\beta})|\omega|}}{1+e^{-2\gamma|\omega|}}.\label{ch4denbounda11} 
\eea
The spin of the state containing this boundary string can be calculated using $S^z=\frac{N}{2}-M$, where
\bea \label{ch4nroots2}M=1+\int_{-\pi}^{\pi}\rho_{\ket{0}_{\beta A_1}}(\lambda) d\lambda. \eea
We obtain $S^z_{\ket{0}_{\beta A_1}}=0$, $\beta=L,R$.  Hence there are two states with $S^z=0$ that correspond to the presence of the boundary strings $\lambda_{bsL}$ and $\lambda_{bsR}$. The energy of the boundary string can be calculated using (\ref{ch4energy}).  We have 

\bea \label{ch4boundena1}
E_{\lambda_{bs\beta}}= -\frac{2\sinh^2\gamma}{\cosh\gamma+\cosh\gamma(1-\tilde{\epsilon}_{\beta})}-2\sinh\gamma\int_{-\pi}^{\pi}a(\lambda-\pi,1)\Delta\rho_{\beta}(\lambda)d\lambda
\eea
Using (\ref{ch4denbounda11}) and evaluating the integral one obtains,

\bea \label{ch4bounden22}E_{\lambda_{bs\beta}}=-\sinh\gamma \sum_{\omega=-\infty}^{\infty} (-1)^{\omega}\frac{e^{-\gamma(1-\tilde{\epsilon}_{\beta})|\omega|}}{\cosh\gamma|\omega|}=-m_{\beta}.\eea
Hence the ground state is either $\ket{0}_{L, A_1}$ or $\ket{0}_{R, A_1}$ depending on the values of $h_L,h_R$.

\subsection{$C_1$ Odd and even number of sites}
In this region both $h_L,h_R$ take the following values: $h_{c1}<h_L,h_R<h_{c2}$. By starting with Bethe reference state with all spin down, and considering the state with all real $\lambda_j$, we obtain the following logarithmic form of Bethe  equations

\bea\nonumber
(2N+1) a(\lambda,1)-(l_1 a(\lambda,1+\tilde{\epsilon}_{L})+l_2 a(\lambda-\pi,1+\tilde{\epsilon}_{L}))-(r_1 a(\lambda,1+\tilde{\epsilon}_{R})+ r_2 a(\lambda-\pi,1+\tilde{\epsilon}_{R}))\\+a(\lambda-\pi,1)-2\pi\delta(\lambda)-2\pi\delta(\lambda-\pi)=2\pi\rho(\lambda)+\sum_{\sigma=\pm}\int a(\lambda+\sigma \mu,2)\rho(\mu)d\mu . \eea
By following the same procedure as above we obtain

\bea \nonumber\hat{\rho}_{(\ket{-\frac{1}{2}}_{C_1}}(\omega)=\frac{(2N+1)e^{-\gamma|\omega|} + (-1)^{\omega}e^{-\gamma|\omega|}-(1+(-1)^{\omega})}{4\pi(1+e^{-2\gamma|\omega|})}\\
-\frac{(l_1+l_2(-1)^{\omega})e^{-\gamma (1+\tilde{\epsilon}_{L})|\omega|}+(r_1+r_2(-1)^{\omega})e^{-\gamma (1+\tilde{\epsilon}_{R})|\omega|}}{4\pi(1+e^{-2\gamma|\omega|})}, \label{ch4denoddc11} \eea
where the parameters $l_1,l_2,r_1,r_2$ take the values given in (Tab:\ref{ch4tableap1}) for different values of $h_L,h_R$.

 \begin{table}[h!]
\centering
\caption{Values of the parameters in (\ref{ch4denoddc11}) corresponding to various ranges of the boundary magnetic fields}

\begin{tabular}{c|c|c|c|c}
\hline
\hline
  & \;\footnotesize{$h_{c1}<h_L<\sinh\gamma$}\;&\;\footnotesize{$h_{c1}<h_L<\sinh\gamma$} \; &\;\footnotesize{$\sinh\gamma<h_L<h_{c2}$}&\; \footnotesize{$\sinh\gamma<h_L<h_{c2}$}\\

&\; \footnotesize{$h_{c1}<h_R<\sinh\gamma$} \;& \;\footnotesize{$\sinh\gamma<h_R<h_{c2}$}\;&\;\footnotesize{$h_{c1}<h_R<\sinh\gamma$}\;&\;\footnotesize{$\sinh\gamma<h_R<h_{c2}$}\;\\
\hline
 $l_1$&  0 & 0  &1&1  \\
  $l_2$&1& 1& 0&0\\
  $r_1$&0& 1& 0& 1\\
  $r_2$ &1&0&1&0\\
 \hline
\hline
\end{tabular}
\label{ch4tableap1}
\end{table}  

The total spin $S^z$ can be obtained by using $S^z=\frac{N}{2}$, where $M$ is given by (\ref{ch4nroots1}). We obtain $S^z_{(\ket{-\frac{1}{2}}_{C_1})}=-\frac{1}{2}$. To obtain the lowest energy state corresponding to even number of sites, we need to add a propagating hole (spinon) to the state with all real roots corresponding to all spin down reference state. We obtain 

\bea\nonumber
(2N+1) a(\lambda,1)-(l_1 a(\lambda,1+\tilde{\epsilon}_{L})+l_2 a(\lambda-\pi,1+\tilde{\epsilon}_{L}))-(r_1 a(\lambda,1+\tilde{\epsilon}_{R})+r_2 a(\lambda-\pi,1+\tilde{\epsilon}_{R}))+a(\lambda-\pi,1)\\-2\pi\delta(\lambda)-2\pi\delta(\lambda-\pi)-2\pi\delta(\lambda-\theta)-2\pi\delta(\lambda+\theta)=2\pi\rho(\lambda)+\sum_{\sigma=\pm}\int a(\lambda+\sigma \mu,2)\rho(\mu)d\mu  \eea
By following the same procedure as above we obtain

\bea \hat{\rho}_{\ket{-1}_{C_1}}(\omega)= \hat{\rho}_{\ket{-\frac{1}{2}}_{C_1}}(\omega)+ \Delta \hat{\rho}_{\theta}(\omega),\eea
where
\bea \Delta \hat{\rho}_{\theta}(\omega)= -\frac{1}{4\pi}\sum_{\omega} \frac{\cos(\theta \omega)}{\cosh(\gamma \omega)}e^{\gamma|\omega|}.
\eea
The total spin of this state is  $S^z=-1$. We denote this state by by $\ket{-1}_{C_1}$. The energy of the spinon is

\bea E_{\theta}= -4\pi \sum_{\omega}\hat{a}(\omega,1)\Delta \hat{\rho}_{\theta}(\omega).\eea
After simplification we obtain

\bea \label{ch4energyhole}E_{\theta}=\sinh\gamma\sum_{\omega=-\infty}^{\infty}\frac{\cos(\theta\omega)}{\cosh(\gamma\omega)},\eea
where \bea m<E_{\theta}<M, \hspace{6mm} m= \sinh\gamma\sum_{\omega=-\infty}^{\infty}\frac{(-1)^{\omega})}{\cosh(\gamma\omega)}, \hspace{6mm} M= \sinh\gamma\sum_{\omega=-\infty}^{\infty}\frac{1}{\cosh(\gamma\omega)}.\eea

\subsection{$F_1$ Odd number of sites}

The region $F_1$ corresponds to the following values of the boundary magnetic fields: $h_{c2}<h_L,h_R$. This corresponds to $\epsilon_{\alpha}=-\tilde{\epsilon}_{\alpha}$, with $\tilde{\epsilon}_{\alpha}<1$, $\alpha=L,R$. Starting with the Bethe equations corresponding to all spin up reference state and considering the state with all real roots, we have

\bea\nonumber(2N+1) a(\lambda,1)-\sum_{\alpha=L,R}a(\lambda,1-\tilde{\epsilon}_{\alpha})+a(\lambda-\pi,1)-2\pi\delta(\lambda)-2\pi\delta(\lambda-\pi)\\=2\pi\rho(\lambda)+\sum_{\sigma=\pm}\int a(\lambda+\sigma \mu,2)\rho(\mu)d\mu. \eea
Following the usual procedure we obtain the following density distribution

\bea\nonumber \hat{\rho}_{\ket{\frac{1}{2}}_{F_1}}(\omega)=\frac{(2N+1)e^{-\gamma|\omega|} + (-1)^{\omega}e^{-\gamma|\omega|}-(1+(-1)^{\omega})-\sum_{\alpha=L,R}e^{-\gamma (1-\tilde{\epsilon}_{\alpha})|\omega|}}{4\pi(1+e^{-2\gamma|\omega|})} \\\label{ch4denoddf11}.\eea
The total spin $S^z$ of this state is $S^z=\frac{1}{2}$. We denote this state by $\ket{\frac{1}{2}}_{F_1}$. By starting with the Bethe equations corresponding to all spin down reference state we have

\bea\nonumber
(2N+1) a(\lambda,1)-\sum_{\alpha=L,R}a(\lambda,1+\tilde{\epsilon}_{\alpha})+a(\lambda-\pi,1)-2\pi\delta(\lambda)-2\pi\delta(\lambda-\pi)\\=2\pi\rho(\lambda)+\sum_{\sigma=\pm}\int a(\lambda+\sigma \mu,2)\rho(\mu)d\mu. \label{ch4logbef11}\eea
Following the same procedure as above, we obtain the following distribution for a state with all real $\lambda$ 

\bea \hat{\rho}_{\ket{-\frac{1}{2}}_{F_1}}(\omega)=\frac{(2N+1)e^{-\gamma|\omega|} + (-1)^{\omega}e^{-\gamma|\omega|}-(1+(-1)^{\omega})-\sum_{\alpha=L,R}e^{-\gamma (1+\tilde{\epsilon}_{\alpha})|\omega|}}{4\pi(1+e^{-2\gamma|\omega|})}.\label{ch4denoddf12} \eea
The total spin $S^z$ of this state is $S^z=-\frac{1}{2}$. We denote this state by $\ket{-\frac{1}{2}}_{F_1}$. Using (\ref{ch4energy}) we can calculate the energy difference between the two states $\ket{\frac{1}{2}}_{F_1}$ and $\ket{-\frac{1}{2}}_{F_1}$. We have \bea \label{ch4diffenf11}E_{\ket{\frac{1}{2}}_{F_1}}-E_{\ket{-\frac{1}{2}}_{F_1}}= h_L+h_R-2\sinh\gamma\int_{-\pi}^{\pi}a(\lambda,1)\;\delta\rho_{\tiny{\ket{\frac{1}{2}},\ket{-\frac{1}{2}}}}(\lambda)d\lambda,
\eea
where $\delta\rho_{\tiny{\ket{\frac{1}{2}},\ket{-\frac{1}{2}}}}(\lambda)$ is the difference in the density distributions of the states $\ket{\frac{1}{2}}_{F_1}$ and $\ket{-\frac{1}{2}}_{F_1}$. The expression (\ref{ch4diffenf11}) can be written as
 
 \bea  E_{\ket{\frac{1}{2}}_{F_1}}-E_{\ket{-\frac{1}{2}}_{F_1}}= h_L+h_R+4\pi\sinh\gamma\sum_{\omega=-\infty}^{\infty} \hat{a}(\omega,1)\Delta\hat{\rho}_{\tiny{\ket{\frac{1}{2}},\ket{-\frac{1}{2}}}}(\omega). \eea
 Using (\ref{ch4denoddf11}) and (\ref{ch4denoddf12}) in the above expression we obtain
 \bea  E_{\ket{\frac{1}{2}}_{F_1}}-E_{\ket{-\frac{1}{2}}_{F_1}}= h_L+h_R+ \sinh\gamma\sum_{\alpha=L,R} \sum_{\omega=-\infty}^{\infty}\; \frac{\sinh(\gamma \tilde{\epsilon}_{\alpha}|\omega|)}{\cosh(\gamma\omega)}e^{-\gamma|\omega|}. \eea
 This can be written as \bea E_{\ket{\frac{1}{2}}_{F_1}}-E_{\ket{-\frac{1}{2}}_{F_1}}=m'_L+m'_R,\eea where

\bea m'_{\alpha}=h_{\alpha}+ \sinh\gamma \sum_{\omega=-\infty}^{\infty}\; \frac{\sinh(\gamma \tilde{\epsilon}_{\alpha}|\omega|)}{\cosh(\gamma\omega)}e^{-\gamma|\omega|}. \eea

Hence, the ground state is $\ket{-\frac{1}{2}}_{F_1}$.

\subsection{$F_1$: Even number of sites}
For even number of sites the lowest energy state is obtained by starting with Bethe equations corresponding to all spin down reference state and considering a state with all real roots and a spinon. We have

\bea\nonumber
(2N+1) a(\lambda,1)-\sum_{\alpha=L,R}a(\lambda,1+\tilde{\epsilon}_{\alpha})+a(\lambda-\pi,1)-2\pi\delta(\lambda)-2\pi\delta(\lambda-\pi)\\-2\pi\delta(\lambda-\theta)-2\pi\delta(\lambda+\theta)=2\pi\rho(\lambda)+\sum_{\sigma=\pm}\int a(\lambda+\sigma \mu,2)\rho(\mu)d\mu \eea
By following the same procedure as above we obtain

\bea \hat{\rho}_{\ket{-1}_{\theta, F_1}}(\omega)= \hat{\rho}_{\ket{-\frac{1}{2}}_{F_1}}(\omega)+ \Delta \hat{\rho}_{\theta}(\omega).\eea
The total spin of this state is $S^z=-1$. We denote this state by $\ket{-1}_{F_1}$. The Bethe equations corresponding to all spin up reference state contain two boundary string solutions $\lambda'_{bs \alpha}=\pm i\gamma(1-\tilde{\epsilon}_{\alpha})$, $\alpha=L,R$. Considering a state with all real roots and either of the boundary strings $\lambda'_{bs \alpha}$, we have

\bea \nonumber 2N \varphi(\lambda_j,1)-\sum_{\alpha=L,R}\varphi(\lambda_j,1-\tilde{\epsilon}_{\alpha})+\varphi(\lambda_j,1)+\varphi^{\prime}(\lambda_j,1)-\varphi(\lambda,(3-\tilde{\epsilon}_{\beta}))\\-\varphi(\lambda,(1+\tilde{\epsilon}_{\beta}))=2\pi I_j+ \sum_{\sigma=\pm}\sum_{k\neq j}\varphi(\lambda_j+\sigma \lambda_k,2),
\eea
where $\beta$ is either $L$ or $R$. Differentiating the above equation with respect to $\lambda$ and taking the Fourier transform we obtain

\bea\label{ch4denboundf11} \tilde{\rho}_{\ket{0'}_{\beta F_1}} (\omega)=\tilde{\rho}_{\ket{\frac{1}{2}}_{F_1}}(\omega)+\Delta\tilde{\rho'}_{\beta} (\omega), \;\;\Delta\tilde{\rho'}_{\beta} (\omega)=-\frac{1}{4\pi}\frac{e^{-\gamma(3-\tilde{\epsilon}_{\beta})|\omega|}+e^{-\gamma(1+\tilde{\epsilon}_{\beta})|\omega|}}{1+e^{-2\gamma|\omega|}}.
\eea
The spin of the state containing this boundary string can be calculated using $S^z=\frac{N}{2}-M$, where
\bea \label{ch4nroots22}M=1+\int_{-\pi}^{\pi}\rho_{\ket{0}_{\beta F_1}}(\lambda) d\lambda. \eea
We obtain $S^z_{\ket{0'}_{\beta F_1}}=0$, $\beta=L,R$.  Hence there are two states with $S^z=0$ that correspond to the presence of the boundary strings $\lambda'_{bsL}$ and $\lambda'_{bsR}$. The energy of the boundary string can be calculated using (\ref{ch4energy}).  We have 

\bea \label{ch4boundenf1}
E_{\lambda'_{bs\beta }}= -\frac{2\sinh^2\gamma}{\cosh\gamma+\cosh\gamma(1-\tilde{\epsilon}_{\beta})}-2\sinh\gamma\int_{-\pi}^{\pi}a(\lambda-\pi,1)\Delta\rho_{\beta '}(\lambda)d\lambda. 
\eea
Using (\ref{ch4denboundf11}) and evaluating the integral one obtains,

\bea E_{\lambda'_{bs\beta }}=-\frac{2\sinh^2\gamma}{\cosh\gamma+\cosh\gamma(1-\tilde{\epsilon}_{\beta})}+\sinh\gamma \sum_{\omega=-\infty}^{\infty} e^{-2\gamma|\omega|}\frac{\cosh\gamma(1-\tilde{\epsilon}_{\beta})|\omega|}{\cosh(\gamma|\omega |)}=-m'_{\beta}.\label{ch4bounden23}\eea
Hence there exists two states $\ket{0}_{\beta F_1}, \beta=L,R$ with total spin $S^z=0$. These two states contain the bound state at the right and left edges respectively whose energy is greater than $M$. Starting with reference state with all up spin and considering a state with real Bethe roots, the boundary strings $\lambda'_{bs L}, \lambda'_{bs R}$ and a spinon with rapidity $\theta$ we obtain the following distribution

\bea\label{ch4denboundf12} \tilde{\rho}_{\ket{0}_{\theta F_1}} (\omega)=\tilde{\rho}_{\ket{\frac{1}{2}}_{F_1}}(\omega)+\sum_{\beta=L,R}\Delta\tilde{\rho'}_{\beta} (\omega)+\Delta \hat{\rho}_{\theta}(\omega).\eea
The two states $\ket{-1}_{\theta, F_1}$ and $\ket{0}_{\theta F_1}$ are degenerate (in thermodynamic limit).

\subsection{$E_1$ Odd number of sites}

The region $E_1$ corresponds to the following values of the boundary magnetic fields: $h_{c2}<h_R$, $0<h_L<h_{c1}$. This corresponds to $\epsilon_{R}=-\tilde{\epsilon}_{R}$, $\epsilon_{L}=i\pi-\tilde{\epsilon}_{L}$with $\tilde{\epsilon}_{\alpha}<1$, $\alpha=L,R$. Starting with the Bethe equations corresponding to all spin up reference state and considering the state with all real roots, we have

\bea\nonumber(2N+1) a(\lambda,1)-a(\lambda,1-\tilde{\epsilon}_{R})-a(\lambda-\pi,1-\tilde{\epsilon}_{L})+a(\lambda-\pi,1)\\-2\pi\delta(\lambda)-2\pi\delta(\lambda-\pi)=2\pi\rho(\lambda)+\sum_{\sigma=\pm}\int a(\lambda+\sigma \mu,2)\rho(\mu)d\mu. \eea
Following the usual procedure we obtain the following density distribution

\bea \nonumber\hat{\rho}_{\ket{\frac{1}{2}}_{E_1}}(\omega)=\frac{(2N+1)e^{-\gamma|\omega|} + (-1)^{\omega}e^{-\gamma|\omega|}-(1+(-1)^{\omega})}{4\pi(1+e^{-2\gamma|\omega|})}\\-\frac{e^{-\gamma (1-\tilde{\epsilon}_{R})|\omega|}+(-1)^{\omega}e^{-\gamma (1-\tilde{\epsilon}_{L})|\omega|}}{4\pi(1+e^{-2\gamma|\omega|})}. \label{ch4denodde11}\eea
The total spin $S^z$ of this state is $S^z=\frac{1}{2}$. We denote this state by $\ket{\frac{1}{2}}_{E_1}$. By starting with the Bethe equations corresponding to all spin down reference state we have

\bea\nonumber
(2N+1) a(\lambda,1)-a(\lambda,1+\tilde{\epsilon}_{R})-a(\lambda-\pi,1+\tilde{\epsilon}_{L})+a(\lambda-\pi,1)\\-2\pi\delta(\lambda)-2\pi\delta(\lambda-\pi)=2\pi\rho(\lambda)+\sum_{\sigma=\pm}\int a(\lambda+\sigma \mu,2)\rho(\mu)d\mu. \label{ch4logbee11}\eea
Following the same procedure as above, we obtain the following distribution for a state with all real $\lambda$ 

\bea\nonumber \hat{\rho}_{\ket{-\frac{1}{2}}_{E_1}}(\omega)=\frac{(2N+1)e^{-\gamma|\omega|} + (-1)^{\omega}e^{-\gamma|\omega|}-(1+(-1)^{\omega})}{4\pi(1+e^{-2\gamma|\omega|})}\\
-\frac{e^{-\gamma (1+\tilde{\epsilon}_{R})|\omega|}+(-1)^{\omega}e^{-\gamma (1+\tilde{\epsilon}_{L})|\omega|}}{4\pi(1+e^{-2\gamma|\omega|})}. \label{ch4denodde12}\eea
The total spin $S^z$ of this state is $S^z=-\frac{1}{2}$. We denote this state by $\ket{-\frac{1}{2}}_{E_1}$. Using (\ref{ch4energy}) we can calculate the energy difference between the two states $\ket{\frac{1}{2}}_{E_1}$ and $\ket{-\frac{1}{2}}_{E_1}$. We have \bea \label{ch4diffene11}E_{\ket{\frac{1}{2}}_{E_1}}-E_{\ket{-\frac{1}{2}}_{E_1}}= h_L+h_R-2\sinh\gamma\int_{-\pi}^{\pi}a(\lambda,1)\;\delta\rho_{\tiny{\ket{\frac{1}{2}},\ket{-\frac{1}{2}}}}(\lambda)d\lambda.
\eea
Here $\delta\rho_{\tiny{\ket{\frac{1}{2}},\ket{-\frac{1}{2}}}}(\lambda)$ is the difference in the density distributions of the states $\ket{\frac{1}{2}}_{E_1}$ and $\ket{-\frac{1}{2}}_{E_1}$. The expression (\ref{ch4diffene11}) can be written as
 
 \bea  E_{\ket{\frac{1}{2}}_{E_1}}-E_{\ket{-\frac{1}{2}}_{E_1}}= h_L+h_R+4\pi\sinh\gamma\sum_{\omega=-\infty}^{\infty} \hat{a}(\omega,1)\Delta\hat{\rho}_{\tiny{\ket{\frac{1}{2}},\ket{-\frac{1}{2}}}}(\omega). \eea
 Using (\ref{ch4denodde11}) and (\ref{ch4denodde12}) in the above expression we obtain
 \bea \nonumber E_{\ket{\frac{1}{2}}_{E_1}}-E_{\ket{-\frac{1}{2}}_{E_1}}= h_L+h_R+ \sinh\gamma \sum_{\omega=-\infty}^{\infty}\; \frac{\sinh(\gamma \tilde{\epsilon}_{R}|\omega|)}{\cosh(\gamma\omega)}e^{-\gamma|\omega|}\\+ \sinh\gamma \sum_{\omega=-\infty}^{\infty}\; (-1)^{\omega}\frac{\sinh(\gamma \tilde{\epsilon}_{L}|\omega|)}{\cosh(\gamma\omega)}e^{-\gamma|\omega|}.  \eea
 This can be written as \bea E_{\ket{\frac{1}{2}}_{E_1}}-E_{\ket{-\frac{1}{2}}_{E_1}}=m_L+m'_R.\eea Hence, $\ket{-\frac{1}{2}}_{E_1}$ is the ground state.

\subsection{$E_1$: Even number of sites}

The Bethe equations corresponding to all spin up reference state contain two boundary string solutions $\lambda_{bs R '}=\pm i\gamma(1-\tilde{\epsilon}_{R})$, $\lambda_{bs L}=\pi\pm i\gamma(1-\tilde{\epsilon}_{L})$ . The ground state is obtained by adding $\lambda_{bs R '}$ to the state $\ket{\frac{1}{2}}_{E_1}$. Adding $\lambda_{bs R '}$  to the Bethe equations and taking logarithm we obtain

\bea \nonumber 2N \varphi(\lambda_j,1)-\varphi(\lambda_j-\pi,1-\tilde{\epsilon}_{L})-\varphi(\lambda_j,1-\tilde{\epsilon}_{R})+\varphi(\lambda_j,1)+\varphi^{\prime}(\lambda_j,1)\\-\varphi(\lambda,(3-\tilde{\epsilon}_{R}))-\varphi(\lambda,(1+\tilde{\epsilon}_{R}))=2\pi I_j+ \sum_{\sigma=\pm}\sum_{k\neq j}\varphi(\lambda_j+\sigma \lambda_k,2).
\eea
Differentiating the above equation with respect to $\lambda$ and taking the Fourier transform we obtain

\bea\nonumber \tilde{\rho}_{\ket{0'}_{R, E_1}} (\omega)=\tilde{\rho}_{\ket{\frac{1}{2}}_{E_1}}(\omega)+\Delta\tilde{\rho'}_{R} (\omega), \;\;\Delta\tilde{\rho'}_{R} (\omega)=-\frac{1}{4\pi}\frac{e^{-\gamma(3-\tilde{\epsilon}_{R})|\omega|}+e^{-\gamma(1+\tilde{\epsilon}_{R})|\omega|}}{1+e^{-2\gamma|\omega|}}.
\label{ch4denboune1}\eea
The spin of the state containing this boundary string can be calculated using $S^z=\frac{N}{2}-M$, where
\bea \label{ch4nrootse12}M=1+\int_{-\pi}^{\pi}\rho_{\ket{0}_{R' E_1}}(\lambda) d\lambda. \eea
Using this we obtain $S^z_{\ket{0'}_{R, E_1}}=0$.

\subsection{$B_1$: Odd number of sites}

Region $B_1$ corresponds to the following values of the boundary fields: $h_{c1}<h_R<h_{c2}$, $0<h_L<h_{c1}$. This region can be further divided into two regions depending on whether $h_{c1}<h_R<\sinh\gamma$ and $\sinh\gamma<h_R<h_{c2}$.

\subsubsection{$h_{c1}<h_R<\sinh\gamma$}

In the case of $h_{c1}<h_R<\sinh\gamma$, the Bethe equations for all down reference state take the same form as that in the region $A_1$. The density distribution is again given by (\ref{ch4denodda12}) with total spin $S^z=-\frac{1}{2}$.

\subsubsection{$\sinh\gamma<h_R<h_{c2}$}

In the case of $h_{c2}>h_R>\sinh\gamma$, the Bethe equations for all down reference state take the same form as that in the region $E_1$. The density distribution is again given by (\ref{ch4denodde12}) with total spin $S^z=-\frac{1}{2}$.






\subsection{$B_1$: Even number of sites}

\subsubsection{$h_{c1}<h_R<\sinh\gamma$}

In this case we have $\epsilon_R=-\tilde{\epsilon}_R+i\pi$, $\tilde{\epsilon}_R>1$. The logarithmic form of Bethe equations corresponding to all spin up reference state take the following form 

\bea\nonumber
(2N+1) a(\lambda,1)+a(\lambda-\pi,\tilde{\epsilon}_{R}-1)-a(\lambda-\pi,1-\tilde{\epsilon}_{L})+a(\lambda-\pi,1)\\-2\pi\delta(\lambda)-2\pi\delta(\lambda-\pi)=2\pi\rho(\lambda)+\sum_{\sigma=\pm}\int a(\lambda+\sigma \mu,2)\rho(\mu)d\mu. \label{ch4logbeb11}\eea
Following the same procedure as above we obtain 

\bea \nonumber\hat{\rho}_{(\ket{0}_{B_1,h_R<\sinh\gamma}}(\omega)=\frac{(2N+1)e^{-\gamma|\omega|} + (-1)^{\omega}e^{-\gamma|\omega|}-(1+(-1)^{\omega})}{4\pi(1+e^{-2\gamma|\omega|})}\\
+\frac{(-1)^{\omega}(e^{-\gamma (\tilde{\epsilon}_{R}-1)|\omega|}-e^{-\gamma (1-\tilde{\epsilon}_{L})|\omega|})}{4\pi(1+e^{-2\gamma|\omega|})}. \label{ch4denoddb11}\eea
The total spin $S^z$ can be found using $S^z=\frac{N}{2}-M$ where $M$ is given by (\ref{ch4nroots1}). We obtain $S^z_{(\ket{0}_{B_1,h_R<\sinh\gamma}}=0$.

\subsubsection{$\sinh\gamma<h_R<h_{c2}$}

In this case we have $\epsilon_R=-\tilde{\epsilon}_R$, $\tilde{\epsilon}_R>1$. The logarithmic form of Bethe equations corresponding to all spin up reference state take the following form 

\bea\nonumber
(2N+1) a(\lambda,1)+a(\lambda,\tilde{\epsilon}_{R}-1)-a(\lambda-\pi,1-\tilde{\epsilon}_{L})+a(\lambda-\pi,1)\\-2\pi\delta(\lambda)-2\pi\delta(\lambda-\pi)=2\pi\rho(\lambda)+\sum_{\sigma=\pm}\int a(\lambda+\sigma \mu,2)\rho(\mu)d\mu. \label{ch4logbeb12}\eea
Following the same procedure as above we obtain 

\bea\nonumber \hat{\rho}_{(\ket{0}_{B_1,h_R>\sinh\gamma}}(\omega)=\frac{(2N+1)e^{-\gamma|\omega|} + (-1)^{\omega}e^{-\gamma|\omega|}-(1+(-1)^{\omega})}{4\pi(1+e^{-2\gamma|\omega|})}\\
+\frac{e^{-\gamma (\tilde{\epsilon}_{R}-1)|\omega|}-(-1)^{\omega}e^{-\gamma (1-\tilde{\epsilon}_{L})|\omega|}}{4\pi(1+e^{-2\gamma|\omega|})}. \label{ch4denoddb112}\eea
The total spin $S^z$ can be found using $S^z=\frac{N}{2}-M$ where $M$ is given by (\ref{ch4nroots1}). We obtain again $S^z_{(\ket{0}_{B_1},h_R>\sinh\gamma)}=0$.

\subsection{$D_1$: Odd number of sites}

Region $D_1$ corresponds to the following values of the boundary fields: $h_{c1}<h_L<h_{c2}$, $h_R>h_{c2}$. This region can be further divided into two regions depending on whether $h_{c1}<h_L<\sinh\gamma$ and $\sinh\gamma<h_L<h_{c2}$.

\subsubsection{$h_{c1}<h_L<\sinh\gamma$}

In the case of $h_{c1}<h_L<\sinh\gamma$, the Bethe equations for all down reference state take the same form as that in the region $E_1$. The density distribution is again given by (\ref{ch4denodde12}) with total spin $S^z=-\frac{1}{2}$.

\bea \hat{\rho}_{\ket{-\frac{1}{2}}_{E_1}}(\omega)\equiv \hat{\rho}_{\ket{-\frac{1}{2}}_{D_1,h_L<\sinh\gamma }}. \eea

\subsubsection{$\sinh\gamma<h_L<h_{c2}$}

In the case of $\sinh\gamma<h_L<h_{c2}$, we have for all down reference state. In the case of $h_{c2}>h_L>\sinh\gamma$, the Bethe equations for all down reference state take the same form as that in the region $F_1$. The density distribution is again given by (\ref{ch4denoddf12}) with total spin $S^z=-\frac{1}{2}$. 

\bea \hat{\rho}_{\ket{-\frac{1}{2}}_{F_1}}(\omega)\equiv \hat{\rho}_{\ket{-\frac{1}{2}}_{D_1,h_L>\sinh\gamma }}. \eea

\subsection{$D_1$: Even number of sites}

\subsubsection{$h_{c1}<h_L<\sinh\gamma$}

In this case we have $\epsilon_L=-\tilde{\epsilon}_L+i\pi$, $\tilde{\epsilon}_L>1$. The logarithmic form of Bethe equations corresponding to all spin up reference state take the following form 

\bea\nonumber
(2N+1) a(\lambda,1)+a(\lambda-\pi,\tilde{\epsilon}_{L}-1)-a(\lambda,1-\tilde{\epsilon}_{R})+a(\lambda-\pi,1)\\-2\pi\delta(\lambda)-2\pi\delta(\lambda-\pi)=2\pi\rho(\lambda)+\sum_{\sigma=\pm}\int a(\lambda+\sigma \mu,2)\rho(\mu)d\mu. \label{ch4logbed11}\eea
Following the same procedure as above we obtain 

\bea\nonumber \hat{\rho}_{(\ket{0}_{D_1,h_L<\sinh\gamma}}(\omega)=\frac{(2N+1)e^{-\gamma|\omega|} + (-1)^{\omega}e^{-\gamma|\omega|}-(1+(-1)^{\omega})}{4\pi(1+e^{-2\gamma|\omega|})}\\
+\frac{(-1)^{\omega}(e^{-\gamma (\tilde{\epsilon}_{L}-1)|\omega|}-e^{-\gamma (1-\tilde{\epsilon}_{R})|\omega|})}{4\pi(1+e^{-2\gamma|\omega|})}. \label{ch4denoddd11}\eea
The total spin $S^z$ can be found using $S^z=\frac{N}{2}-M$ where $M$ is given by (\ref{ch4nroots1}). We obtain $S^z_{\ket{0}_{D_1,h_R<\sinh\gamma}}=0$. Starting from the Bethe reference state with all spin down and considering the state with all real Bethe roots and a spinon we have 

\bea \hat{\rho}_{\ket{-1}_{\theta, D_1, h_L<\sinh\gamma}}(\omega)= \hat{\rho}_{\ket{-\frac{1}{2}}_{E_1}}(\omega)+ \Delta \hat{\rho}_{\theta}(\omega).\eea
The total spin of this state is $S^z=-1$. We denote this state by $\ket{-1}_{\theta, D_1, h_L<\sinh\gamma}$. The energy difference between the states $\ket{-1}_{\theta, D_1, h_L<\sinh\gamma}$ and $\ket{0}_{D_1,h_R<\sinh\gamma}$ is 

 \bea  \nonumber E_{\ket{0}_{D_1,h_L<\sinh\gamma}}-E_{\ket{-1}_{\theta, D_1, h_L<\sinh\gamma}}= h_L+h_R-E_{\theta}
 + \sinh\gamma \sum_{\omega=-\infty}^{\infty}\; \frac{\sinh(\gamma \tilde{\epsilon}_{R}|\omega|)}{\cosh(\gamma\omega)}e^{-\gamma|\omega|}- \sinh\gamma \sum_{\omega=-\infty}^{\infty}\; (-1)^{\omega}e^{-\gamma\tilde{\epsilon}_L|\omega|}. \\\eea
 After simplification we obtain

\bea  E_{\ket{0}_{D_1,h_L<\sinh\gamma}}-E_{\ket{-1}_{\theta, D_1, h_L<\sinh\gamma}} =m'_R-E_{\theta}. \eea
Hence the ground state is $\ket{-1}_{\theta, D_1, h_L<\sinh\gamma}$.

\subsubsection{$\sinh\gamma<h_L<h_{c2}$}

In this case we have $\epsilon_R=-\tilde{\epsilon}_L$, $\tilde{\epsilon}_L>1$. The logarithmic form of Bethe equations corresponding to all spin up reference state take the following form 

\bea\nonumber
(2N+1) a(\lambda,1)+a(\lambda,\tilde{\epsilon}_{L}-1)-a(\lambda,1-\tilde{\epsilon}_{R})+a(\lambda-\pi,1)\\-2\pi\delta(\lambda)-2\pi\delta(\lambda-\pi)=2\pi\rho(\lambda)+\sum_{\sigma=\pm}\int a(\lambda+\sigma \mu,2)\rho(\mu)d\mu. \label{ch4logbed12}\eea
Following the same procedure as above we obtain 

\bea \nonumber\hat{\rho}_{(\ket{0},B_1,h_L>\sinh\gamma)}(\omega)=\frac{(2N+1)e^{-\gamma|\omega|} + (-1)^{\omega}e^{-\gamma|\omega|}-(1+(-1)^{\omega})}{4\pi(1+e^{-2\gamma|\omega|})}\\+\frac{e^{-\gamma (\tilde{\epsilon}_{L}-1)|\omega|}-e^{-\gamma (1-\tilde{\epsilon}_{R})|\omega|}}{4\pi(1+e^{-2\gamma|\omega|})} .\label{ch4denoddd112}\eea
The total spin $S^z$ can be found using $S^z=\frac{N}{2}-M$ where $M$ is given by (\ref{ch4nroots1}). We obtain again $S^z_{(\ket{0}_{D_1},h_R>\sinh\gamma)}=0$. Starting from the Bethe reference state with all spin down and considering the state with all real Bethe roots and a spinon we have

\bea \hat{\rho}_{\ket{-1}_{\theta, D_1, h_L>\sinh\gamma}}(\omega)= \hat{\rho}_{\ket{-\frac{1}{2}}_{F_1}}(\omega)+ \Delta \hat{\rho}_{\theta}(\omega).\eea
The total spin of this state is $S^z=-1$. We denote this state by $\ket{-1}_{\theta, D_1, h_L>\sinh\gamma}=-1$. The energy difference between the states $\ket{-1}_{\theta, D_1, h_L>\sinh\gamma}$ and $\ket{0}_{D_1,h_R>\sinh\gamma}$ is 

 \bea  \nonumber E_{\ket{0}_{D_1,h_L>\sinh\gamma}}-E_{\ket{-1}_{\theta, D_1, h_L>\sinh\gamma}}= h_L+h_R-E_{\theta}
 + \sinh\gamma \sum_{\omega=-\infty}^{\infty}\; \frac{\sinh(\gamma \tilde{\epsilon}_{R}|\omega|)}{\cosh(\gamma\omega)}e^{-\gamma|\omega|}- \sinh\gamma \sum_{\omega=-\infty}^{\infty}\; e^{-\gamma\tilde{\epsilon}_L|\omega|}.\\ \eea
 After simplification we obtain

\bea  E_{\ket{0}_{D_1,h_L>\sinh\gamma}}-E_{\ket{-1}_{\theta, D_1, h_L>\sinh\gamma}} =m'_R-E_{\theta}. \eea
Hence the ground state is $\ket{-1}_{\theta, D_1, h_L>\sinh\gamma}$.

\subsection{$A_2$: Odd and even number of sites}

The region $A_2$ corresponds to the following values of the boundary magnetic fields: $0<h_R< h_{c1}$, $-h_{c1}<h_L<0$. In this region the logarithmic form of the Bethe equations can be obtained from (\ref{ch4logbea11}) by the transformation $\tilde{\epsilon}_L\rightarrow -\tilde{\epsilon}_L$. We have 

\bea\nonumber
(2N+1) a(\lambda,1)-a(\lambda-\pi,1-\tilde{\epsilon}_{R})-a(\lambda-\pi,1+\tilde{\epsilon}_{L})+a(\lambda-\pi,1)\\-2\pi\delta(\lambda)-2\pi\delta(\lambda-\pi)=2\pi\rho(\lambda)+\sum_{\sigma=\pm}\int a(\lambda+\sigma \mu,2)\rho(\mu)d\mu. \label{ch4logbeA2} \eea
Taking Fourier transform we obtain

\bea\nonumber \hat{\rho}_{\ket{\frac{1}{2}}_{A_2}}(\omega)=\frac{(2N+1)e^{-\gamma|\omega|} + (-1)^{\omega}e^{-\gamma|\omega|}-(1+(-1)^{\omega})}{4\pi(1+e^{-2\gamma|\omega|})}\\-\frac{(-1)^{\omega}e^{-\gamma (1-\tilde{\epsilon}_{R})|\omega|}+(-1)^{\omega}e^{-\gamma (1+\tilde{\epsilon}_{L})|\omega|}}{4\pi(1+e^{-2\gamma|\omega|})}. \label{ch4denodda21}\eea
The number of Bethe roots can be obtained by using the relation \bea \label{ch4nroots12}M= \int_{-\pi}^{\pi} \rho(\lambda)d\lambda. \eea 
The total spin $S^z$ of the state can be found using the relation $S^z=\frac{N}{2}-M$. Using (\ref{ch4denodda21}) in the above relations we find that the total spin $S^z$ of the state described by the distribution $\hat{\rho}_{\ket{\frac{1}{2}}_{A_2}}(\omega)$ is $S^z =\frac{1}{2}$.  We denote this state by $\ket{\frac{1}{2}}_{A_2}$.
By starting with the Bethe equations corresponding to all spin down reference state we have

\bea\nonumber
(2N+1) a(\lambda,1)-a(\lambda-\pi,1+\tilde{\epsilon}_{R})-a(\lambda-\pi,1-\tilde{\epsilon}_{L})+a(\lambda-\pi,1)\\-2\pi\delta(\lambda)-2\pi\delta(\lambda-\pi)=2\pi\rho(\lambda)+\sum_{\sigma=\pm}\int a(\lambda+\sigma \mu,2)\rho(\mu)d\mu.  \label{ch4logbea22} \eea
Following the same procedure as above, we obtain the following distribution for a state with all real $\lambda$ 

\bea\nonumber \hat{\rho}_{\ket{-\frac{1}{2}}_{A_1}}(\omega)=\frac{(2N+1)e^{-\gamma|\omega|} + (-1)^{\omega}e^{-\gamma|\omega|}-(1+(-1)^{\omega})}{4\pi(1+e^{-2\gamma|\omega|})}\\
-\frac{(-1)^{\omega}e^{-\gamma (1+\tilde{\epsilon}_{R})|\omega|}+(-1)^{\omega}e^{-\gamma (1-\tilde{\epsilon}_{L})|\omega|}}{4\pi(1+e^{-2\gamma|\omega|})}. \label{ch4denodda22}\eea
The total spin $S^z$ of this state is $S^z=-\frac{1}{2}$. We denote this state by $\ket{-\frac{1}{2}}_{A_2}$. Using (\ref{ch4energy}) we can calculate the energy difference between the two states $\ket{\frac{1}{2}}_{A_2}$ and $\ket{-\frac{1}{2}}_{A_2}$. We have \bea \label{ch4diffena21}E_{\ket{\frac{1}{2}}_{A_2}}-E_{\ket{-\frac{1}{2}}_{A_2}}= -h_L+h_R-2\sinh\gamma\sum_{\alpha=L,R}\int_{-\pi}^{\pi}a(\lambda,1)\;\delta\rho_{\tiny{\ket{\frac{1}{2}},\ket{-\frac{1}{2}}}}(\lambda)d\lambda,
\eea
where $\delta\rho_{\tiny{\ket{\frac{1}{2}},\ket{-\frac{1}{2}}}}(\lambda)$ is the difference in the density distributions of the states $\ket{\frac{1}{2}}$ and $\ket{-\frac{1}{2}}$. The expression (\ref{ch4diffena21}) can be written as
 
 \bea  E_{\ket{\frac{1}{2}}_{A_2}}-E_{\ket{-\frac{1}{2}}_{A_2}}= -h_L+h_R+4\pi\sinh\gamma\sum_{\omega=-\infty}^{\infty} \hat{a}(\omega,1)\Delta\hat{\rho}_{\tiny{\ket{\frac{1}{2}},\ket{-\frac{1}{2}}}}(\omega). \eea
 Using (\ref{ch4denodda21}) and (\ref{ch4denodda22}) in the above expression we obtain
 \bea \nonumber E_{\ket{\frac{1}{2}}_{A_2}}-E_{\ket{-\frac{1}{2}}_{A_2}}= -h_L+h_R+ \sinh\gamma (-1)^{\omega}\; \frac{\sinh(\gamma \tilde{\epsilon}_{R}|\omega|)}{\cosh(\gamma\omega)}e^{-\gamma|\omega|}\\-\sinh\gamma (-1)^{\omega}\; \frac{\sinh(\gamma \tilde{\epsilon}_{L}|\omega|)}{\cosh(\gamma\omega)}e^{-\gamma|\omega|},  \eea
 which can be written as
 
 \bea E_{\ket{\frac{1}{2}}_{A_2}}-E_{\ket{-\frac{1}{2}}_{A_2}}=m_R-m_L. \eea
 Hence the ground state for odd number of sites is $\ket{\pm\frac{1}{2}}_{A_2}$ depending on the values of $h_L,h_R$. The Bethe equations corresponding to all spin up reference state have two boundary string solutions $\lambda_{bs R}=\pi\pm i\gamma(1-\tilde{\epsilon}_{R})$, $\lambda_{bs L'}=\pi\pm i\gamma(1+\tilde{\epsilon}_{L})$. Adding $\lambda_{bs R}$ to the state $\ket{\frac{1}{2}}_{A_2}$  leads to the state with following root distribution

\bea\label{ch4denbound1a2} \tilde{\rho}_{\ket{0}_{\beta A_2}} (\omega)=\tilde{\rho}_{\ket{\frac{1}{2}}_{A_2}}(\omega)+\Delta\tilde{\rho}_{R} (\omega). \eea
The spin of the state containing this boundary string can be calculated using $S^z=\frac{N}{2}-M$, where
\bea \label{ch4nroots23}M=1+\int_{-\pi}^{\pi}\rho_{\ket{0}_{R A_2}}(\lambda) d\lambda. \eea
Using this we obtain $S^z_{\ket{0}_{R A_2}}=0$. The energy of the boundary string is given by (\ref{ch4bounden2}). Adding the boundary string $\lambda_{bs L'}$ to the state $\ket{\frac{1}{2}}_{A_2}$ we obtain 
 
 \bea\label{ch4denbound1a22} \tilde{\rho}_{\ket{0}_{L' A_2}} (\omega)=\tilde{\rho}_{\ket{\frac{1}{2}}_{A_2}}(\omega)+\Delta\tilde{\rho}_{L'} (\omega), \eea
where 

\bea\Delta\tilde{\rho}_{L'} (\omega)=-\frac{1}{4\pi}(-1)^{\omega}\frac{e^{-\gamma(3+\tilde{\epsilon}_{L})|\omega|}+e^{-\gamma(1-\tilde{\epsilon}_{L})|\omega|}}{1+e^{-2\gamma|\omega|}}.
\eea

The spin of the state containing this boundary string can be calculated using $S^z=\frac{N}{2}-M$, where
\bea \label{ch4nroots24}M=1+\int_{-\pi}^{\pi}\rho_{\ket{0}_{L' A_2}}(\lambda) d\lambda. \eea
We obtain $S^z_{\ket{0}_{L' A_2}}=0$. The energy of the boundary string $\lambda_{bs L'}$ is given by

\bea \label{ch4bounden24}E_{\lambda_{bsL'}}=-\sinh\gamma \sum_{\omega=-\infty}^{\infty} (-1)^{\omega}\frac{e^{-\gamma(1+\tilde{\epsilon}_{L})|\omega|}}{\cosh\gamma|\omega|}=m_L.\eea
The energy difference between the states $\ket{0}_{L' A_2}$ and $\ket{0}_{R A_2}$ can be calculated similar to the previous section, we obtain  
 
 \bea E_{\ket{0}_{L' A_2}}-E_{\ket{0}_{R A_2}} =  m_L+m_R. \eea
 Hence the ground state for even number of sites is $\ket{0}_{R A_2}$.

 \subsection{$C_2$: Even and odd number of sites}
In this region both $h_L,h_R$ take the following values: $h_{c1}<h_R<h_{c2}, -h_{c2}<h_L<-h_{c1}$. This region can be further split into four sub regions depending on whether the absolute values of the boundary fields are greater than or less than $\sinh\gamma$. The solution in each of these regions is constructed below. By starting with Bethe reference state with all spin down, and considering the state with all real $\lambda_j$, we obtain the following logarithmic form of Bethe  equations

\bea\nonumber
(2N+1) a(\lambda,1)-(l_1 a(\lambda,1+\tilde{\epsilon}_{L})+l_2 a(\lambda-\pi,1+\tilde{\epsilon}_{L}))+(r_1 a(\lambda,\tilde{\epsilon}_{R}-1)+r_2 a(\lambda-\pi,\tilde{\epsilon}_{R}-1))\\+a(\lambda-\pi,1)-2\pi\delta(\lambda)-2\pi\delta(\lambda-\pi)=2\pi\rho(\lambda)+\sum_{\sigma=\pm}\int a(\lambda+\sigma \mu,2)\rho(\mu)d\mu.  \eea
By following the same procedure as above we obtain

\bea \nonumber\hat{\rho}_{(\ket{0}_{\uparrow C_2}}(\omega)=\frac{(2N+1)e^{-\gamma|\omega|} + (-1)^{\omega}e^{-\gamma|\omega|}-(1+(-1)^{\omega})}{4\pi(1+e^{-2\gamma|\omega|})}\\-\frac{(l_1+l_2(-1)^{\omega})e^{-\gamma (1+\tilde{\epsilon}_{L})|\omega|}-(r_1+r_2(-1)^{\omega})e^{-\gamma (\tilde{\epsilon}_{R}-1)|\omega|}}{4\pi(1+e^{-2\gamma|\omega|})}. \label{ch4denoddc21} \eea
The total spin $S^z$ can be found using $S^z=\frac{N}{2}-M$ where $M$ is given by (\ref{ch4nroots1}). We obtain $S^z_{(\ket{0}_{C_2})}=0$. By starting with Bethe equations corresponding to all spin down reference state we obtain a state with total spin $S^z-0$ described by the distribution

\bea \nonumber\hat{\rho}_{(\ket{0}_{\downarrow C_2}}(\omega)=\frac{(2N+1)e^{-\gamma|\omega|} + (-1)^{\omega}e^{-\gamma|\omega|}-(1+(-1)^{\omega})}{4\pi(1+e^{-2\gamma|\omega|})}\\+\frac{(l_1+l_2(-1)^{\omega})e^{-\gamma (\tilde{\epsilon}_{L}-1)|\omega|}-(r_1+r_2(-1)^{\omega})e^{-\gamma (1+\tilde{\epsilon}_{R})|\omega|}}{4\pi(1+e^{-2\gamma|\omega|})}, \label{ch4denoddc22} \eea
where the parameters $l_1,l_2,r_1,r_2$ take the values given in (Tab:\ref{ch4tableap2}) for different values of $h_L,h_R$.

 \begin{table}[h!]
\centering
\caption{Values of the parameters in (\ref{ch4denoddc22}) corresponding to various ranges of the boundary magnetic fields}

\begin{tabular}{c|c|c|c|c}
\hline
\hline
  & \;\scriptsize{$-h_{c1}>h_L>-\sinh\gamma$}\;&\;\scriptsize{$-h_{c1}>h_L>-\sinh\gamma$} \; &\;\scriptsize{$-\sinh\gamma>h_L>-h_{c2}$}&\; \scriptsize{$-\sinh\gamma>h_L>-h_{c2}$}\\

&\; \scriptsize{$h_{c1}<h_R<\sinh\gamma$} \;& \;\scriptsize{$\sinh\gamma<h_R<h_{c2}$}\;&\;\scriptsize{$h_{c1}<h_R<\sinh\gamma$}\;&\;\scriptsize{$\sinh\gamma<h_R<h_{c2}$}\\
\hline
 $l_1$&  0 & 0  &1&1  \\
  $l_2$&1& 1& 0&0\\
  $r_1$&0& 1& 0& 1\\
  $r_2$ &1&0&1&0\\
 \hline
\hline
\end{tabular}
\label{ch4tableap2}
\end{table}

  The two distributions (\ref{ch4denoddc21}), (\ref{ch4denoddc22}) describe the same state $\ket{0}_{C_2}$. To obtain the lowest energy state corresponding to odd number of sites, we need to add a spinon to the state with all real roots corresponding to all spin up reference state. We obtain 

\bea\nonumber
(2N+1) a(\lambda,1)-(l_1 a(\lambda,1+\tilde{\epsilon}_{L})+l_2 a(\lambda-\pi,1+\tilde{\epsilon}_{L}))+(r_1 a(\lambda,\tilde{\epsilon}_{R}-1)+r_2 a(\lambda-\pi,1+\tilde{\epsilon}_{R}))+a(\lambda-\pi,1)\\-2\pi\delta(\lambda)-2\pi\delta(\lambda-\pi)-2\pi\delta(\lambda-\theta)-2\pi\delta(\lambda+\theta)=2\pi\rho(\lambda)+\sum_{\sigma=\pm}\int a(\lambda+\sigma \mu,2)\rho(\mu)d\mu.  \eea
By following the same procedure as above we obtain

\bea \hat{\rho}_{\ket{\frac{1}{2}}_{C_2}}(\omega)= \hat{\rho}_{\ket{0}_{\uparrow C_2}}(\omega)+ \Delta \hat{\rho}_{\theta}(\omega).\eea
By adding the spinon to the state with all real roots corresponding to all spin down reference state. We obtain 

  \bea \hat{\rho}_{\ket{-\frac{1}{2}}_{C_2}}(\omega)= \hat{\rho}_{\ket{0}_{\downarrow C_2}}(\omega)+ \Delta \hat{\rho}_{\theta}(\omega).\eea  
 The spin of these states can be obtained by using $S^z=\frac{N}{2}$, where $M$ is given by (\ref{ch4nroots1}). We obtain $S^z_{(\ket{\frac{1}{2}}_{c_2})}=\frac{1}{2},\;S^z_{(\ket{-\frac{1}{2}}_{c_2})}=-\frac{1}{2}$.

 \subsection{$F_2$: Even and odd number of sites}

The region $F_2$ corresponds to the following values of the boundary magnetic fields: $h_{c2}<h_R$, $h_L<-h_{c2}$. This corresponds to $\epsilon_{R}=-\tilde{\epsilon}_{R}$, $\epsilon_{L}=\tilde{\epsilon}_{L}$ with $|\tilde{\epsilon}_{\alpha}|<1$, $\alpha=L,R$. Making the transformation $\tilde{\epsilon}_{L}\rightarrow -\tilde{\epsilon}_{L}$ and starting with the Bethe equations corresponding to all spin up reference state, and considering the state with all real roots, we have

\bea\nonumber(2N+1) a(\lambda,1)-a(\lambda,1-\tilde{\epsilon}_{R})-a(\lambda,1+\tilde{\epsilon}_{L})+a(\lambda-\pi,1)\\-2\pi\delta(\lambda)-2\pi\delta(\lambda-\pi)=2\pi\rho(\lambda)+\sum_{\sigma=\pm}\int a(\lambda+\sigma \mu,2)\rho(\mu)d\mu. \eea
Following the usual procedure we obtain the following density distribution

\bea \nonumber\hat{\rho}_{\ket{\frac{1}{2}}_{F_2}}(\omega)=\frac{(2N+1)e^{-\gamma|\omega|} + (-1)^{\omega}e^{-\gamma|\omega|}-(1+(-1)^{\omega})}{4\pi(1+e^{-2\gamma|\omega|})}\\
-\frac{e^{-\gamma (1-\tilde{\epsilon}_{R})|\omega|}+e^{-\gamma (1+\tilde{\epsilon}_{L})|\omega|}}{4\pi(1+e^{-2\gamma|\omega|})}. \label{ch4denoddf21}\eea
The total spin $S^z$ of this state is $S^z=\frac{1}{2}$. We denote this state by $\ket{\frac{1}{2}}_{F_2}$. By starting with the Bethe equations corresponding to all spin down reference state we have

\bea\nonumber
(2N+1) a(\lambda,1)-a(\lambda,1-\tilde{\epsilon}_{L})-a(\lambda,1+\tilde{\epsilon}_{R})+a(\lambda-\pi,1)\\-2\pi\delta(\lambda)-2\pi\delta(\lambda-\pi)=2\pi\rho(\lambda)+\sum_{\sigma=\pm}\int a(\lambda+\sigma \mu,2)\rho(\mu)d\mu. \label{ch4logbef21}\eea
Following the same procedure as above, we obtain the following distribution for a state with all real $\lambda$ 

\bea\nonumber\hat{\rho}_{\ket{-\frac{1}{2}}_{F_1}}(\omega)=\frac{(2N+1)e^{-\gamma|\omega|} + (-1)^{\omega}e^{-\gamma|\omega|}-(1+(-1)^{\omega})}{4\pi(1+e^{-2\gamma|\omega|})} \\
-\frac{e^{-\gamma (1-\tilde{\epsilon}_{L})|\omega|}+e^{-\gamma (1+\tilde{\epsilon}_{R})|\omega|}}{4\pi(1+e^{-2\gamma|\omega|})}.  \label{ch4denoddf22}\eea
The total spin $S^z$ of this state is $S^z=-\frac{1}{2}$. We denote this state by $\ket{-\frac{1}{2}}_{F_2}$. Using (\ref{ch4energy}) we can calculate the energy difference between the two states $\ket{\frac{1}{2}}_{F_2}$ and $\ket{-\frac{1}{2}}_{F_2}$. We have \bea \label{ch4diffenf21}E_{\ket{\frac{1}{2}}_{F_2}}-E_{\ket{-\frac{1}{2}}_{F_2}}= -h_L+h_R-2\sinh\gamma\int_{-\pi}^{\pi}a(\lambda,1)\;\delta\rho_{\tiny{\ket{\frac{1}{2}},\ket{-\frac{1}{2}}}}(\lambda)d\lambda.
\eea
Here $\delta\rho_{\tiny{\ket{\frac{1}{2}},\ket{-\frac{1}{2}}}}(\lambda)$ is the difference in the density distributions of the states $\ket{\frac{1}{2}}_{F_2}$ and $\ket{-\frac{1}{2}}_{F_2}$. The expression (\ref{ch4diffenf21}) can be written as
 
 \bea  E_{\ket{\frac{1}{2}}_{F_2}}-E_{\ket{-\frac{1}{2}}_{F_2}}= -h_L+h_R+4\pi\sinh\gamma\sum_{\omega=-\infty}^{\infty} \hat{a}(\omega,1)\Delta\hat{\rho}_{\tiny{\ket{\frac{1}{2}},\ket{-\frac{1}{2}}}}(\omega). \eea
 Using (\ref{ch4denoddf21}) and (\ref{ch4denoddf22}) in the above expression we obtain
 \bea  E_{\ket{\frac{1}{2}}_{F_2}}-E_{\ket{-\frac{1}{2}}_{F_2}}= -h_L+h_R+ \sinh\gamma \sum_{\omega=-\infty}^{\infty}\; \frac{\sinh(\gamma \tilde{\epsilon}_{R}|\omega|)-\sinh(\gamma \tilde{\epsilon}_{L}|\omega|)}{\cosh(\gamma\omega)}e^{-\gamma|\omega|}. \eea
 This can be written as \bea E_{\ket{\frac{1}{2}}_{F_2}}-E_{\ket{-\frac{1}{2}}_{F_2}}=m'_R-m'_L.\eea

 The Bethe equations corresponding to all spin up reference state contain two boundary string solutions $\lambda'_{bs R}=\pm i\gamma(1-\tilde{\epsilon}_{R})$, $\lambda'_{bs L'}=\pm i\gamma(1+\tilde{\epsilon}_{L})$ . The ground state for even number of sites contains the boundary string $\lambda'_{bs R}$ in addition to all real Bethe roots. We obtain

\bea \nonumber 2N \varphi(\lambda_j,1)-\varphi(\lambda_j,1-\tilde{\epsilon}_{R})-\varphi(\lambda_j,1+\tilde{\epsilon}_{L})+\varphi(\lambda_j,1)+\varphi^{\prime}(\lambda_j,1)\\-\varphi(\lambda,(3-\tilde{\epsilon}_{\beta}))-\varphi(\lambda,(1+\tilde{\epsilon}_{R}))=2\pi I_j+ \sum_{\sigma=\pm}\sum_{k\neq j}\varphi(\lambda_j+\sigma \lambda_k,2).
\eea
 Differentiating the above equation with respect to $\lambda$ and taking the Fourier transform we obtain

\bea \tilde{\rho}_{\ket{0'}_{R, F_2}} (\omega)=\tilde{\rho}_{\ket{\frac{1}{2}}_{F_2}}(\omega)+\Delta\tilde{\rho'}_{R} (\omega), \;\;\Delta\tilde{\rho'}_{R} (\omega)=-\frac{1}{4\pi}\frac{e^{-\gamma(3-\tilde{\epsilon}_{R})|\omega|}+e^{-\gamma(1+\tilde{\epsilon}_{R})|\omega|}}{1+e^{-2\gamma|\omega|}}.
\label{ch4denboundf21}\eea
The spin of the state containing this boundary string can be calculated using $S^z=\frac{N}{2}-M$, where
\bea \label{ch4nroots25}M=1+\int_{-\pi}^{\pi}\rho_{\ket{0'}_{R, F_2}}(\lambda) d\lambda. \eea Hence, we obtain $S^z_{\ket{0'}_{R,F_2}}=0$. For odd number of sites, the ground state is obtained by adding a spinon to the state $\ket{0'}_{R, F_2}$.

 \subsection{$E_2$ Even and odd number of sites}

The region $E_2$ corresponds to the following values of the boundary magnetic fields: $h_{c2}<h_R$, $0>h_L>-h_{c1}$. This corresponds to $\epsilon_{R}=-\tilde{\epsilon}_{R}$, $\epsilon_{L}=-i\pi+\tilde{\epsilon}_{L}$with $|\tilde{\epsilon}_{\alpha}|<1$, $\alpha=L,R$. We use the transformation $\tilde{\epsilon}_{L}\rightarrow -\tilde{\epsilon}_{L}$. Starting with the Bethe equations corresponding to all spin up reference state and considering the state with all real roots, we have

\bea\nonumber(2N+1) a(\lambda,1)-a(\lambda,1-\tilde{\epsilon}_{R})-a(\lambda-\pi,1+\tilde{\epsilon}_{L})+a(\lambda-\pi,1)\\-2\pi\delta(\lambda)-2\pi\delta(\lambda-\pi)=2\pi\rho(\lambda)+\sum_{\sigma=\pm}\int a(\lambda+\sigma \mu,2)\rho(\mu)d\mu. \eea
Following the usual procedure we obtain the following density distribution

\bea\nonumber \hat{\rho}_{\ket{\frac{1}{2}}_{E_2}}(\omega)=\frac{(2N+1)e^{-\gamma|\omega|} + (-1)^{\omega}e^{-\gamma|\omega|}-(1+(-1)^{\omega})}{4\pi(1+e^{-2\gamma|\omega|})}\\
-\frac{e^{-\gamma (1-\tilde{\epsilon}_{R})|\omega|}+(-1)^{\omega}e^{-\gamma (1+\tilde{\epsilon}_{L})|\omega|}}{4\pi(1+e^{-2\gamma|\omega|})}. \label{ch4denodde21}\eea
The total spin $S^z$ of this state is $S^z=\frac{1}{2}$. We denote this state by $\ket{\frac{1}{2}}_{E_2}$. By starting with the Bethe equations corresponding to all spin down reference state we have

\bea\nonumber
(2N+1) a(\lambda,1)-a(\lambda,1+\tilde{\epsilon}_{R})-a(\lambda-\pi,1-\tilde{\epsilon}_{L})+a(\lambda-\pi,1)\\-2\pi\delta(\lambda)-2\pi\delta(\lambda-\pi)=2\pi\rho(\lambda)+\sum_{\sigma=\pm}\int a(\lambda+\sigma \mu,2)\rho(\mu)d\mu. \label{ch4logbee21}\eea Following the same procedure as above, we obtain the following distribution for a state with all real $\lambda$ 

\bea\nonumber \hat{\rho}_{\ket{-\frac{1}{2}}_{E_2}}(\omega)=\frac{(2N+1)e^{-\gamma|\omega|} + (-1)^{\omega}e^{-\gamma|\omega|}-(1+(-1)^{\omega})}{4\pi(1+e^{-2\gamma|\omega|})}\\
-\frac{e^{-\gamma (1+\tilde{\epsilon}_{R})|\omega|}+(-1)^{\omega}e^{-\gamma (1-\tilde{\epsilon}_{L})|\omega|}}{4\pi(1+e^{-2\gamma|\omega|})}. \label{ch4denodde22}\eea
The total spin $S^z$ of this state is $S^z=-\frac{1}{2}$. We denote this state by $\ket{-\frac{1}{2}}_{E_2}$. Using (\ref{ch4energy}) we can calculate the energy difference between the two states $\ket{\frac{1}{2}}_{E_2}$ and $\ket{-\frac{1}{2}}_{E_2}$. We have \bea \label{ch4diffene21}E_{\ket{\frac{1}{2}}_{E_2}}-E_{\ket{-\frac{1}{2}}_{E_2}}= -h_L+h_R-2\sinh\gamma\int_{-\pi}^{\pi}a(\lambda,1)\;\delta\rho_{\tiny{\ket{\frac{1}{2}},\ket{-\frac{1}{2}}}}(\lambda)d\lambda.
\eea
Here $\delta\rho_{\tiny{\ket{\frac{1}{2}},\ket{-\frac{1}{2}}}}(\lambda)$ is the difference in the density distributions of the states $\ket{\frac{1}{2}}_{E_2}$ and $\ket{-\frac{1}{2}}_{E_2}$. The expression (\ref{ch4diffene21}) can be written as
 
 \bea  E_{\ket{\frac{1}{2}}_{E_2}}-E_{\ket{-\frac{1}{2}}_{E_2}}= -h_L+h_R+4\pi\sinh\gamma\sum_{\omega=-\infty}^{\infty} \hat{a}(\omega,1)\Delta\hat{\rho}_{\tiny{\ket{\frac{1}{2}},\ket{-\frac{1}{2}}}}(\omega). \eea
 Using (\ref{ch4denodde21}) and (\ref{ch4denodde22}) in the above expression we obtain
 \bea \nonumber E_{\ket{\frac{1}{2}}_{E_2}}-E_{\ket{-\frac{1}{2}}_{E_2}}= -h_L+h_R+ \sinh\gamma \sum_{\omega=-\infty}^{\infty}\; \frac{\sinh(\gamma \tilde{\epsilon}_{R}|\omega|)}{\cosh(\gamma\omega)}e^{-\gamma|\omega|}\\- \sinh\gamma \sum_{\omega=-\infty}^{\infty}\; (-1)^{\omega}\frac{\sinh(\gamma \tilde{\epsilon}_{L}|\omega|)}{\cosh(\gamma\omega)}e^{-\gamma|\omega|},  \eea
 which can be written as \bea E_{\ket{\frac{1}{2}}_{E_2}}-E_{\ket{-\frac{1}{2}}_{E_2}}=-m_L+m'_R.\eea 
The state $\ket{-\frac{1}{2}}_{E_2}$ is the ground state for odd number of sites case. The Bethe equations corresponding to all spin up reference state contain two boundary string solutions $\lambda'_{bs R}=\pm i\gamma(1-\tilde{\epsilon}_{R})$, $\lambda_{bs L'}=\pi\pm i\gamma(1+\tilde{\epsilon}_{L})$. The ground state for even number of sites contains the boundary string $\lambda'_{bs R}$ in addition to all real Bethe roots. We obtain

 \bea\label{ch4denbounde21} \tilde{\rho}_{\ket{0'}_{R, E_2}} (\omega)=\tilde{\rho}_{\ket{\frac{1}{2}}_{E_2}}(\omega)+\Delta\tilde{\rho'}_{R} (\omega).\eea 
The spin of the state containing this boundary string can be calculated using $S^z=\frac{N}{2}-M$, where
\bea \label{ch4nroots26}M=1+\int_{-\pi}^{\pi}\rho_{\ket{0'}_{R, E_2}}(\lambda) d\lambda. \eea
Hence, we obtain $S^z_{\ket{0'}_{R,E_2}}=0$.

 \subsection{$B_2$: Even and odd number of sites}
Region $B_2$ corresponds to the following values of the boundary fields: $h_{c1}<h_R<h_{c2}$, $-h_{c1}<h_L<0$. This region can be further divided into two regions depending on whether $h_{c1}<h_R<\sinh\gamma$ and $\sinh\gamma<h_R<h_{c2}$.

\subsubsection{$h_{c1}<h_R<\sinh\gamma$}

In this case we have $\epsilon_R=-\tilde{\epsilon}_R+i\pi$, $\tilde{\epsilon}_R>1$.  Making the transformation $\tilde{\epsilon}_L\rightarrow -\tilde{\epsilon}_L$, the logarithmic form of Bethe equations corresponding to all spin up reference state take the following form 

\bea\nonumber
(2N+1) a(\lambda,1)+a(\lambda-\pi,\tilde{\epsilon}_{R}-1)-a(\lambda-\pi,1+\tilde{\epsilon}_{L})+a(\lambda-\pi,1)\\-2\pi\delta(\lambda)-2\pi\delta(\lambda-\pi)=2\pi\rho(\lambda)+\sum_{\sigma=\pm}\int a(\lambda+\sigma \mu,2)\rho(\mu)d\mu. \label{ch4logbeb13}\eea
Following the same procedure as above we obtain the ground state for even number of sites 

\bea\nonumber \hat{\rho}_{(\ket{0},B_2,h_R<\sinh\gamma)}(\omega)=\frac{(2N+1)e^{-\gamma|\omega|} + (-1)^{\omega}e^{-\gamma|\omega|}-(1+(-1)^{\omega})}{4\pi(1+e^{-2\gamma|\omega|})}\\
+\frac{(-1)^{\omega}(e^{-\gamma (\tilde{\epsilon}_{R}-1)|\omega|}-e^{-\gamma (1+\tilde{\epsilon}_{L})|\omega|})}{4\pi(1+e^{-2\gamma|\omega|})}. \label{ch4denoddb21}\eea
The total spin $S^z$ can be found using $S^z=\frac{N}{2}-M$ where $M$ is given by (\ref{ch4nroots1}). We obtain $S^z_{(\ket{0}_{B_2},h_R<\sinh\gamma)}=0$. For odd number of sites,  the Bethe equations for all down reference state take the same form as that in the region $A_2$. The density distribution for the ground state is again given by (\ref{ch4denodda22}) with total spin $S^z=-\frac{1}{2}$.

\subsubsection{$\sinh\gamma<h_R<h_{c2}$}

In this case we have $\epsilon_R=-\tilde{\epsilon}_R$, $\tilde{\epsilon}_R>1$. The logarithmic form of Bethe equations corresponding to all spin up reference state take the following form 

\bea\nonumber
(2N+1) a(\lambda,1)+a(\lambda,\tilde{\epsilon}_{R}-1)-a(\lambda-\pi,1+\tilde{\epsilon}_{L})+a(\lambda-\pi,1)\\-2\pi\delta(\lambda)-2\pi\delta(\lambda-\pi)=2\pi\rho(\lambda)+\sum_{\sigma=\pm}\int a(\lambda+\sigma \mu,2)\rho(\mu)d\mu. \label{ch4logbeb14}\eea
Following the same procedure as above we obtain the ground state for even number of sites

\bea\nonumber \hat{\rho}_{(\ket{0},B_2,h_R>\sinh\gamma)}(\omega)=\frac{(2N+1)e^{-\gamma|\omega|} + (-1)^{\omega}e^{-\gamma|\omega|}-(1+(-1)^{\omega})}{4\pi(1+e^{-2\gamma|\omega|})}\\
+\frac{e^{-\gamma (\tilde{\epsilon}_{R}-1)|\omega|}-(-1)^{\omega}e^{-\gamma (1+\tilde{\epsilon}_{L})|\omega|}}{4\pi(1+e^{-2\gamma|\omega|})}. \label{ch4denoddb212}\eea
The total spin $S^z$ can be found using $S^z=\frac{N}{2}-M$ where $M$ is given by (\ref{ch4nroots1}). We obtain again $S^z_{(\ket{0}_{B_1},h_R>\sinh\gamma)}=0$. For even number of sites, the Bethe equations for all down reference state take the same form as that in the region $E_2$. The density distribution for the ground state is again given by (\ref{ch4denodde22}) with total spin $S^z=-\frac{1}{2}$.

\subsection{$D_2$: Even and odd number of sites}

Region $D_2$ corresponds to the following values of the boundary fields: $-h_{c1}>h_L>-h_{c2}$, $h_R>h_{c2}$. This region can be further divided into two regions depending on whether $-h_{c1}>h_L>-\sinh\gamma$ and $-\sinh\gamma>h_L>-h_{c2}$. We make the transformation $\tilde{\epsilon}_{L}\rightarrow -\tilde{\epsilon}_{L}$.

\subsubsection{$-h_{c1}>h_L>-\sinh\gamma$}

Starting with the Bethe equations corresponding to all spin down reference state and considering the state with all real roots, we have

\bea\nonumber
(2N+1) a(\lambda,1)-a(\lambda,1+\tilde{\epsilon}_{R})+a(\lambda-\pi,\tilde{\epsilon}_{L}-1)+a(\lambda-\pi,1)\\-2\pi\delta(\lambda)-2\pi\delta(\lambda-\pi)=2\pi\rho(\lambda)+\sum_{\sigma=\pm}\int a(\lambda+\sigma \mu,2)\rho(\mu)d\mu. \label{ch4logbed21}\eea
Following the same procedure as above, we obtain the following distribution for a state with all real $\lambda$ 

\bea\nonumber\hat{\rho}_{\ket{0}_{D_2}}(\omega)=\frac{(2N+1)e^{-\gamma|\omega|} + (-1)^{\omega}e^{-\gamma|\omega|}-(1+(-1)^{\omega})}{4\pi(1+e^{-2\gamma|\omega|})} \\
-\frac{e^{-\gamma (1+\tilde{\epsilon}_{R})|\omega|}-(-1)^{\omega}e^{-\gamma (\tilde{\epsilon}_{L}-1)|\omega|}}{4\pi(1+e^{-2\gamma|\omega|})}.  \label{ch4denoddd22}\eea
We obtain $S^z_{\ket{0}_{D_2,|h_L|<\sinh\gamma}}=0$.  The ground state for even number of sites is given by  $\ket{0}_{D_2,|h_L|<\sinh\gamma}$.
\subsubsection{$\sinh\gamma<h_L<h_{c2}$}

Starting with the Bethe equations corresponding to all spin down reference state and considering the state with all real roots, we have

\bea\nonumber
(2N+1) a(\lambda,1)-a(\lambda,1+\tilde{\epsilon}_{R})+a(\lambda,\tilde{\epsilon}_{L}-1)+a(\lambda-\pi,1)\\-2\pi\delta(\lambda)-2\pi\delta(\lambda-\pi)=2\pi\rho(\lambda)+\sum_{\sigma=\pm}\int a(\lambda+\sigma \mu,2)\rho(\mu)d\mu. \label{ch4logbed23}\eea
Following the same procedure as above, we obtain the following distribution for a state with all real $\lambda$ 

\bea\nonumber\hat{\rho}_{\ket{0}_{D_2}}(\omega)=\frac{(2N+1)e^{-\gamma|\omega|} + (-1)^{\omega}e^{-\gamma|\omega|}-(1+(-1)^{\omega})}{4\pi(1+e^{-2\gamma|\omega|})} \\
-\frac{e^{-\gamma (1+\tilde{\epsilon}_{R})|\omega|}-e^{-\gamma (\tilde{\epsilon}_{L}-1)|\omega|}}{4\pi(1+e^{-2\gamma|\omega|})}.  \label{ch4denoddd24}\eea
Hence, we obtain $S^z_{\ket{0}_{D_2,|h_L|>\sinh\gamma}}=0$.  The ground state for even number of sites is given by  $\ket{0}_{D_2,|h_L|>\sinh\gamma}$. For odd number of sites case, we obtain the ground state by adding a spinon to $\ket{0}_{D_2,|h_L|<\sinh\gamma}$, $\ket{0}_{D_2,|h_L|>\sinh\gamma}$ in the respective cases.

\end{widetext}

\end{document}